\documentclass[%
reprint,
 amsmath,amssymb,
 aps,
]{revtex4-2}

\usepackage{graphicx}% Include figure files
\usepackage{dcolumn}% Align table columns on decimal point
\usepackage{bm}% bold math
\usepackage[usenames,dvipsnames]{xcolor}
\usepackage{tcolorbox}
\usepackage{tabularx}
\usepackage{array}
\usepackage{colortbl}
\tcbuselibrary{skins}

\usepackage{appendix}

\usepackage{multirow}

\usepackage{ragged2e}

\begin{document}

\preprint{APS/123-QED}

\title{Fractional Topology in Interacting 1D Superconductors}

\author{Frederick del Pozo$^{*}$}
\author{Lo\" ic Herviou$^{\dagger}$}
\author{Karyn Le Hur$^{*}$} 
\affiliation{\vspace{1em}$^{*}$Centre de Physique Théorique, École Polytechnique (IP Paris), 91128 Palaiseau, France}
\affiliation{$^{\dagger}$SB IPHYS, EPFL,
Rte de la Sorge 
CH-1015 Lausanne}

\date{\today}

\begin{abstract}
We investigate the topological phases of two one-dimensional (1D) interacting superconducting wires and propose topological markers directly measurable from ground state correlation functions. These quantities remain powerful tools in the presence of couplings and interactions. We show with the \emph{density matrix renormalization group} that the \emph{double critical Ising (DCI)} phase discovered in \cite{Herviou_2016} is a fractional topological phase with gapless Majorana modes in the bulk, and a one-half topological invariant per wire. Using both numerics and quantum field theoretical methods, we show that the phase diagram remains stable in the presence of an inter-wire hopping amplitude $t_{\bot}$ at length scales below $\sim 1/t_{\bot}$. A large inter-wire hopping amplitude results in the emergence of two integer topological phases, stable also at large interactions. They host one edge mode per boundary shared between both wires. At large interactions, the two wires are described by Mott physics, with the $t_{\bot}$ hopping amplitude resulting in a paramagnetic order.
\end{abstract}

\maketitle

%\tableofcontents

\section{\label{sec:level1}Introduction\protect}

Quantum Mechanics 
% is certainly counter intuitive. One would expect the interference of many particles to wash out the quantum properties of many-body dynamics as a result of decoherence effects, yet quantum mechanics survives for a large range of interacting systems. This 
gives rise to important phenomena such as Bose-Einstein condensation and its charged analog superconductivity, wherein a coherence builds up between the constituents in the bulk of the material. Such phases of matter are governed by quantum phase transitions, wherein even at zero temperature quantum fluctuations become relevant enough to drive, or destabilize, an underlying ordering. Already since the 70s, it was found that such transitions can occur without the inherent breaking of symmetries, and, instead, the phases of matter are distinguished by winding numbers or more generally by \emph{topological invariants} \cite{KT,TKNN}. Whilst these are usually found to be integer values \cite{QHE}, in the presence of interactions fractional topological states of matter have also been predicted as for example in the fractional quantum hall effect (FQHE) \cite{Laughlin}, associated with the direct observation of fractional charges \cite{Stormer, Kapfer_2019}.\\

Due to their robustness against perturbations and impurities, topological materials have seen a growing interest in recent years. Particularly the applications of topological superconductors and insulators in quantum circuits and computers \cite{Alicea_2012} drive both experimental and theoretical developments. The key reason these topological materials are particularly promising is the existence of exotic anyonic edge modes, which have been proposed as candidates to build resilient, large-scale quantum computers \cite{Alicea_2011}. Since the seminal work by Kitaev \cite{Kitaev_2001}, one-dimensional spinless superconductors are predicted to host elusive Majorana fermions \cite{Majorana_OG} on their edges, which are essentially the real and imaginary parts of a complex Dirac fermion. Majoranas are as elusive as they are sought for. Whilst their potential applications to realizing low-error quantum computation \cite{Google_Maj,Alicea_2011} have stimulated vigorous research, the scientific community has yet to reach a consensus on whether or not Majorana edge modes have been detected experimentally. Yet, there is recent hope that their discovery is on the horizon \cite{HopeMajorana}, paving the way for further interesting applications, for example with Majorana wire heterostructures. As demonstrated in \cite{Alicea_2011}, networks of spinless p-wave superconducting nano-wires offer a promising platform to realise anyonic exchange statistics. Due to the proximity of the wires in these heterostructures, hopping and super-conductive pairing terms between different wires will naturally occur in realistic setups. Already two coupled wires, i.e. ladders, are found to have a rich phase diagram without interactions \cite{ladder1, ladder2}. Also in the quasi two-dimensional limit, additional hopping and pairing terms can have a substantial impact on the macroscopic properties, as was demonstrated for example in \cite{Yang_2020} for a quasi-two-dimensional grid of Majorana wires. It was shown that in the presence of cross-wire couplings, it is possible to design \emph{(p+ip)} superconductivity by threading appropriate fluxes through each unit cell. \\

The theory for proximity-induced topological superconductivity (SC) focuses strongly on non-interacting electron models \cite{Alicea_2011}. However, realistic materials will necessarily be exposed to both internal and external interactions, which may give rise to previously unknown transitions \cite{Katsura_2015, Alicea_interacting}. For example, it was found in \cite{Ledermann_2000} that due to the interplay between interactions and inter-wire hopping amplitude a transition to a superconducting state can occur, for two chains of spinless fermions without SC-pairing. Another fascinating effect often found in interacting systems is the fractionalization of the underlying degrees of freedom, for example, the fractionalization of charge in the FQHE \cite{Kapfer_2019} or also in the case of quantum wires \cite{Chargefract_leHur}. Another instance of fractionalization was discovered in the case of two interacting Kitaev wires \cite{Herviou_2016}, wherein the gapless \emph{double critical Ising} phase was found to host free Majoranas in the bulk. Against the backdrop of topological materials in modern technology, developing a deeper understanding of the DCI phase and its Majorana physics is therefore relevant. Recent advances in unraveling the phase diagram of two interacting Bloch spheres \cite{Hutchinson_2021, lehur_new} revealed also the emergence of a fractional topological phase at large interactions. In fact, for the single Kitaev wire, there exists a well-established map onto the Bloch sphere through the Bogoliubov-De Gennes
representation \cite{Sato_2017,lehur_new}. Thus these recent findings stimulate us to investigate the DCI phase further, and in particular its link to the fractional Bloch sphere phase in \cite{Hutchinson_2021}.\\

The extensions beyond non-interacting chains are plentiful and there is a lot of ground to cover. In this paper we focus in particular on deepening our understanding of the DCI phase, its link to fractional topology as well as the effects of an inter-wire hopping amplitude $t_{\bot}$ on the phase diagram found in \cite{Herviou_2016}. We first propose in Sec. \ref{sec:Cherns} topological invariants for the one-dimensional superconducting Kitaev chains, which by mapping onto the Bloch sphere are found to be direct analogs of the TKNN invariant \cite{TKNN}. We show that these quantities, defined through two-point correlation functions, can be extended to two or more wires and that they remain powerful tools and topological markers to study the phases of coupled wires also in the presence of interactions. We evaluate these correlation functions using Density Matrix Renormalization Group (DMRG) calculations \cite{PhysRevLett.69.2863,PhysRevB.48.10345, Schollw_ck_2011}. In Sec. \ref{subsec:sw_critical_theory}, we elaborate on the $C=1/2$ topological invariant(s) associated with the critical theory of the Majorana fermions at the topological quantum phase transition for one Kitaev wire, and we establish a relation with the Bloch sphere. In Sec. \ref{sec:DCI}, we first review briefly the regions in the phase diagram of two interacting superconducting wires \cite{Herviou_2016}. We present an approach to understanding the underlying quantum field theory (QFT) of both the single and double critical Ising (DCI) phases \cite{Herviou_2016} in terms of chiral bulk modes, intimately related to the critical theory of a single Kitaev wire. Then, in Sec. \ref{sec:Fractional}, we show with DMRG calculations that our topological markers take on fractional values of $C = 1/2$ in the DCI phase, which establishes a clear link between the topological phases of two interacting Bloch spheres \cite{Hutchinson_2021}. Furthermore, we argue that this also strengthens the notion that the DCI phase is described by two $c = 1/2$ critical models per wire where now $c$ refers to the central charge \cite{PhysRevLett.56.746, tsvelik_2003, Bosonization_and_strongly, itzykson_drouffe_1989} of the model. \\

Finally in Sec. \ref{sec:TP} we also investigate both numerically and analytically the phase diagram in \cite{Herviou_2016} in the presence of an inter-wire hopping amplitude $t_{\bot}$, similar to the ladders studied in \cite{ladder1}, to determine the stability of the coupled-wire phases towards more experimentally realistic setups. It is perhaps relevant to mention here that the $t_{\bot}$ term does not have a simple correspondence on the Bloch spheres' model since mapping fermions onto spins-$\frac{1}{2}$ through the Jordan-Wigner transformation will result in ``strings" accompanying this operator in the spin language. Therefore, it is worthwhile studying the effect of such a perturbation in the two-wires' model. The strength of the $t_{\perp}$ term compared to the Coulomb interaction may be adjusted by fixing the distance between the two wires (a hopping term is supposed to decay exponentially with distance from a Wentzel-Kramers-Brillouin picture). We find that for perturbative values of $t_{\bot}$ the DCI phase and phase diagram remain robust. We study the phase diagram for larger values of the inter-wires hopping term, as well as interactions. We also address the case for prominent interactions between both wires, which  gives rise to Mott physics \cite{Herviou_2016} in the $|g| \rightarrow \infty$ limit. Our results are finally summarized in \ref{sec:Conclusion}. In Appendices, we present additional information on DMRG and quantum field theory.

\section{Topological markers from correlation functions}\label{sec:Cherns}
The aim of this section is to define topological invariants for Kitaev p-wave spin-polarized superconducting wires \cite{Kitaev_2001}, and express these quantities in terms of real-space correlation functions. First, we remind the reader of the Kitaev wire \cite{Kitaev_2001}, and associated Bogoliubov-De Gennes formalism, which due to the particle-hole symmetry of the Kitaev wire allows a direct mapping onto a single Bloch sphere \cite{Le_Hur_2022,lehur_new, Sato_2017}. From this, a definition of a Chern number {\emph{\`a la}} \cite{Hutchinson_2021} leads directly to the desired real-space correlation functions. Then, by similar considerations, we argue that in the case of two weakly interacting wires in the sense of \cite{Herviou_2016}, a mapping on two-coupled Bloch spheres remains sensible \cite{lehur_new}. Whilst the quantization of the invariants is no longer guaranteed, we show numerically using DMRG that these Chern numbers remain sensible markers to characterize and investigate the topological phases in the presence of interactions.

\subsection{Review of the Kitaev wire and its topological phase diagram}
\begin{figure}[h!]
    \centering
    \includegraphics[width = 0.5\textwidth]{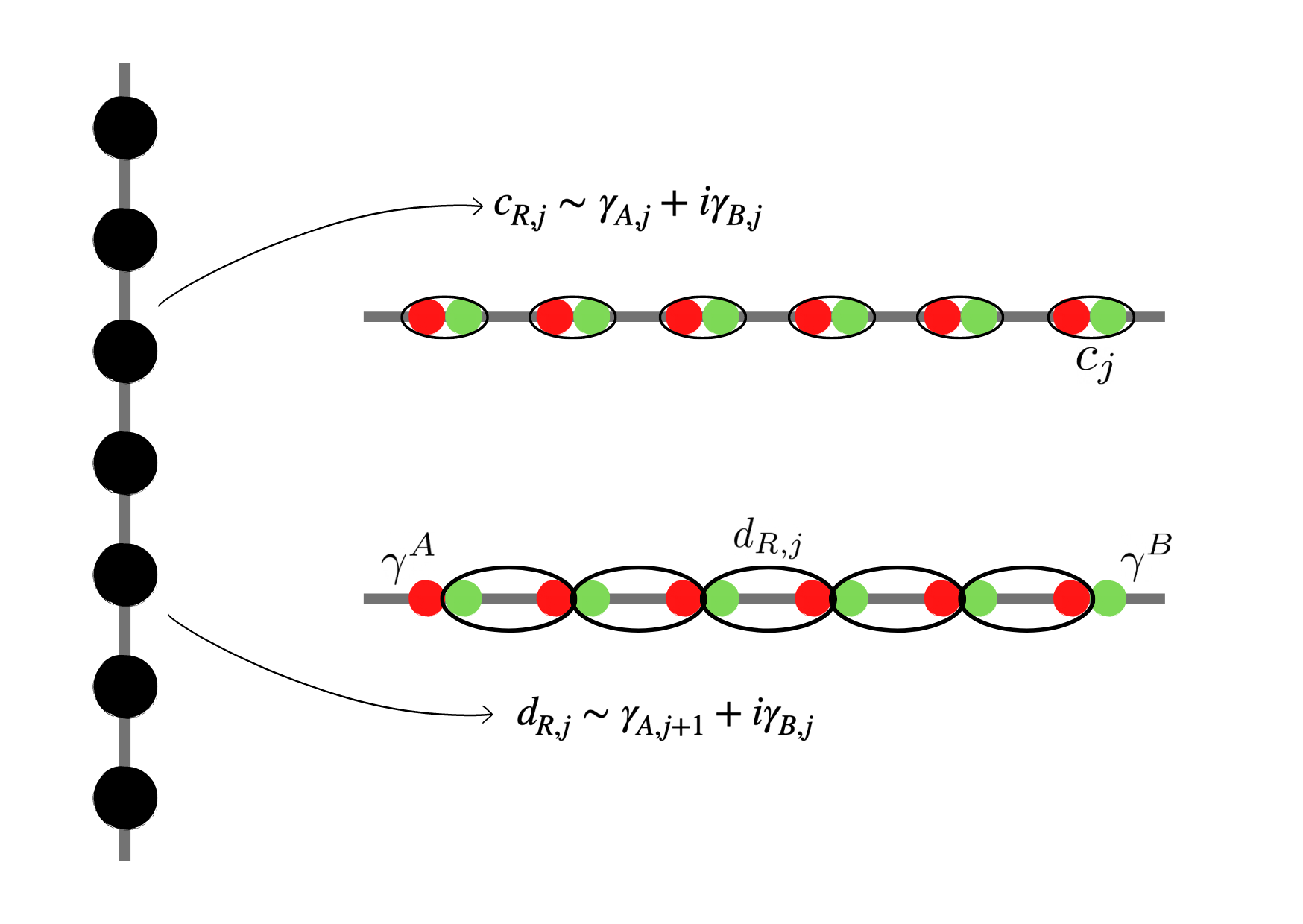}
    \caption{ The two distinct patterns of Majorana fermions in the trivial (upper) and topological (lower) phases of a Kitaev wire. We refer to Ref. \cite{Alicea_2012} for a more in-depth discussion { and review of the topological phases of the Kitaev wire.}}
    \label{fig:Patterns}
\end{figure}

% {\color{black} 
% One central result of this section is to introduce a relation between the topological properties 
% and correlation functions of the ground state for one-dimensional superconducting wires associated with Eqs. (\ref{Combined_pole_equations}). This will allow us to evaluate the topological markers of the wire(s) using DMRG. Then, within this section,s we also motivate the correspondence between two wires and two interacting Bloch spheres.}\\ 
In this article we investigate the topological phases of (interacting and coupled) Kitaev wires of spinless fermions. The Kitaev wire is defined by the following Hamiltonian \cite{Kitaev_2001, Alicea_2012}
\begin{equation}\label{Ham_c}
\begin{aligned}
              H_{K} = -&\mu\sum_{j} n_{j} - t\sum_{j} c^{\dagger}_{j}c_{j+1}+ \text{h.c.}\\  +& \triangle e^{i\varphi}\sum_{j} c^{\dagger}_{j}c^{\dagger}_{j+1} + \text{h.c.}.
\end{aligned}
\end{equation}
Here $\mu$ is the chemical potential acting globally on both wires, $t$ is the hopping amplitude,  $\Delta e^{i\varphi}$ the superconducting pairing-strength with phase $\varphi$ and $n_{c,j} = c^{\dagger}_{j}c_{j}$. The single wire has two topologically distinct extended phases. A topological phase transition occurs at the critical chemical potentials $\mu = \pm 2t$. In his seminal work \cite{Kitaev_2001}, Kitaev developed an intuitive picture of these phases, by decomposing the original $c$-fermions into their Majorana constituents \cite{Sato_2017}
\begin{equation}\label{Majdecomp}
    c_{j} = \frac{1}{2} \left(\gamma_{A,j} + i\gamma_{B,j}\right).
\end{equation}
The Hamiltonian \eqref{Ham_c}, for $\varphi = 0$, takes the following form when expressed in terms of the Majorana fermions
\begin{equation}\label{Ham_maj}
\begin{aligned}
    H_{K} = \ &\sum_{j} i\frac{t - \Delta}{2} \gamma_{B,j}\gamma_{A,j+1} + i\frac{t + \Delta}{2} \gamma_{B,j+1}\gamma_{A,j} \\ -&i\frac{ \mu}{2} \sum_{j} \gamma_{A,j}\gamma_{B,j}. 
\end{aligned}
\end{equation}
The two topologically distinct phases can then be understood in the two \emph{patterns} of Majoranas which emerge, depending on the chemical potential. For a pictorial representation of the two distinct patterns, see figure \ref{fig:Patterns}.
In the trivial phases with $|\mu| > 2t$, the on-site pairing of Majoranas is favoured, thus in the $t =\Delta = 0$ limit we find the upper (trivial) pattern in \eqref{Majdecomp}. In the topological phase $|\mu| <2t$, the nearest-neighbour pairing will dominate, such that for $t = \Delta$ and $\mu = 0$ the lower (topological) pattern in \eqref{Majdecomp} emerges. This can be understood in terms of the following non-local fermions
\begin{equation}\label{left-Kitaev}
    d_{R,j} = \frac{1}{2}\left( \gamma_{A,j+1}+ i\gamma_{B,j}\right).
\end{equation}
For open boundary conditions (OBCs) this results in zero-energy ``dangling edge modes" which make the ground state doubly degenerate. \\

Whilst the emergence of edge modes and ground state degeneracy are both hallmarks of a topological transition, it is the definition of global and robust invariant which makes the direct link to topology in the mathematical sense.
% Generally speaking, a phase transition represents a change in the underlying properties of a many-body system, manifesting itself in different physical properties. Examples of this are the melting of ice to its liquid state or the ordering of the spins in Heisenberg ferromagnets in an external magnetic field.  A common feature of these examples is a change in symmetry as a result of the phase transition. This is not a necessary feature of phase transitions, as demonstrated in the 70s by Berezinskii, Kosterlitz, and Thouless \cite{KT}. They found that there exist other classes of transitions determined instead by a \emph{topological} invariant. 
This link is perhaps best understood when considering the seminal work in \cite{TKNN}, where the $\mathbb{Z}$ valued \emph{TKNN invariant} (or First Chern number) is defined from the Berry connection \cite{Berry} ${a}^{n}_j(\mathbf{k})=i\langle n, \mathbf{k}| \frac{\partial}{\partial k_j}| n, \mathbf{k}\rangle$ of the eigenstates of each band \cite{Sato_2017,Kaufmann_2016}
\begin{equation}\label{TKNN}
c_{1} = \sum_{\text{n'th filled band}} \int_{\mathbb{T}^{2}} d k^{2}\left(\frac{\partial a_{y}^{n}}{\partial k_{x}}-\frac{\partial a_{x}^{n}}{\partial k_{y}}\right).
\end{equation}
{The Chern number is calculated on the ground state by summing over the $n$ lowest filled bands $|n, \mathbf{k}\rangle$. It is quantized and can only take integer values, and is a direct analogue of the Gauss-Bonnet theorem for surfaces: The expression in brackets defines a curvature 2-form or tensor $\Omega_{k_{x},k_{y}}$ on the Brillouin zone which is a 2-torus $\mathbb{T}^{2}$ due to periodic boundary conditions (PBCs) in both directions.
% Due to the periodic boundary conditions (PBCs) the momenta $k_{x}$ and $k_{y}$ are good quantum numbers, and thus the ground states are labeled as $|n, \mathbf{k}\rangle$. The full Hamiltonian $H$ can be reduced to the single-particle Hamiltonian $\mathcal{H}_{k_{x},k_{y}}$, where $k_{x},k_{y}$ are treated as external parameters. 
The TKNN-invariant defined above is not suitable for distinguishing all classes of topological systems. For example, the Chern number \eqref{TKNN} is not invariant under \emph{time-reversal-symmetry} (TRS), and hence vanishes in such systems. Instead, for example in the two-dimensional \emph{quantum spin Hall} phase, Kane and Mele discovered \cite{Kane_2005} that the appropriate topological number is a $\mathbb{Z}_{2}$ invariant. This has been generalized also to other systems \cite{Fu_2007}, and is now often referred to as the \emph{Fu-Kane-Mele} invariant. For non-interacting systems the different possible topological phases have been classified by dimension and symmetries \cite{Chiu_2016} in various ``periodic tables", \emph{cf.} Altand-Zirnbauer classification \cite{Altland_1997} or Kitaev \cite{Kitaev_2009}. The Kitaev wire described by \eqref{Ham_c} is characterized by a $\mathbb{Z}_{2}$ invariant. \\

In momentum space and the Bogoliubov-De Gennes Hamiltonian representation (i.e. spinless
Nambu basis) defined through } $\psi^{\dagger}_{k} = \left( c^{\dagger}_{k}, c_{-k}\right)$, the single-particle  Hamiltonian in \eqref{Ham_c} can be written as the $2\times 2$ Matrix 
\begin{equation}\label{Ham_Nambu}
    H_{K} = \sum_{k} \psi^{\dagger}_{k}
    \begin{pmatrix}\epsilon_{k} & \Delta_{k} \\ 
  \Delta^{*}_{k} & -\epsilon_{k}
\end{pmatrix}\psi_{k} = \sum_k \psi^{\dagger}_{k}  \mathcal{H}_{k} \psi_k,
\end{equation}
where we define $\epsilon_{k} = -\left(\frac{\mu}{2} + t\cos\left(ka\right)\right)$ and SC pairing $\Delta_{k} = i\triangle e^{i\varphi} \sin\left(ka\right)$. We choose the Fourier transform on the $N$-site chain with $x = ja$ with the convention
\begin{equation}
    c_{j} = \frac{1}{\sqrt{N}}\sum_{k \in BZ} e^{ikx}c_{k}.
\end{equation}
The two distinct topological phases can then be determined from the signs of the kinetic term $\epsilon_{k}$ at the high-symmetry points $k = 0$ and $k = \pi/a$ \cite{Sato_2017}, labeled by $\delta_{ak = 0,\pi}$ respectively. The trivial phases result in $\delta_{0} = \delta_{\pi} = \pm 1$. The topological ones instead have $\delta_{0} = - \delta_{\pi} = \pm 1$, depending on the signs of $t$ and $\mu$. This defines a $\mathbb{Z}_{2}$-invariant $\nu$, which in this case is introduced in the well-known way \cite{Kitaev_2001,Sato_2017}
\begin{equation}\label{Z2Kitaev}
\nu = \left(-1\right)^{\delta_{0}\delta_{\pi}}.
\end{equation} 
% The $\mathbb{Z}_{2}$ invariant defined in \eqref{Z2Kitaev} is the appropriate topological number for the (short-range) Kitaev chain. 
Long-range hopping and pairing terms may result in higher topological invariants \cite{Herviou_2017}, which are then the \emph{winding numbers} $m$. However, as we consider only nearest-neighbor hopping and pairing terms, \emph{cf.} eq. (\ref{Ham_c}), the winding numbers are restricted to $0$, $\pm 1$ (in agreement with the structure of Majorana fermions at the edges). The $\mathbb{Z}_2$ number in \eqref{Z2Kitaev} is thus enough to classify the system.

\subsection{Chern number from real-space correlation functions}\label{subsec_cherns_for_single_wire}
% In the short-range case both the $\mathbb{Z}_{2}$ invariant $\nu$ and integer winding numbers $m$ coincide.

We show in the following how to define a (\emph{particle}) Chern number $C = 0,1$, which is directly measurable from real-space correlation functions of the wire.  \\

Central to our argument is the duality between the momentum space Hamiltonian in \eqref{Ham_Nambu} and a Bloch sphere interacting with an external field \cite{Le_Hur_2022,lehur_new}, \emph{cf.} the discussion in appendix \ref{Appendix:Topo_BS}. From the momentum space Hamiltonian \eqref{Ham_Nambu}, the single-particle (Bogoliubov-De Gennes) Hamiltonian $\mathcal{H}_{k}$ is a complex $2\times2$ matrix. By introducing (pseudo-)spin $\vec{S}_{k}$ and $\vec{d}_{k}$ vector
\begin{equation}
    \vec{S} = \begin{pmatrix}
        c^{\dagger}_{k}c^{\dagger}_{-k} + c_{-k}c_{k} \\ 
   - i\left(c^{\dagger}_{k}c^{\dagger}_{-k} - c_{-k}c_{k}\right)\\
        c^{\dagger}_{k}c_{k}- c_{-k}c^{\dagger}_{-k}
    \end{pmatrix} , \    \vec{d}_{k} = \begin{pmatrix}-\frac{\Delta_{k}+\Delta_{k}^*}{2} \\ \frac{\Delta_{k} - \Delta^{*}_{k}}{2i} \\ -\epsilon_{k}
    \end{pmatrix},
\end{equation}
we may write the single-particle Hamiltonian reminiscent of a spin in a magnetic field \cite{Sato_2017}
\begin{equation}
    \mathcal{H}_{k} = -\vec{d}_{k}\cdot \vec{S}_{k}.
\end{equation}
Here $\vec{d}_{k}$ acts like a ``magnetic field" on the pseudo-spin $\vec{S}_{k}$, \emph{cf.} discussion in \ref{Appendix:Topo_BS}. Together with the super-conducting phase $\varphi$, which remains a free parameter,
we map the vector $\vec{d}_{k}$ onto the two-sphere, interpreting the momentum label $ka$ and super-conducting phase $\varphi$ as the angular coordinates $\vartheta$ and $\tilde{\varphi}$ in \eqref{SpinBfield}. 
Through the definitions in Appendix \ref{Appendix:Topo_BS} on the sphere this results in \cite{Le_Hur_2022, lehur_new} 
the identifications 
\begin{eqnarray}
\label{angle}
 \cos\left({\vartheta}_{k}\right) &=& \frac{2t\cos(ka)+\mu}{E(ka)}, \\ \nonumber
 \sin\left({\vartheta}_{k}\right)e^{-i\tilde{\varphi}} &=& -\frac{i\Delta e^{i\varphi} 2\sin(ka)}{E(ka)}
 \end{eqnarray}
 where $E(ka) = \sqrt{\left(\mu + 2t\cos\left(ka\right)\right)^{2} + 4\triangle^{2}\sin\left(ka\right)^{2}}$. To be more precise, we introduce the $\vec{d}$-vector as $\vec{d}=|\vec{d}|(\cos{\vartheta}\sin\tilde{\varphi},\sin{\vartheta}\sin\tilde{\varphi},\cos\vartheta)$
 with the energetic correspondence $E(ka)=2|\vec{d}|$. There is a correspondence between the two eigenstates of the spin-$\frac{1}{2}$ particle and the definitions of the quasiparticles in the wire model. Here, $\tilde{\varphi}$ represents the azimuthal angle on the sphere. 
Fixing $\mu = 0$ and $t = \Delta$ we observe that $\vartheta_{k} = ka$ (and $e^{-i \tilde{\varphi}} = -i e^{i \varphi}$). On the sphere, ${\vartheta}_{k}\in [0;\pi]$ such 
that this will effectively correspond to a half Brillouin zone on the lattice due to the particle-hole symmetry (PHS) of the Kitaev model. Conversely if $\mu \gg t = \Delta$, then ${\vartheta}_{k} = \text{const.}$. \\

The $2\times 2$ Hamiltonian is diagonalized by the Bogoliubov quasi-particles $\eta^{\dagger}_{k} = u^{*}_{k}c_{k}^{\dagger} + v^{*}_{k}c_{-k}$ similarly as in the Bardeen Cooper and Schrieffer model \cite{PhysRev.112.1900}. For a general phase $\varphi$, we can e.g. define the quasiparticle operator $\eta^{\dagger}$ 
associated to an occupied quasiparticle state and to the lowest energy eigenstate of the spin-$\frac{1}{2}$ in (\ref{Bloch_GS}). %The Bogoliubov de Gennes (BdG) transformation can then be defined through the lowest energy eigenvector of \eqref{Ham_Nambu} with Hamiltonian $-E(ka)\eta^{\dagger}_k \eta_k$. This results in 
The Bogoliubov de Gennes (BdG) transformation then diagonalizes the Hamiltonian, yielding two quasi-particles $\eta^{\pm}$ corresponding to the upper and lower band. As particle-hole symmetry relates both, the label $\pm$ can be dropped and the BdG transformation defined through the lowest energy eigenvector of \eqref{Ham_Nambu} 
with Hamiltonian $-E(ka)\eta^{\dagger}_k \eta_k$. This results in 
\begin{equation}
\label{eta}
    \eta^{\dagger}_{k} = \cos\left({\vartheta}_{k}/2\right) c^{\dagger}_{k} + ie^{-i\varphi}\sin\left({\vartheta}_{k}/2\right)c_{-k}.
\end{equation}
The $|BCS\rangle$ ground state can be defined for the filled energy states as $\eta_{k}\eta^{\dagger}_{k} |BCS\rangle = 0$. It has the following explicit expression \cite{Herviou_2016}
\begin{equation}\label{BCS_WF}
\begin{aligned}
|\mathrm{BCS}\rangle = &\Big( \left( \delta _ { \mu < - 2 t } + ( 1 - \delta _ { \mu < - 2 t } ) c _ { 0 } ^ { \dagger } \right) \times\\
& \prod_{k>0}^{k<\frac{\pi}{a}}\left(\sin\left({\vartheta}_k / 2\right)-ie^{+i\varphi}\cos \left({\vartheta}_k / 2\right) c_k^{\dagger} c_{-k}^{\dagger}\right) \\ & \times\left(\delta_{\mu<2 t}+\left(1-\delta_{\mu<2 t}\right) c_\pi^{\dagger}\right)\Big) |0\rangle .
\end{aligned}
\end{equation}
The points $k=0$ and $ka=\mp \pi$ require some care where the pairing function goes to zero. We have adjusted the $\delta$ functions to correspond to filled or empty states according to the value of the chemical potential in agreement with the matrix (\ref{Ham_Nambu}).
From the Bogoliubov-De Gennes quasi-particle basis, the link to the Bloch sphere eigenvector $|GS\rangle$ in Appendix \ref{Appendix:Topo_BS} for each $k$ label is made by taking $c^{\dagger}_{k} \longrightarrow |\uparrow\rangle$ and $c_{-k} \longrightarrow |\downarrow\rangle$ \cite{Le_Hur_2022,lehur_new}.\\
% From the Bogoliubov-De Gennes quasi-particle basis, the link to the Bloch sphere eigenvector $|GS\rangle$ in Appendix \ref{Appendix:Topo_BS} for each $k$ label can be made \cite{Le_Hur_2022,lehur_new}: acting on the Fermi sea $|FS\rangle$ \cite{https://doi.org/10.1002/andp.19985100401}, it follows $c^{\dagger}_{k}|FS\rangle = |\uparrow\rangle$ or a hole $c_{-k}|FS\rangle = |\downarrow\rangle$ for $k\geq 0$.}

As was first shown in \cite{Hutchinson_2021} for (coupled-) Bloch spheres, the quantized Chern number in \eqref{TKNN} can be written from the polarizations of the spin at the two poles of the Bloch sphere:
\begin{equation}\label{Cherns_JH}
    C = \frac{1}{2}\left( \langle S^{z}\left(\vartheta = 0\right)\rangle - \langle S^{z}\left(\vartheta = \pi\right)\rangle\right).
\end{equation}
% The link to the \emph{TKNN}-invariant can be made very generally, and purely from geometric arguments independently of the exact ground state wave function. 
% As we detail in the next Section \ref{subsec_cherns_for_single_wire},
% this $\mathbb{Z}$ formulation of the topological number is also related through the $\mathbb{Z}_2$ topological number defined above when measuring the product of the spin magnetizations at the poles \cite{lehur_new}. 
For two spheres the well-definedness of partial Chern numbers \footnote{ie. for each individual sphere} was demonstrated for a wide range of interactions, and resulted in the discovery of a fractional geometric phase at comparably large interactions \cite{Hutchinson_2021}. By mapping $\left(k,\varphi\right)$ onto the Bloch sphere (two-sphere), we can define a Chern number $C$  \emph{\`a la} \cite{Hutchinson_2021} by simply evaluating the $z-$Spin operator at the ``poles" $k = 0$ and $ka = \pi$ 
\begin{equation}\label{Cherns_nambu}
    C = \frac{1}{2}\left(\langle S^{z}_{ka = 0}\rangle- \langle S^{z}_{ka = \pi}\rangle \right).
\end{equation}
Acting with $S^{z}$ on the BCS ground state we explicitly find
$\langle S^{z}_{k}\rangle = \cos\left(\vartheta_{k}\right)$, which is the Ehrenfest theorem for a spin half particle in a radial magnetic field \cite{Le_Hur_2022,lehur_new}.
For $\mu \gg t$ and thus $\vartheta_{k} = const.$, we immediately find $C = 0$. In the $\mu = 0$ we have $\vartheta_{k} = ka$, and hence $C = \frac{1}{2} \left( \cos\left(0\right) - \cos\left(\pi\right)\right) = 1$.
% which is \emph{not} in contradiction to the notion that the short-range Kitaev wire has \emph{total} Chern number $C = 0$, due to its particle-hole symmetry. 
This splitting results in two bands, the (quasi-)particles and holes, such that $C_{tot} = C_{p} + C_{h}$. From the PHS it also follows that $C_{p} = -C_{h}$,  which can be readily seen by the formula $ C_{h} = \frac{1}{2}\left( \langle S^{z}_{ka = \pi}\rangle - \langle S^{z}_{ka = 2\pi}\rangle \right)$ and the fact that $ka = 2\pi$ is identified with $k = 0$. Thus the total Chern number $C_{tot} = 0$, in agreement with the literature. We also recover the same $\mathbb{Z}_{2}$ invariant as defined in equation \eqref{Z2Kitaev} \cite{lehur_new}, by taking the product $\langle S^{z}_{ka = 0}\rangle\cdot\langle S^{z}_{ka = \pi}\rangle$.\\
%In what follows we present how such topological numbers for the Kitaev wire can be defined, and offer a novel approach based on real-space correlation functions of the ground state.

We now show that the particle Chern number $C$ defined in \eqref{Cherns_nambu} can recast as a physically and experimentally sensible quantity, as it can be expressed in terms of real-space correlation functions of the wire.

\subsection{Measuring topology from correlation functions}\label{subsec:measuring_topology}

The first step is to represent the $z$-spin operators $S^{z}_{ka = 0,\pi}$ in terms of the real-space (spinless-) fermionic operators $c_{j}$ and $c^{\dagger}_{j}$. For this, we perform a Fourier transform of the pseudo-spin $ S_{k}^{z}$ yielding
\begin{equation}
\begin{aligned}
    \mathfrak{S}^{z}_{i} \equiv& \frac{1}{\sqrt{N}}\sum_{k} e^{ikx_{i}}  S^{z}_{k}  \\  
    = & \frac{1}{\sqrt{N}^{3}} \sum_{k\in BZ; j,r} e^{ik\left(x_{i} + x_{j}- x_{r}\right)} c^{\dagger}_{j}c_{r} \\ -& \frac{1}{\sqrt{N}^{3}} \sum_{k\in BZ; j,r} e^{-ik\left( x_{r} - x_{j}-x_{i}\right)} c_{j}c^{\dagger}_{r}.
\end{aligned}
\end{equation}
Performing the sum over the momentum label $k\in BZ$ reduces the above expression to
\begin{equation}\label{SzFrak}
      \mathfrak{S}^{z}_{i} = \frac{1}{\sqrt{N}}\sum_{j}\left( c^{\dagger}_{j}c_{j+i} - c_{j}c^{\dagger}_{j+i}\right).
\end{equation}
As can be seen, for $i>0$ these operators are intrinsically related to the amplitudes of $i$'th neighbour hopping, whilst for $i = 0$ it is found to be simply $\frac{1}{N}\sum_{j} \left(2n_{j} - 1\right)$. Since the momenta $k = 0$ and $k = \pi/a$ are special in the sense that $e^{ikx} = 1$ and $e^{ikx} = \left(-1\right)^{x}$ respectively, the ``backwards" Fourier transform is especially simple to perform, and results in 
\begin{equation}
    S_{ka = 0}^{z} = \frac{1}{\sqrt{N}}\sum_{i} \mathfrak{S}^{z}_{i} \ , \  S_{ ka = \pi}^{z} = \frac{1}{\sqrt{N}}\sum_{i} \left(-1\right)^{i} \mathfrak{S}^{z}_{i}.
\end{equation}
From these two (independent) equations we define the Chern number $C$, as well as its ``dual" $\overline{C}$ as
\begin{equation}\label{Combined_pole_equations}
\begin{aligned}
          &\frac{1}{2}\left\langle \left(S_{k = 0}^{z} - S_{k = \frac{\pi}{a}}^{z}\right)\right\rangle\equiv C = \frac{1}{\sqrt{N}} \sum_{i=0}^{N/2} \langle \mathfrak{S}^{z}_{2i+1}\rangle \\  &\frac{1}{2}\left\langle \left(S_{k=0}^{z} + S_{k = \frac{\pi}{a}}^{z}\right)\right\rangle \equiv \overline{C} = \frac{1}{\sqrt{N}}\sum_{i=0}^{N/2} \left\langle \mathfrak{S}^{z}_{2i}\right\rangle.
\end{aligned}
\end{equation}

Here the Chern number $C$ is the \emph{relative polarization}, and $\overline{C}$ the \emph{absolute polarization}, \emph{dual} to $C$ by measuring the degree of alignment of the spins. Equation \eqref{Combined_pole_equations} defines topological markers in real space. This sets them apart from other, momentum space topological numbers, making them directly accessible numerically for example with DMRG. Whilst other topological markers have previously been defined over non-local correlation functions in real space \cite{nonlocal1, nonlocal2}, the quantities defined in \eqref{Combined_pole_equations} repackage that information into a physically intuitive, and conceptual observable. As an example, we now show how to obtain the correct pattern of Majoranas shown in figure \ref{fig:Patterns} from $C$ and $\overline{C}$.\\

For the single Kitaev wire, the ground state is simply given by the \emph{BCS} wave function, from which it is straightforward to calculate the real-space correlation functions. In terms of the $\vartheta_{k}$ angles, the expectation values of $\langle c^{\dagger}_{r+j}c_{j} + \text{h.c.}\rangle$ can be calculated directly and one finds
\begin{equation}\label{FrakSzsinglewire}
\begin{aligned}
        \langle \mathfrak{S}^{z}_{j>0} \rangle =& \frac{1}{N}\sum_{k\in BZ} \cos\left(k\cdot j\right) \left(1-\cos\left( \vartheta_{k}\right)\right) \\ 
        \langle \mathfrak{S}^{z}_{j = 0} \rangle =&  \frac{1}{N}\sum_{j\in BZ} \left(1-\cos\left(\vartheta_{k}\right)\right) - 1.
\end{aligned}
\end{equation}
For the $j=0$ case one needs to account for the anti-commutation relations in \eqref{SzFrak}. When the spectrum has a gap, the correlations decay exponentially with a correlation length $\xi$, and thus we only need of order $\xi$ terms to obtain a robust topological marker. The following two examples correspond to the extreme limit where $\xi$ is minimal: The two patterns presented in figure \ref{fig:Patterns} above are exact in the limits of $t = \Delta = 0$, and $\mu = 0$ respectively \cite{Kitaev_2001}, for which $\vartheta_{k}$ simplifies considerably. In the trivial ($\mu \gg t$) case $\vartheta_{k} = \pi$, such that the only non-zero expectation value in \eqref{FrakSzsinglewire} is $\langle \mathfrak{S}^{z}_{j = 0} \rangle = 1$. This is equivalent to a density of fermions $n_{c} = 1$ and the wire is filled with on-site bound Majoranas for each lattice site, \emph{cf.} figure \ref{fig:Patterns}.
For the topological case at $\mu = 0$ we found that $\vartheta_{k} = ka$, thus only $\langle\mathfrak{S}^{z}_{j = 1}\rangle \neq 0$. This implies on the one hand that $2n_{c} - 1 = 0$, i.e. \emph{half-filling}. In addition to the non-local fermion \eqref{left-Kitaev}, we introduce the \emph{left-binding} fermions $d_{L,j} = \frac{1}{2} \left(\gamma_{A,j} - i\gamma_{B,j+1}\right)$. Together with $d_{R,j}$ one finds 
\begin{equation}
    \mathfrak{S}^{z}_{j =1 } = \frac{1}{N}\sum_{j} \left(d^{\dagger}_{R,j}d_{R,j} - d^{\dagger}_{L,j}d_{L,j}\right) = n_{R} - n_{L}
\end{equation}
Since the densities $n_{R/L} \in [0,1]$, the Chern number $C = 1$ implies $n_{R} = 1$ and $n_{L} = 0$, which is exactly the non-trivial pattern in figure  \ref{fig:Patterns}.
% Interpreting the symbols $R$ and $L$ as right-mvoing
% and left-moving particles, then $C=1$ is indeed equivalent to a particle moving one-way from North to South pole or from $ka=0$ to $ka=\pi$.
For general $|\mu| < \infty$, the function $\vartheta_{k}$ is more complicated and thus $\cos\left(\vartheta_{k}\right)$ smears out and one must include more $\mathfrak{S}^{z}_{j}$ correlators. In the critical case $\mu = \pm 2t$ this ``smearing" becomes maximal, and all values of $j$ become comparably relevant - one of the key features of critical behaviour. 

\subsection{Reproducing the phase diagram of the single Kitaev wire with DMRG}

For one non-interacting wire the BCS ground state is known exactly. However, as we are interested in interacting systems, we present already in the following results obtained with the \emph{Density Matrix Renormalization Group} (DMRG) algorithm \cite{PhysRevB.48.10345}. DMRG, first introduced by White \cite{PhysRevLett.69.2863} in the early 90s, provides a powerful tool to approximate the ground state of a Hamiltonian, and has become a standardized and well-established tool in condensed matter physics.
% In recent years it has also been recast in the language of tensor networks and  Matrix Product States (MPS) \cite{Schollw_ck_2011}. 
In this work, we use the library ITensor in Julia \cite{https://doi.org/10.48550/arxiv.2007.14822} to perform our computations \footnote{For an extensive list of papers using the ITensor library see http://itensor.org/docs.cgi?page=papers}. The performance of the DMRG algorithm depends strongly on having a low degree of entanglement between the subsystems (or blocks), which is why it is optimized for short-range interacting, one-dimensional systems with open boundary conditions (OBCs). However, by defining the Chern numbers from the poles of the Bloch sphere representation $ka = 0$ and $\pi$, we intrinsically assume the well-definedness of a BZ and thus of PBCs.
Additional optimization of the algorithm can be achieved by exploiting symmetries of the system, as this makes the Hamiltonian block diagonal which reduces complexity considerably. Whilst the Kitaev wire does not preserve particle number $N = \sum_{i}n_{i}$ due to the superconducting gap $\Delta$, the fermion parity defined as 
$\hat{P} \equiv \left(-1\right)^{P} = e^{i\pi \sum_{i}n_{i}} = \Pi_{i}\left(1-2n_{i}\right)$ is conserved, since only cooper pairs can be annihilated or created. Therefore $\hat{P}$ and $H$ commute, and the $|GS\rangle$ can be labeled by its parity, or equivalently by $P$. From now on, we will label the two distinct parity sectors by this binary notation $P = 0$ (even) and $P = 1$ (odd). For more details about the DMRG for two coupled wires, we refer to Appendix \ref{Appendix_Numerics}. \\

Apart from performance, there is another, more subtle reason why it is important to impose the conservation of symmetries in the DMRG algorithm explicitly: The DMRG searches for ground states of $H$ by optimizing an initial guess $|GS\rangle = |\psi_{0}\rangle$, and subsequently improving on it with each iteration. However, consider two degenerate GSs labeled by $a$ and $b$ i.e $E\left(|GS,a\rangle\right) = E\left(|GS,b\rangle\right)$. Then our initial guess $|\psi_{0}\rangle$ fixes the superposition of these degenerate states, i.e. the DMRG is insensitive to the precise mixture of the two sectors. The \emph{physical} GS lives in the space spanned by $|GS,a/b\rangle$, i.e.
\begin{equation}\label{degen}
    |GS\rangle = \chi_{a}|GS,a\rangle + \chi_{b}|GS,b\rangle .
\end{equation}
The initial guess $|\psi_{0}\rangle$ will therefore only sample $\left(\chi_{a}, \chi_{b}\right) = \left(\chi_{a,0},\chi_{b,0}\right)$. The precise degree of mixing $\chi_{a/b}$ will therefore be necessary to determine the \emph{physical} GS. We return to this discussion later on in more detail, when examining the critical phases of Kitaev wires.\\

\begin{figure}[h!]
    \centering
    \includegraphics[width = 0.5\textwidth]{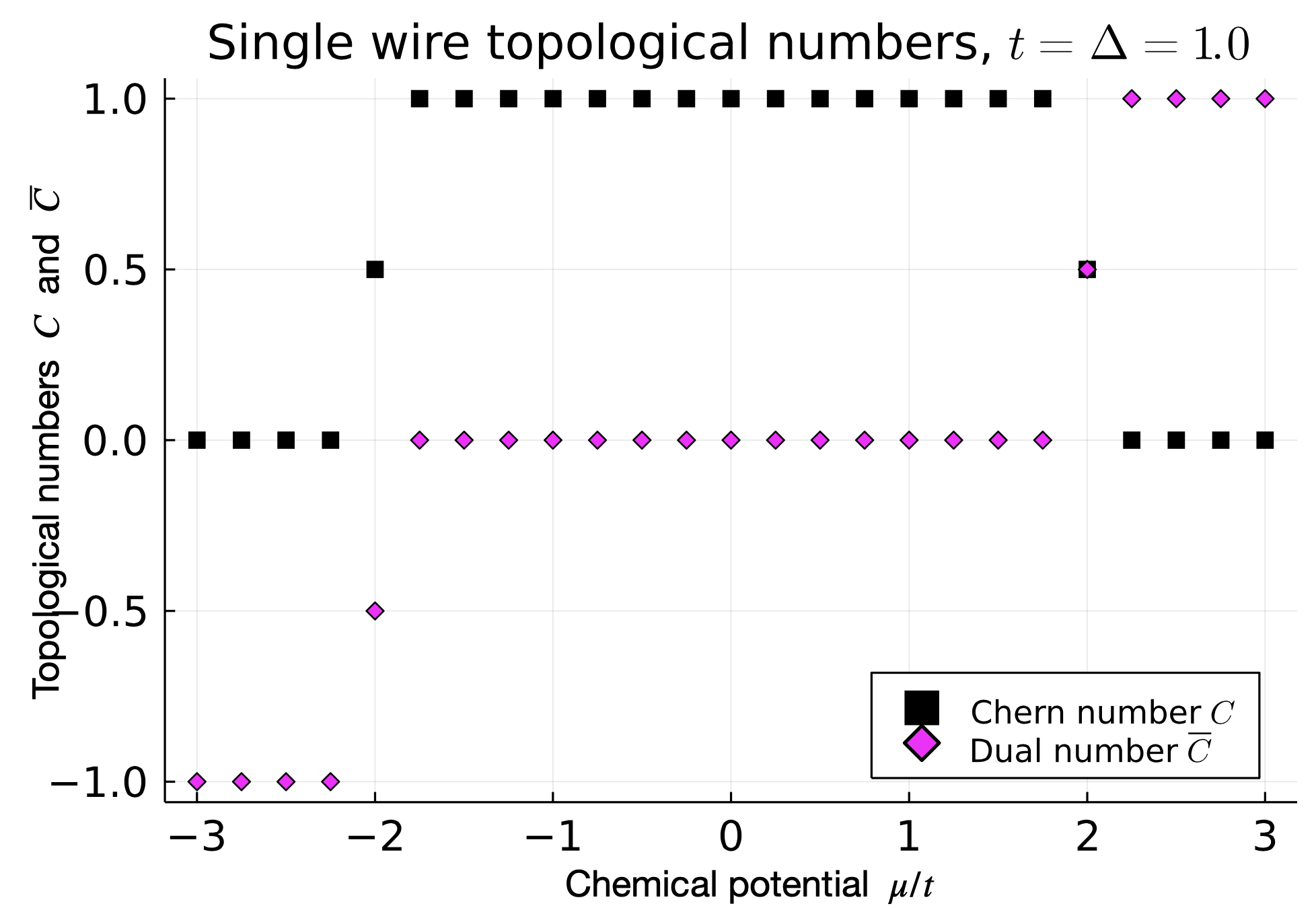}
    \caption{DMRG calculations for $N = 60$ site wire with PBCs. Transitions from \emph{trivial} to \emph{topological} and vice-versa are at $\mu = \pm 2t$. GS of the trivial phases has parity $P = 0$ whilst the topological phase has $P = 1$. At the transition point the GS is degenerate and in an equal superposition of $P=0$ and $P=1$, leading to $C = 1/2$.}
    \label{fig:SW_toponumbers1}
\end{figure}

DMRG provides the GSs for both parity sectors, from which we identify the \emph{true} one as the lowest-lying state. ITensor gives us the GS as a Matrix Product State (MPS), from which the correlation functions in \eqref{FrakSzsinglewire} are extracted. 
% It is then very straightforward to obtain the Chern numbers $C$ and $\overline{C}$ for various parameters $\mu$ and $t = \Delta$. 
As can be seen in figure \ref{fig:SW_toponumbers1}, the numerical results match the theoretical predictions also far from the limiting cases $\mu = 0$ and $\mu \gg |t|$. Interestingly, the dual number $\overline{C}$ can distinguish between the two trivial phases \footnote{and the two Quantum Critical Points (QCPs), as will be discussed in detail in \ref{sec:Fractional}.}. This feature persists and is very promising for applications. We highlight now already the two Quantum Critical Points (QCPs) $\mu = \pm 2t$, which seem to interpolate between the trivial and topological phases with $C = 1/2$. We return to this in more depth in Sec. \ref{sec:Fractional}.

% \subsection{Extending the Bloch sphere/{\color{black} wire} correspondence}\label{Extending_correspondence}

\subsection{Reviewing the phase diagram of two interacting Kitaev wires}\label{Extending_correspondence}

Recurring themes when contemplating applications for topological matter in emergent technologies are both scalable quantum computers as well as quantum circuits. Especially for the latter case, super-conducting wires like the Kitaev chain could be of particular interest, not least due to the dangling Majorana edge modes. As recently demonstrated \cite{Yang_2020}, a quasi-two-dimensional grid of Kitaev wires admits $\left(p+ip\right)$-superconductivity when threading appropriate fluxes through each unit cell. \\

Following the example in Ref. \cite{Herviou_2016}, we introduce two Kitaev chains
\begin{equation}\label{Ham_coupled}
    H_{0} = -\sum_{\sigma,i}\left(\frac{\mu}{2} n^{\sigma}_{c,i} + t c^{\sigma\dagger}_{i}c^{\sigma}_{i+1}  -\triangle c^{\sigma\dagger}_{i}c^{\sigma\dagger}_{i+1} + \text{h.c.}\right),
\end{equation}
with label $\sigma$ to distinguish the two chains. The proximity of wires in realistic heterostructures will certainly make coupling between two neighbouring wires relevant. One prominent force will be the electrostatic one, or Coulomb interaction, between the charge carriers of both wires. Investigating the effects of Hubbard-like interactions as a first approximation to this is the central theme hereafter. Following the example of \cite{Herviou_2016}, we now include
\begin{equation}\label{Hint_wires}
 H_{int} =   g \sum_{i} \left(n^{1}_{i}-\frac{1}{2}\right)\left(n^{2}_{i}-\frac{1}{2}\right).
\end{equation}
The presence of such (strong) interactions renders the classification of topological phases from their symmetries no longer complete \cite{Fidkowski_2010}, and exotic phases may emerge
The phase diagram in the presence of such an interaction, obtained in \cite{Herviou_2016}, is presented in the figure below.

% We briefly touched on the non-critical phases of \cite{Herviou_2016} in section \ref{integer_phases}, and thus believe it to be judicious to place the critical DCI phase in the phase diagram, \emph{cf.} figure \ref{fig:trajectories_tp05}. \\ 
\begin{figure}[h!]
    \centering
    \includegraphics[width = 0.49\textwidth]{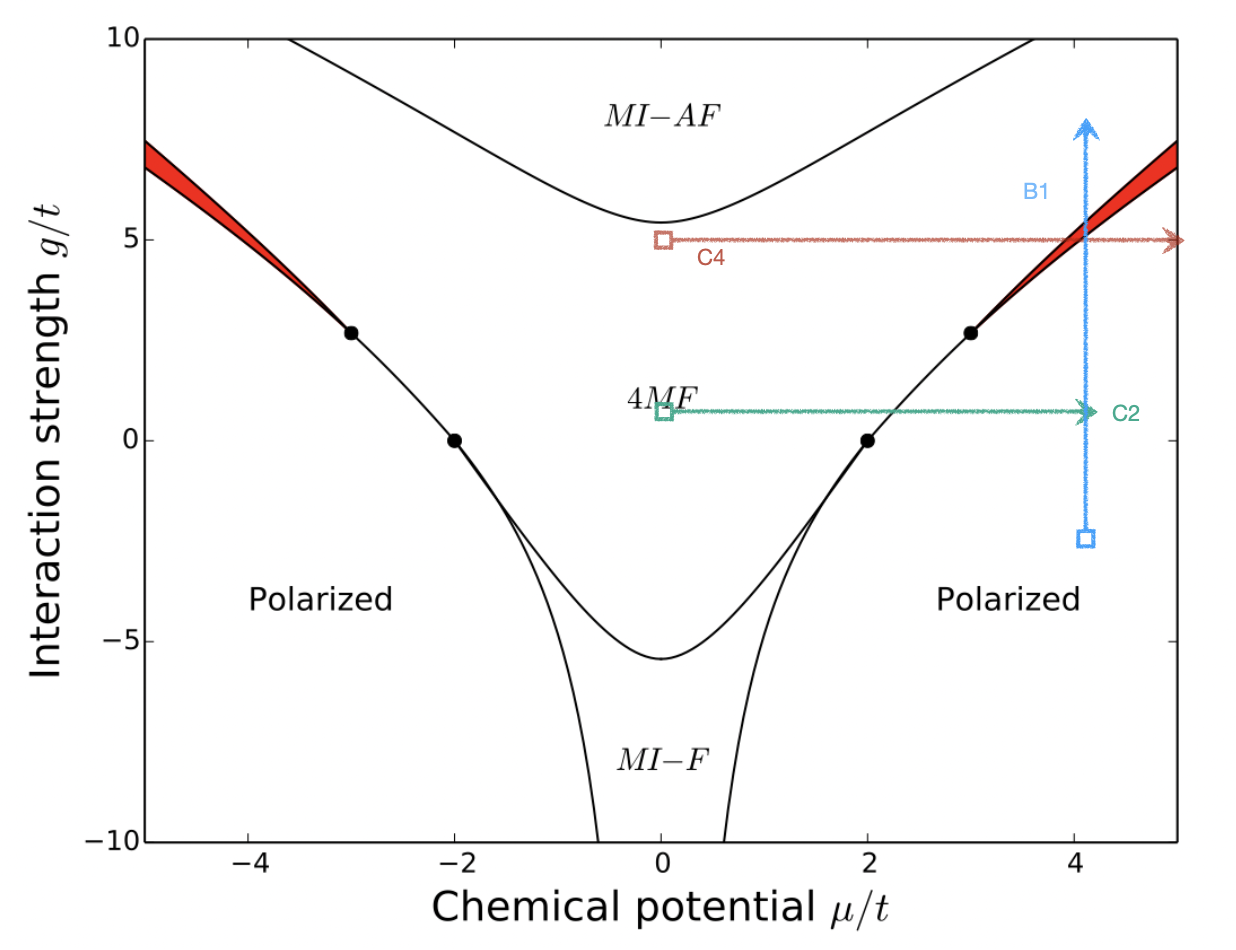}
    \caption{Phase diagram for two interacting Kitaev wires, from \cite{Herviou_2016}. Gapless DCI phases are in red, whilst \emph{4MF} and \emph{polarized} denote doubly topological and doubly trivial phases respectively. At high $|g|$ we find \emph{MI-AF} and $\emph{MI-FM}$, i.e. Mott insulating (anti-)ferromagnetic phases. Coloured lines (B1, C2 and C4) are \emph{trajectories} relevant when studying the effects of an inter-wire hopping term $t_{\bot}$ in Sec. \ref{sec:TP}.} 
    \label{fig:trajectories_tp05}
\end{figure}

A key observation is, that next to the doubly topological (\emph{4MF}) and doubly trivial (\emph{polarized}) phases which one expects from coupling two Kitaev wires, additional phases appear both at large interactions and at half-filling. On the one hand a transition to Mott physics at large $|g|$ is observed, and (anti-)ferromagnetic ordering in the $|g| \longrightarrow \infty$ limit emerges. On the other hand, far from half-filling, a gapless critical phase also opens up: the \emph{double critical Ising phase} (DCI). An important quantity to classify critical theories is the so-called central charge \cite{PhysRevLett.56.746}. A central charge $c=1$ may refer to a critical Dirac fermion or a bosonic Luttinger theory in $(1+1)$D. As found in \cite{Herviou_2016}, the DCI phase is characterized by a total central charge $c = 1$.
% , for both wires, or a fractional $c=\frac{1}{2}$ per wire. 
This can be identified with two gapless Majorana fermions in the bulk, or similarly critical Ising modes in two dimensions \cite{Karyn1999}.

From the link between the single Bloch sphere and the single Kitaev wire \cite{Sato_2017}, it is then natural to ask whether an extension of this ``duality" is possible for two coupled chains and two coupled wires. In fact, as was investigated in \cite{Hutchinson_2021}, coupling two Bloch spheres in $z-$direction through a term $\sim S^{z,1}S^{z,2}$ gives rise to a phase diagram not very different to figure \ref{fig:trajectories_tp05}. Both doubly topological and trivial phases, as well as distinct high-$|g|$ phases emerge. Additionally, also a fractional topological $C^{1} = C^{2} = 1/2$ phase in the presence of strong interactions is observed, when a $\mathbb{Z}_{2}$ (exchange) symmetry between the spheres is present \cite{Hutchinson_2021}. It is important to emphasize here that this interaction acts directly on the Bloch sphere which then means the reciprocal space of an associated topological lattice model. Various forms of interactions have been shown to stabilize the $C^{1} = C^{2} = 1/2$. In particular, a simple scenario to realize this phase is a local interaction in the reciprocal space. An important prerequisite to observe this fractional phase is the presence of an adjustable $MS^{z,i}$ term which from the link between the single Kitaev wire and the Bloch sphere is induced by the presence of a global chemical potential acting on the two wires. \\ 
% {\color{green} For an Ising interaction of the form $gS^{z}_{1}S^{z}_{2}$ the fractional phase is predicted to precisely occur for $d_z-M<g<d_z+M$. Here, we should make this correspondence perhaps more quantitative including the modified analytical phase diagram that was built from the energetics of the wires' model with interaction (\ref{Hint}) to anticipate that everything works fine and thatthings are clear.}

In Appendix \ref{Appendix:Spheres_Wires} we extend and develop the link between two interacting wires \cite{Herviou_2016} and Bloch spheres \cite{Hutchinson_2021},
by re-writing the low-energy Hamiltonian \eqref{Hint_wires} in momentum space. We find that the two-interacting Bloch spheres are in a sense momentum space duals to the two interacting wires. Therefore, at least for the phases of two-interacting wires in \cite{Herviou_2016} which are connected to the $g = 0$ line, the topological numbers defined in \eqref{Combined_pole_equations} are appropriate and reveal their distinct topology. The existence of a mixed (non-local) momentum term in \eqref{Hint_wire_sphere} may suggest that for larger interaction parameters the mapping is \emph{not} precisely exact onto the two spheres. However, this term can be reabsorbed into the $\sim \mu$ chemical potential contribution for each wire. Due to the wire symmetry, this will act as a uniform renormalization $\mu = \mu\left(g\right)$ for both wires, which then supports the idea that the spheres/wires duality is robust even towards larger interactions, as long as the Bogoliubov quasi-particle basis is the correct Hilbert space basis to describe the wires. Below, we quantify the effects of these non-local interactions via DMRG systematically in the phase diagram of the two-wires model. That way we will also verify that the fractional topological numbers introduced for the two-spheres' model \cite{Hutchinson_2021,lehur_new} are also useful to describe the DCI phase of the two-wires' model.

\subsection{Integer topological phases for two interacting Kitaev superconducting wires}\label{integer_phases}

First, we evaluate the topological numbers $C^i$ and $\overline{C}^i$ for the two wires, navigating through the phases with integer topological numbers \cite{Herviou_2016}.
As introduced prior, the phase diagram for low $|g|$ presents three distinct phases \cite{Herviou_2016}:  Two polarized phases, and the topological \emph{4MF} phase, \emph{cf.} figure \ref{fig:trajectories_tp05}. By continuity to the non-interacting limit, these phases are the extensions of the topological and trivial phases of two independent Kitaev wires. With open boundary conditions (OBCs), the \emph{4MF} phase thus has \emph{four} dangling Majorana edge modes \cite{Herviou_2016}.
\begin{figure}[h!]
    \centering
    \includegraphics[width = 0.5\textwidth]{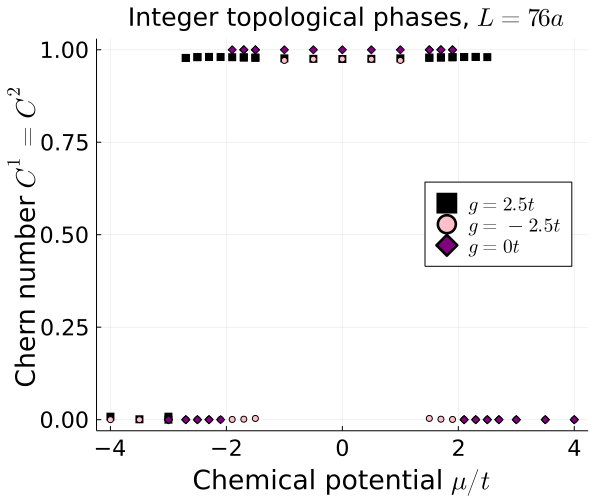}
    \caption{Chern numbers $C^{\sigma}$ for $g = 0$ and $t = \Delta=1.0$ and $L = 76a$ per wire with PBCs. A clear transition from polarized to \emph{4MF} phases is visible.} 
    \label{fig:Integer_1}
\end{figure} 

\begin{figure}[h!]
    \centering
    \includegraphics[width = 0.5\textwidth]{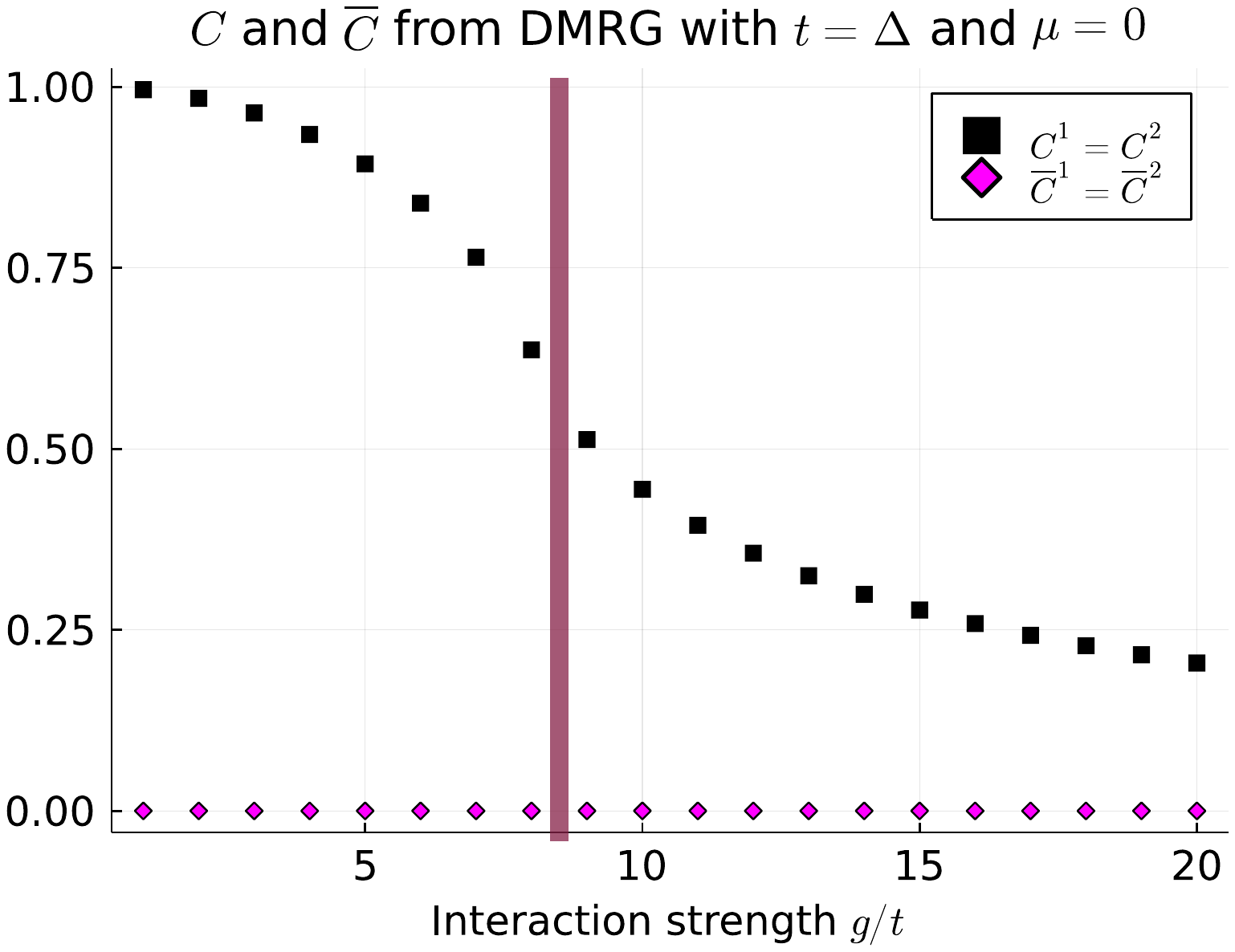}
    \caption{$C$ and $\overline{C}$ for $N = 50$ sites/wire. $C$ is no longer quantized, approaching zero smoothly for large $|g|$. The red line, indicating the gap-closing in figure \ref{fig:energies_hg}, is in qualitative agreement with both the inflection point above, and the transition point from the toplogical (\emph{4MF}) to Mott-insulating antiferromagnetic phase \emph{MI-AF} in \cite{Herviou_2016}, \emph{cf.} figure \eqref{fig:trajectories_tp05}.}
    \label{fig:cherns_hg}
\end{figure}

\begin{figure}[h!]
    \centering
    \includegraphics[width = 0.5\textwidth]{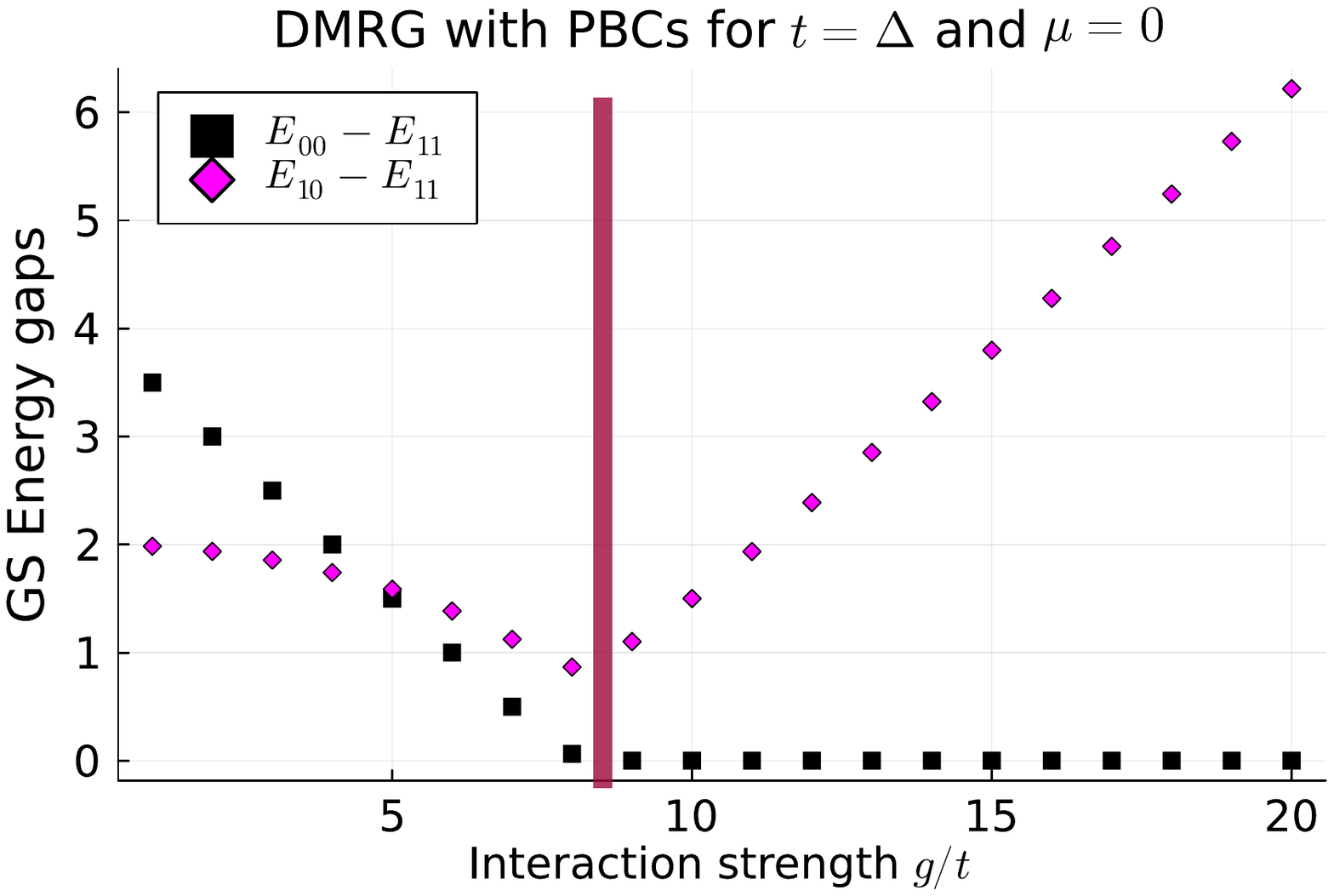}
    \caption{ Energy gaps obtained from DMRG for two interacting Kitaev chains, with $N = 50$ sites per wire. The spectrum becomes degenerate at the gap-closing point $g \approx 7t$, highlighted by the \emph{red} line, and signals a phase transition. }
    \label{fig:energies_hg}
\end{figure}

We evaluate the topological numbers defined in \eqref{Combined_pole_equations} numerically for the doubly topological \emph{4MF} phase, the two trivial \emph{polarized} regions as well as the large $|g|$ Mott phases for strong interactions \cite{Herviou_2016}. We additionally take into account the conservation of parity within each wire, also in the presence of the interaction \eqref{Hint_wires}, resulting in \emph{four} parity sectors, labeled by $\left(P_{1},P_{2}\right)$. We label the GS and GS energies by their parities $|\text{GS}\rangle_{(P_{1},P_{2})}$ and $E_{\text{P1P2}}$. Figures \ref{fig:Integer_1} and \ref{fig:Integer_12} confirm that the \emph{doubly trivial} (polarized) phases are characterized by  $C^{1} = C^{2} = 0$ and $\overline{C}^{1} = \overline{C}^{2} = \pm 1$. As for the single shown in figure \ref{fig:SW_toponumbers1}, the $\overline{C}^{1/2}$ are sensitive to the trivial phases and can thus be used to identify the various ``equivalent" regions in figure \ref{fig:trajectories_tp05}. Similarly \emph{4MF} phase has $C^{1} = C^{2} = 1$ and $\overline{C}^{1} = \overline{C}^{2} = 0$. 
% Including an interaction strength $g \neq 0$ affects the quantization of $C$ and $\overline{C}$.
In figure \ref{fig:cherns_hg}, the quantization of $C$ is shown to be no longer exact for large $g$,  potentially indicating proximity to a Mott transition. In the two-spheres model \cite{Hutchinson_2021}, a transition to a topologically trivial phase $C^{1} + C^{2} = 0$ (as for the duals) is expected, which does not imply the individual topological markers to be zero as well.
The energy spectrum for $g \gtrapprox 7t$ is degenerate between the $(P_{1},P_{2}) = (0,0)$ and $(1,1)$ sectors \emph{cf.} figure \ref{fig:energies_hg}.
Comparing this to both the topological numbers in figure \ref{fig:cherns_hg}, and the phase diagram in figure \ref{fig:trajectories_tp05}, we note the qualitative agreement between the gap-closing point and both the transition point between the topological \emph{4MF} phase and the Mott-insulating phase \emph{MI-AF}, as well as the inflection point of the curve in figure \ref{fig:cherns_hg}. Whilst $C$ and $\overline{C}$ are no longer quantized at such large interactions $g$, they can still be used to investigate the phase diagram of two (or more) strongly interacting wires. \\

In the following two sections we extend our analysis to the critical phases of one and two, interacting and coupled Kitaev wires. We both add to the previous work \cite{Herviou_2016}, as well as enhance the phase diagram by numerically extracting the topological markers introduced in \eqref{Combined_pole_equations}. Finally, in Sec. \ref{sec:TP} we investigate the effects of an inter-wire hopping and characterize the phase diagram with the topological markers defined in \eqref{Combined_pole_equations}.

\section{The Kitaev wire at criticality}\label{subsec:sw_critical_theory}

Here we elaborate on the quantum phase transition in the Kitaev model through bulk chiral Majorana modes introduced for example in Ref. \cite{Herviou_2016}.
By investigating the critical theory of the single wire, we develop both the theoretical tools of these chiral modes and benchmark the DMRG by extracting the topological numbers for the single wire. The formalism will be useful to develop further QFT related to the two-wires model in the next section.

\subsection{Quantum Field Theory for the chiral bulk modes}\label{sec_continQFT_chiral}

At the QCP $\mu = 2t$, the single Kitaev wire is described by a single \emph{critical Ising} model. To see this, we formulate the Hamiltonian \eqref{Ham_c} in the Majorana representation. We quantify the distance to the QCP by
\begin{equation}\label{deltamu}
    \delta \mu = \mu + 2t,
\end{equation}
and thus rewrite \eqref{Ham_maj} in a more suggestive manner
\begin{equation}\label{Ham_maj_crit}
    H_{K} = it\sum_{j} \gamma_{B,j}\left(\gamma_{A,j+1} - \gamma_{A,j}\right)  -\frac{i \delta \mu}{2} \sum_{j} \gamma_{A,j}\gamma_{B,j}.
\end{equation}

Introducing the finite-difference derivative for the $A$ Majoranas and setting $\delta\mu = 0$, the above reduces to
\begin{equation}
    H_{K} = ita\sum_{j} \gamma_{B,j}\partial_{j} \gamma_{A,j},
\end{equation}
with $a\partial_{j}f_{j} = f_{j+1} - f_{j}$ and $a$ the short-distance cut-off (i.e. the lattice spacing). The continuum limit for the $A$ and $B$ fields is defined as $\gamma_{A/B}\left(x = ja\right) = \frac{\gamma_{A/B,j}}{\sqrt{a}}$, such that for $a\longrightarrow 0$ the Hamiltonian becomes

\begin{equation}\label{Hamcritp_ba}
    H_{K} \stackrel{a \longrightarrow 0}{\longrightarrow} \ ita\int_{x}\text{d} x \ \gamma_{B}(x)\partial_{x} \gamma_{A}(x).
\end{equation}

We now introduce two \emph{chiral} Majorana modes $\gamma_{R,j}$ and $\gamma_{L,j}$ \cite{Herviou_2016}, defined by the relations
\begin{equation}\label{QCP_ham_sw}
\sqrt{2}\gamma_{L/R}\left(x = ja\right) = \gamma_{B}\left(x =ja\right) \pm \gamma_{A}\left(x = ja\right).
\end{equation}
Before proceeding, a justification for why it is appropriate to call these modes \emph{chiral} is in order. Close to $\delta\mu = 0$, the Fermi momentum is found to be $k_{F} \approx 0$. We can then write down the single-particle Hamiltonian $\mathcal{H}_{k}$ around this value introducing Pauli matrices such that
\begin{equation}
    \mathcal{H}_{k} = -\left(\mu+ 2t\right) \cdot \tau^{z} + \triangle \cdot  2ka \cdot \tau^{x} = \triangle \cdot 2ka \cdot \tau^{x}, 
\end{equation}
where in the last step we dropped $\tau^{z}$ due to $\delta \mu \sim 0$. Decomposing $\psi_{k}$ into two Majorana fermions $\gamma_{A/B}\left(ka\right)$ we find that $
\mathcal{H}_{K}$ is diagonalized by $\gamma_{A} \pm \gamma_{B}$, with eigenvalues $\pm 2\triangle \cdot ka$, i.e. a gapless spectrum. The two chiral modes have velocities $v = \pm 2 a\Delta$ \cite{Yang_2020}. Through the chiral $\gamma_{R/L}$ modes into \eqref{Hamcritp}, the continuum $R/L$-Hamiltonian is found to be
\begin{equation}\label{Hamcritp}
    H_{K} \stackrel{a \longrightarrow 0}{\longrightarrow} \ ita\int_{x}\text{d} x \ (\gamma_{L}(x)\partial_{x} \gamma_{L}(x) - \gamma_{R}(x)\partial_{x} \gamma_{R}(x)).
\end{equation}
Interpreting $\gamma_{R/L}$ as the two chiral components of a Majorana field, the Hamiltonian describes a free, one-dimensional Majorana conformal field theory (CFT). We emphasize that a free Majorana CFT (half of a Dirac fermion) counts as $c =1/2$. Whilst generally difficult to obtain experimentally, the central charge has been shown to enter thermodynamic observables, for example, via a universal term in the free energy and the heat capacity $C_{V}$ \cite{PhysRevLett.56.746}. Another way to obtain the central charge is via the entanglement entropy $S_{A}\left(l\right)$ 
\begin{equation}
S_{A}=-\operatorname{Tr}\left(\rho_{A} \log \left(\rho_{A}\right)\right),
\end{equation}
where $l$ is the length of the subsystem. The central charge was found to appear in the coefficient of the logarithmic contribution \cite{Calabrese_2004}, i.e. 
\begin{equation}\label{ententropy_def}
S_{A}(l)=\frac{c}{3} \log \left(\frac{L}{\pi} \sin \left(\frac{l \pi}{L}\right)\right)+\mathcal{O}(1).
\end{equation}
Additional oscillatory terms are present when OBCs are considered \cite{Song_2010}. At $\delta \mu = 0$ the gapless \emph{critical Ising} model with central charge $c=1/2$, \emph{cf.} figure \ref{fig:fractional_sw} is signaled by the closing of the gap at the critical point and change in central charge extracted from the entanglement entropy. We complete this discussion by considering the effect of $\delta \mu \neq 0$: The Hamiltonian in \eqref{Hamcritp} acquires a chiral mass term, given by $i\delta\mu \gamma_{R}\gamma_{L}$ , i.e. the Majorana mode becomes gapped by $m_{C} = \delta\mu$ and has a central charge $c=0$ as expected. 

\subsection{Fractional topology of the QCP}\label{FractionalTop_QCP}

Numerically we use DMRG to find the physical ground state of the system at criticality and extract both the central charge and topological numbers $C$ and $\overline{C}$. The results are summarized in figure (\ref{fig:fractional_sw}). As already mentioned in \ref{subsec_cherns_for_single_wire}, due to the parity conservation of the Hamiltonian we find a ground state spectrum given by $E_{P = 0}$ and $E_{P =1}$, the true ground state, therefore, is determined by the smaller value of both. Generally the gap $\Delta E \equiv E_{P=0}- E_{P=1} $ is non-zero, with the topological phase being determined by $P=1$ and the trivial ones by $P=0$. Precisely in the critical points $\mu = \pm 2t$, $\Delta E$ is zero and the spectrum becomes degenerate between the parity sectors. For more details on the numerics and subtleties regarding DMRG in the critical phase, we refer to the Appendix \ref{Appendix_Numerics}.
 
 As mentioned prior in the discussion around equation \eqref{degen}, the degeneracy implies that the \emph{physical} ground state lies in the subspace spanned by both parity sectors
\begin{equation}\label{PhysGS_swQCP}
    |GS\rangle = \chi_{P=0}|\text{GS}\rangle_{P = 0} + \chi_{P=1}|\text{GS}\rangle_{P = 1},
\end{equation}
with $\chi^{2}_{P=0} + \chi^{2}_{P=1} = 1$. Therefore, any observable $\mathcal{O}$ is calculated as $\langle \mathcal{O}\rangle = \text{tr}\left[\hat{\rho}\mathcal{O}\right]$, where the density matrix is $\hat{\rho} = |GS\rangle\langle GS|$.
The topological markers defined in \eqref{Combined_pole_equations} commute with the fermion parity operator, and are thus diagonal in the parity basis. Due to the fact that the parity sectors associated to $|GS\rangle_{P=0}$ and $|GS\rangle_{P=1}$ are orthogonal, the expectation values of diagonal observables can be calculated using the effective, diagonal density matrix
\begin{equation}\label{rho_eff_ensemble}
    \hat{\rho}_{eff} = \sum_{P = 0,1} \chi^{2}_{P}|\text{GS}\rangle_{P} \langle\text{GS}|_{P}, 
\end{equation}
Numerically we find $C = 0$ and $C = 1$ for the $P = 0$ and $P =1$ GS's respectively, which results in $\langle C\rangle = \chi^{2}_{P = 1}$. 
In the absence of a bias, the physical GS necessarily has $\chi^{2}_{P} = 1/2$, which results in a fractional $C = 1/2$ value. This is coherent with approaches used to measure and define the fractional Chern invariant of Fractional Chern Insulators.  \\ 

We now offer additional arguments and justifications, that the only physically sensible degree of mixing is given by $\chi^{2}_{P=0,1} = 1/2$, by evaluating the GS of the momentum space Hamiltonian around the poles $ka = 0,\pi$. This discussion makes a link between the gap-closing point in the critical phase of a single Kitaev wire, and the entanglement property at one pole in the $C=1/2$ phase for two-interacting spheres \cite{Hutchinson_2021}. The GS parity is determined from the product of all individual site parities, both in real and momentum space:
\begin{equation}
\hat{P} = e^{i\pi \sum_{i}n_{i}} = e^{i\pi \sum_{k}n_{k}} = \prod_{k}\left(1-2n_k\right) \equiv \prod_{k}\hat{P}_{k}.
\end{equation}
We now determine the GS in momentum space for a single wire at the QCP, and consider in detail the poles $ka = 0$ and $ka = \pi$. We fix $t= \Delta$ and $t > 0$ without loss of generality.  
In the trivial ($|\mu| > 2t$) and topological ($|\mu| < 2t$) regimes, the parity is clearly fixed. By being adiabatically connected to the vacuum $|0\rangle$ and fully occupied $\prod_{k} c^{\dagger}_{k}|0\rangle$ states, the two trivial regions have fixed parity: $P = 0$ and $P = N\text{mod}2$ respectively. Similarly, the GS of the topological regime is characterized by the BCS state in the bulk, which conserves the parity of the vacuum $|0\rangle$.  Therefore, only the pole-contributions in \eqref{BCS_WF} may alter the parity. Due to $|\mu|<2t$, one pole will host a particle, whilst the other is empty. Thus, $P = 1$ necessarily. \\

% Due to the parity conservation of $\mathcal{H}_{k}$, and by smoothly connecting the two trivial cases to $\mu = \pm \infty$, or the topological one to $\mu = 0$. In the two trivial cases $P = \left(1\right)^{N}$ or $ = \left(-1\right)^{N}$. In the BCS wave function \eqref{BCS_WF}, the central $0 <ka<\pi$ contributions each at most add a particle-hole pair $c^{\dagger}_{k}c^{\dagger}_{-k}$ to the vacuum $|0\rangle$, i.e. conserving the parity. The parity of the BCS state is thus determined solely from the pole-contributions. For example, with $|\mu| < 2t$, the parity is given by $P = 1$. \\
In the critical phase we instead find that both parity sectors become degenerate, which can be  seen from the GS structure around the poles.
% the wave function must lie in an equal superposition of both parity sectors, i.e. $\chi^{2}_{P} = 1/2$. 
To understand this, we note that at $\mu = \pm 2t$, the gap closes at $ka = \pi$ or $ka = 0$ respectively. Away from these points, the bulk remains gapped, and we yet expect the BCS wave function \eqref{BCS_WF} to accurately describe the GS. The parity is thus again determined solely from the poles, and precisely the GS around the gap-closing point fixes the mixed parity GS. To see this, we write $\mathcal{H}_{k}$ to leading order in $k$
\begin{equation}
    \mathcal{H}_{ka \approx 0} = 
\begin{pmatrix}  -\frac{\mu}{2} - t & \mathcal{O}\left(k\right) \\ \mathcal{O}\left(k\right) & +\frac{\mu}{2} + t\end{pmatrix},
\end{equation}
such that for $\mu = -2t$ the gap closes and the off-diagonal terms are leading order. 
% \begin{equation}
% \mathcal{H}_{ka \approx 0} =
% \begin{pmatrix}  0 + \mathcal{O}\left(k^2\right) & ie^{i\varphi}\Delta ka \\ -ie^{-i\varphi}\Delta ka & 0 + \mathcal{O}\left(k^2\right)\end{pmatrix},
% \end{equation}
% whilst at $ak = \pi$ the leading term still remains the diagonal elements $\sim \pm 4t$:
% \begin{equation}
% \mathcal{H}_{ka \approx 0} =
% \begin{pmatrix} -4t & \mathcal{O}\left(k\right) \\ \mathcal{O}\left(k\right) & 4t\end{pmatrix}
% \end{equation}
%The two quasi-particles diagonalizing the above matrix are $\sqrt{2}\eta_{k}^{\pm} = c_{k} \pm ic^{\dagger}_{-k}$, and we find for low momenta
The two quasi-particles we obtain by diagonalizing the Hamiltonian matrix above are $\sqrt{2}\eta_{k}^{\pm} = c_{k} \pm ic^{\dagger}_{-k}$. For low momenta $ka\approx 0$ we thus find the gapless, linear spectrum in terms of the two (related via PHS) quasi-particles 
\begin{equation}
    \mathcal{H}_{crit, ka \approx 0} = \sum_{k \lessapprox 0} v^{+}_{eff}k \eta^{+\dagger}_{k}\eta^{+}_{k} +\sum_{k \gtrapprox 0} v^{-}_{eff}k \eta^{-\dagger}_{k}\eta^{-}_{k}.
\end{equation}
Here $v^{\pm}_{eff} = \pm 2\Delta + \mathcal{O}\left(k^2\right)$, and we find the GS around the gap-closing point to be $\eta^{-\dagger}_{k}|BCS\rangle$ for $k \gtrapprox0$ and $\eta^{+\dagger}_{k}|BCS\rangle$ for $k \lessapprox0$. 
Similar to Eq. (\ref{eta}), $\eta^{-\dagger}$ corresponds to a quasiparticle operator with a negative eigenergy and with $\varphi=0$. Due to the PHS symmetry, the GS is found to satisfy
\begin{equation}\label{entanglement_prop_wire}
\langle S^{z}_{ka\approx 0}\rangle = \langle c^{\dagger}_{k}c_{k} - c_{-k}c^{\dagger}_{-k}\rangle = 0,
\end{equation}
where the expectation values where taken wrt. to the GS. This property will certainly also be true at $ka = 0$, and can be linked to the degenerate GS wrt. parity by considering explicitly
\begin{equation}\label{entanglement_property_BCS}
    0 = \langle S^{z}_{0}\rangle = \langle c^{\dagger}_{0}c_{0} - c_{0}c^{\dagger}_{0}\rangle = 2\langle n_{0} \rangle -1  = -\langle\hat{P}_{k=0}\rangle.
\end{equation}
Therefore, the total parity is equal to zero, as $\hat{P} = \prod_{k} \hat{P}_{k}$. It therefore follows that it lies in a perfect superposition of ``$+$" and ``$-$" parity sectors, i.e. in binary notation 
\begin{equation}
    |GS\rangle = \frac{1}{\sqrt{2}}\left(|GS\rangle_{P = 0} + |GS\rangle_{P=1}\right).
\end{equation}
At the south pole $ka = \pi$ the GS is determined by the dominant diagonal contribution, i.e. $|GS\rangle = |0\rangle$. Instead of the entanglement property \eqref{entanglement_prop_wire} we find $\langle S_{ka\approx \pi}^{z}\rangle=-1$. Hence, the Chern number, calculated from the poles (and derived geometrically on the Bloch sphere from the integral over the Berry curvature \cite{Hutchinson_2021, Le_Hur_2022,lehur_new}), is given by $C = -1/2$.

% In the absence of a bias the average results in a fractional $C = 1/2$ value. To see this, simply integrate over $0\leq\alpha \equiv \chi^{2}_{P = 0} \leq 1$, thus finding that
% \begin{equation}
%     \langle C\rangle = \int_{0}^{1} \text{d}\alpha C(\alpha) = \int_{0}^{1} \text{d}\alpha 
% \alpha = 1/2.
% \end{equation}
% Similar results hold for $\overline{C}$, by simply replacing $\alpha$ with $\beta \equiv \chi^{2}_{P=1}$. 
% This is equivalent to extracting the observables from the physical GS in equation \eqref{PhysGS_swQCP} lies in the averaged $\langle\alpha\rangle= \langle\beta\rangle = 1/2$ state, i.e. $\langle\chi^{2}_{P=0}\rangle = \langle\chi^{2}_{P=1}\rangle = 1/2$. Since on either side of the QCP, the GS parities are $P = 0$ and $P =1$ respectively, the $1/2$ average is also sensible physically from symmetry. 
This is how we numerically find the $C= \pm1/2$ value in figure \ref{fig:SW_toponumbers1}. \\ 

By similar arguments the same holds for the second topological number $\overline{C} = \pm1/2$. A remarkable result is, that $\overline{C}$ is sensitive to all regions of the phase diagram, even distinguishing both critical points  \emph{cf.} figure \ref{fig:SW_toponumbers1}.
% {\color{black} For the two wires' system, it will not be necessary to perform such an analysis on the momentum space Hamiltonian, as the $\mathbb{Z}_{2}$ (exchange) symmetry between the wires fixes the $\chi^{2}_{P = 0,1}$ parameters to $\frac{1}{2}$ in the corresponding superposition of $|GS\rangle$.} \\
The QCP to has a central charge $c=1/2$, \emph{cf.} figure \ref{fig:fractional_sw}.
The gapless $R/L$-chiral modes defined in \eqref{QCP_ham_sw} also emerge from the momentum-space Hamiltonian around the poles, thus establishing a first duality between the central charge and the Chern number in the critical point: Recall that, for $\mu = -2t$, the GS is $\eta^{-\dagger}_{k}|BCS\rangle$ at $ka = 0$ (North pole). Decomposing $\eta^{\pm,\dagger}_{k}$ into their Majorana constituents we find
\begin{equation}
\begin{aligned}
        \eta^{\pm}_{k} =& \frac{1}{2}\left(\gamma_{A,k} + i\gamma_{B,k}\right) \pm i \frac{1}{2}\left(\gamma_{A,-k} - i\gamma_{B,-k}\right) \\ =& \frac{1}{2}\left(1\pm i\right)\left(\gamma_{A,k} \pm \gamma_{B, k}\right) \propto \gamma_{L/R} ,
\end{aligned}
\end{equation}
where in the second we used that $\gamma_{k} = \gamma_{-k}$ for Majorana fermions coming from $\gamma^{\dagger} = \gamma$. In the last equation we identified the chiral modes in \eqref{QCP_ham_sw}. As these modes are in momentum space and near $ka \approx 0$, they are (approximately) uniformly distributed over the whole wire. Deviating from the QCP leads to a gap in these chiral modes: Non-zero diagonal elements $\sim \delta\mu$ in the Hamiltonian matrix lead to $\sim i\gamma_{R}\gamma_{L}$ terms, equivalent to a mass term in the QFT \emph{cf.} discussion following \eqref{Ham_maj_crit}.\\

\section{Fractional Topology for two interacting wires}\label{sec:DCI}

We now demonstrate how both the chiral QFT and fractional topological numbers $C = \pm 1/2$ and $\overline{C} = \pm 1/2$ extend to the critical regions in the phase diagram of two interacting Kitaev wires. As was discovered in \cite{Herviou_2016}, the critical line extends as a gapless DCI phase, possible due to the effects of strong interactions. Recasting the model in terms of mixed-wire fermions, such an extended critical phase can be predicted from the QFT. The existence of this phase was also verified numerically using DMRG and quantum information techniques \cite{Herviou_2016}. \\ 

In what follows, we expand on the theoretical description of the critical phase, and  introducing another set of mixed wire Fermions $\Gamma_j$ and $\Theta_j$. From this we obtain an alternative QFT description of the DCI phase, in terms of two chiral complex fermions. These are  directly linked to the chiral Majorana modes of the QCP for a single wire. This link to the QCP is then extended, by showing that the topological markers introduced in \eqref{Combined_pole_equations} are $C=\overline{C}=\frac{1}{2}$ per wire. Finally we also address possible measurement protocols, further providing evidence for the duality between the topological properties of two interacting Bloch spheres \cite{Hutchinson_2021} and Kitaev wires \cite{Herviou_2016}.

\subsection{Chiral QFT for two interacting wires from mixed wire fermions}
We now expand on the idea of chiral bulk modes in the DCI phase, introduced for the single wire in Sec. \ref{sec:DCI}. This presents an alternative approach to the QFT description in \cite{Herviou_2016}, offering additional insight into the physical properties of the gapless DCI phase. We present the main results in the following and refer to Appendix \ref{Appendix_DCI_phase} for more details.

We proceed similarly to the single wire and write the Hamiltonian in \eqref{Ham_coupled} in terms of the Majorana fields, introduced prior in Sec. \ref{sec:Cherns}. Each complex $c$-fermion can be written in terms of the two real Majoranas $\gamma_{A}$ and $\gamma_{B}$. Denoting the distance from the $g= 0$ QCPs by 
$\delta \mu$, the Hamiltonian then reads
\begin{equation}
\begin{aligned}
H=&-\frac{i \delta \mu}{2} \sum_{j, \sigma=1,2} \gamma^{\sigma}_{A,j} \gamma^{\sigma}_{B,j}-i t \sum_{j, \sigma} \gamma^{\sigma}_{B,j}\gamma^{\sigma}_{A,j+1}
\\ &-\frac{g}{4} \sum_{j}\gamma^{1}_{A,j}\gamma^{1}_{B,j}\gamma^{2}_{A,j}\gamma^{2}_{B,j}.
\end{aligned}
\end{equation}

We assumed here the limit $t = \Delta$, and use a similar notation as in \eqref{Ham_maj}. The different wires are labeled by $\sigma = 1,2$. As was already discussed in \cite{Herviou_2016}, the degrees of freedom of both wires are mixed in the DCI phase, resulting in a $c= 1$ critical model for the combined wires. Therefore, we introduce complex fermions which mix both sets of $A$ and $B$ Majorana species across both chains:
\begin{equation}\label{Mixed_wire_ferms_def}
\begin{aligned}
\Gamma_{j} \equiv&   \frac{1}{2}\left(\gamma_{A,j}^{1} + i \gamma_{A,j}^{2}\right) \\ 
\Theta_{j} \equiv& \frac{1}{2}\left(\gamma_{B,j}^{2} - i \gamma_{B,j}^{1}\right).
\end{aligned}
\end{equation}
Rewriting the Hamiltonian in terms of these mixed modes we obtain
\begin{equation}\label{gamma_theta_ham_coupled}
\begin{aligned}
H=&-\left(2 t-\frac{\delta \mu}{2}\right) \sum_{j} \left(\Gamma^{\dagger}_{j+1}\Theta_{j} - \Gamma^{\dagger}_{j}\Theta_{j}\right)  + \text{h.c.}\\ &
-\frac{\delta \mu}{2}\sum_{j}\left( \Gamma^{\dagger}_{j+1}\Theta_{j} + \Gamma^{\dagger}_{j}\Theta_{j}\right)  + \text{h.c.}\\ & - \frac{g}{2} \sum_{j} \left(\Gamma^{\dagger}_{j}\Gamma_{j} + \Theta^{\dagger}_{j}\Theta_{j}\right) 
-g\sum_{j} \Gamma^{\dagger}_{j}\Gamma_{j}\Theta^{\dagger}_{j}\Theta_{j}.
\end{aligned}
\end{equation}
For $\delta\mu = g = 0$, i.e. in the double QCP at $\mu = 2t$, the first bracket resembles strongly the $\gamma_{B}\partial_{x}\gamma_{A}$ term in \eqref{QCP_ham_sw}. Then defining similarly continuum-fields for ``left" and ``right" movers $\psi_{L/R}$ as 
\begin{equation}\label{chiral_modes_def}
 \sqrt{2}\psi_{R/L}\left(x\right) = \Theta\left(x\right)\pm i \Gamma\left(x\right),
\end{equation}
one obtains a critical model in the $a \longrightarrow 0$ limit 
\begin{equation}
  H\left(\delta\mu = g = 0\right) = 2ita\int_{x} dx(\psi^{\dagger}_{L}\partial_{x}\psi_{L} - \psi^{\dagger}_{R}\partial_{x}\psi_{R}).
\end{equation}
This describes a model with central charge $c = 1$. In fact, by decomposing $\Gamma$ and $\Theta$ into their Majorana constituents 
\begin{equation}
\begin{aligned}
       \sqrt{2}\psi_{R} =& \frac{1}{2}\left(\gamma^{2}_{B} -i \gamma^{1}_{B} + i\gamma^{1}_{A} - \gamma^{2}_{A}\right) \sim \gamma^{2}_{R} -i\gamma^{1}_{R}, \\
           \sqrt{2}\psi_{L} =& \frac{1}{2}\left(\gamma^{2}_{B} -i \gamma^{1}_{B} - i\gamma^{1}_{A} + \gamma^{2}_{A}\right) \sim \gamma^{2}_{L} -i\gamma^{1}_{L},
\end{aligned}
\end{equation}
we see that the $\psi_{R/L}$ fermions can be written in terms of the chiral Majoranas on each wire, revealing the \emph{Double Critical Ising character of the phase. The two spheres’ model in the C=1/2 phase also reveals
the same Majorana fermions’ structure \cite{lehur_new} as for the one sphere’ model at the quantum phase transition emphasing the correspondence between
spheres and wires.} Defining $v_{F}/a = 2t - \delta \mu/2$, we rewrite \eqref{gamma_theta_ham_coupled} in terms of $\psi_{R/L}$ 
\begin{equation}\label{Final_Gamma_Theta_Ham}
    \begin{aligned}
           H = & -iv_{F} \int_{x} \text{d}x \left(\psi^{\dagger}_{R} \partial_{x} \psi_{R} - \psi^{\dagger}_{L}\partial_{x} \psi_{L}\right) \\  &-  \frac{\delta \mu }{2} \int_{x} \text{d}x  \left( \psi^{\dagger}_{R}\psi_{L} + \psi^{\dagger}_{L}\psi_{R} \right) \\ &- \frac{g}{2}\int_{x} \text{d}x  \left(\psi^{\dagger}_{R}\psi_{R} + \psi^{\dagger}_{L}\psi_{L}\right) - H_{int}.
    \end{aligned}
\end{equation}
The interaction-Hamiltonian is defined as
\begin{equation}
\begin{aligned}
     H_{int} &= \frac{g}{4a} \int_{x}  \text{d}x\left(\psi^{\dagger}_{R}\psi_{R}\psi^{\dagger}_{R}\psi_{R} + \psi^{\dagger}_{L}\psi_{L}\psi^{\dagger}_{L}\psi_{L} \right) \\ &+ \frac{g}{4a}\int_{x}\text{d}x\left(\psi^{\dagger}_{R}\psi_{R}\psi^{\dagger}_{L}\psi_{L}+\psi^{\dagger}_{R}\psi_{L}\psi^{\dagger}_{L}\psi_{R}\right) \\ &+ \frac{gi}{4a}\int_{x}\text{d}x\left(\psi^{\dagger}_{R}\psi_{R}\psi^{\dagger}_{R}\psi_{L}  - \psi^{\dagger}_{L}\psi_{L}\psi^{\dagger}_{L}\psi_{R}\right).
\end{aligned}
\end{equation}
A powerful analytical tool to investigate one-dimensional interacting fermion systems is \emph{bosonization} \cite{Senechal, giamarchi2004quantum, WeaklydisorderedSpinLadders}, wherein fermionic modes are mapped onto bosonic fields. This may simplify interactions, and we introduce the boson fields $\phi$ and $\theta$ through the standard definitions \cite{Haldane}
\begin{equation}\label{bosonization_defs}
\psi_{R/L}\left(x = ja\right) =\frac{U_{R/L}}{\sqrt{2 \pi \alpha}} e^{-i\left(\pm \phi\left(x\right)-\theta\left(x\right)\right)}.
\end{equation}
Here $U_{R/L}$ are the so-called ``Klein factors" ensuring the anti-commutation properties of the fermionic fields $\psi_{R/L}${, and the short-distance cut-off $\alpha$ is of the order of the lattice constant $a$. } When dealing with higher-order terms and interactions, bosonization becomes fairly lengthy. We thus only present the results in the main text and refer to Appendix \ref{Appendix_bosonization} for more intricate details. To lowest order in the short-distance cut-off $\alpha$ one finds \eqref{Final_Gamma_Theta_Ham} to be equivalent to the following bosonic \emph{Luttinger-Liquid} (LL) Hamiltonian with a \emph{Sine-Gordon} potential for $\phi$
\begin{equation}\label{Final_own_bosonized_text}
    H = \int_{x} dx \left(\frac{v_{F}}{2\pi}\left(\frac{1}{K}\left(\partial_{x}\phi\right)^{2} + K\left(\partial_{x}\theta\right)^{2}\right) - g_{\phi}\cos\left(2\phi\right)\right).
\end{equation}
The LL parameter $K$ and $g_{\phi}$ are related to the model parameters as {$v_{F}K^{-1} = a\left(4t - \delta \mu - \frac{g}{\pi}\right)$, $v_{F}K = a\left(4t - \delta \mu\right) $} and $g_{\phi} = \left(\frac{\delta\mu}{2\pi \alpha} +  \frac{ga}{4\pi^{2}\alpha^{2}}\right)$, scaling as energy times $1/a$. We emphasize that the $g_{\phi}$ Sine-Gordon term comes from the channel $\psi^{\dagger}_{R}\psi_{L}\psi^{\dagger}_{L}\psi_{R}$. We note that in the present Hamiltonian \eqref{Final_own_bosonized_text}, the interaction is effectively halved compared to the boson Hamiltonian in \cite{Herviou_2016}, which is explained by the specific choice of mixed fermion fields \footnote{The choice in \cite{Herviou_2016} results in a single wire of a \emph{doubled} number of sites $N$ at constant length $L = Na$. Thus, one finds $2a^{\prime} = a$, which results in a doubled interaction strength $g$.}.
From the RG analysis on the short-distance cutoff $a$  the flow equations for the LL parameter $K$ and interaction $g_{\phi}$ are found as \cite{Herviou_2016}
\begin{equation}\label{gphiflow}
\begin{aligned}
 \frac{d K}{d l}=& -\frac{4 \pi^{2}}{v_{F}^{2}}g_{\phi}^{2} K^{2}\\ 
   \frac{d g_{\phi}}{d l}=&\left(2- K\right) g_{\phi}.
\end{aligned}
\end{equation}

The DCI phase is both gapless and has a total central charge $c = 1$. The Hamiltonian \eqref{Final_own_bosonized_text} has these properties only when the interaction $g_{\phi}$ vanishes. As long as $K < 2$, the interaction parameter is always relevant in the RG sense, \emph{cf.} \eqref{gphiflow}. The coupling $g_{\phi}$ progressively increases with the parameter $l=\ln(L/a)$ related to the effective length $L$ at which we probe the system. In this case, $g_{\phi} = 0$ traces a highly fine-tuned line in the $g-\mu$ phase diagram \cite{Herviou_2016}, which extends down to $g=0$. Here we have $K= 1$ and the two wires are decoupled, thus the critical line belongs to the class of two \emph{critical Ising} models, i.e. with central charge $c$ \emph{per wire} satisfying $2 c = 1$. However, when the LL parameter becomes $K > 2$, then the $\sim \cos\left(2\phi\right)$ operator with scaling dimension $\sim K$, becomes irrelevant in $(1+1)$d. It is then possible for the gapless phase to open up and extend in the phase diagram.  Noting that $K > 2$ is only possible at large interactions and far from half-filling $|\delta \mu| \gg 0$, the applicability of bosonization is limited in the original fermion basis. However, in the mixed-wire basis $\psi_{R/L}$ the DCI phase occurs at half-filling and thus emphasises the usefulness of the formalism.  \\

An extended DCI region was verified numerically in \cite{Herviou_2016} using both the central charge and logarithmic bipartite charge fluctuations \cite{Herviou_2017}. The latter is necessary since we are unable to access the central charges of each wire individually. Thus it is not possible to distinguish the DCI phase from the single critical Dirac fermion, or boson, all three of which have a total $c = 1$.  However, as was shown in \cite{Herviou_2016}, the bipartite charge fluctuations \cite{Herviou_2017} are sensitive to this subtlety. These fluctuations are, just as the entanglement entropy, defined for a subsystem $A$ with length $l$, interacting with the rest of the wire:
\begin{equation}
F_{Q}=\operatorname{Tr}\left(Q_{A}^{2} \rho_{A}\right)-\operatorname{Tr}\left(Q_{A} \rho_{A}\right)^{2}=\left\langle Q_{A}^{2}\right\rangle-\left\langle Q_{A}\right\rangle^{2}.
\end{equation} 
Whilst the fluctuations have a dominant linear term for the two coupled wires, which grows with subsystem system size $l$ in relation to the quantum Fisher, there is also a sub-dominant logarithmic contribution \cite{Herviou_2016}
\begin{equation}
F_{Q}(l)=\frac{\mid \Delta]}{2|\Delta|+2 t} l+\mathcal{O}(\log l).
\end{equation}
This sub-dominant term traduces the underlying modes in the theory: If the degrees of freedom are Majorana fields, then this logarithmic term enters with a \emph{negative} coefficient whilst otherwise it remains positive \cite{Herviou_2016}. It is therefore possible to verify numerically that the DCI phase indeed corresponds to two $c =1/2$ modes, i.e. a \emph{double critical Ising} phase.

\subsection{Fractional topology of the DCI phase}\label{sec:Fractional}
Here, we investigate the critical regions in the phase diagram \cite{Herviou_2016} of two interacting topological superconducting wires using the topological markers \eqref{Combined_pole_equations}. Using DMRG we show that even at comparatively large $g$ values the Chern number of each wire is found to be stable with $C = 1/2$. This unveils the fractional topological nature of the DCI phase. Together with the considerations made in Sec. \ref{Extending_correspondence}, the intrinsic link between the DCI and the fractional $C=1/2$ phase for two coupled Bloch spheres becomes apparent. 
 \\ 

From  Sec. \ref{Extending_correspondence}, the analogy between the two wires and the two entangled Bloch spheres \cite{Hutchinson_2021} implies that a projection of the Chern number onto the subsystems is possible, by measuring the individual polarizations $\langle S^{z,\sigma}_{k}\rangle$ at the poles $k = 0$ and $k = \pi/a$. From \eqref{SzFrak} and \eqref{Combined_pole_equations} this translates to measuring the relevant two-point correlation functions on each wire. We calculated the GS again for each parity sector $(P_{1},P_{2}) = (0,1),(1,0),(0,0),(1,1)$ using DMRG, see figure  \ref{fig:degeneracy_twowires} in the Appendix. Similar to the single wire, in the region of the DCI phase the GSEs become two-fold degenerate. However, in this case, such that the total parity remains odd, i.e. between the parity sectors $(0,1)$ and $(1,0)$. Thus, effectively, the GS parity for each wire is again a super-position of $P = 0$ and $P=1$, just as in the single-wire QCP in Sec. \ref{subsec_cherns_for_single_wire}. \\
\begin{figure}[h!]
    \centering
    \includegraphics[width = 0.5\textwidth]{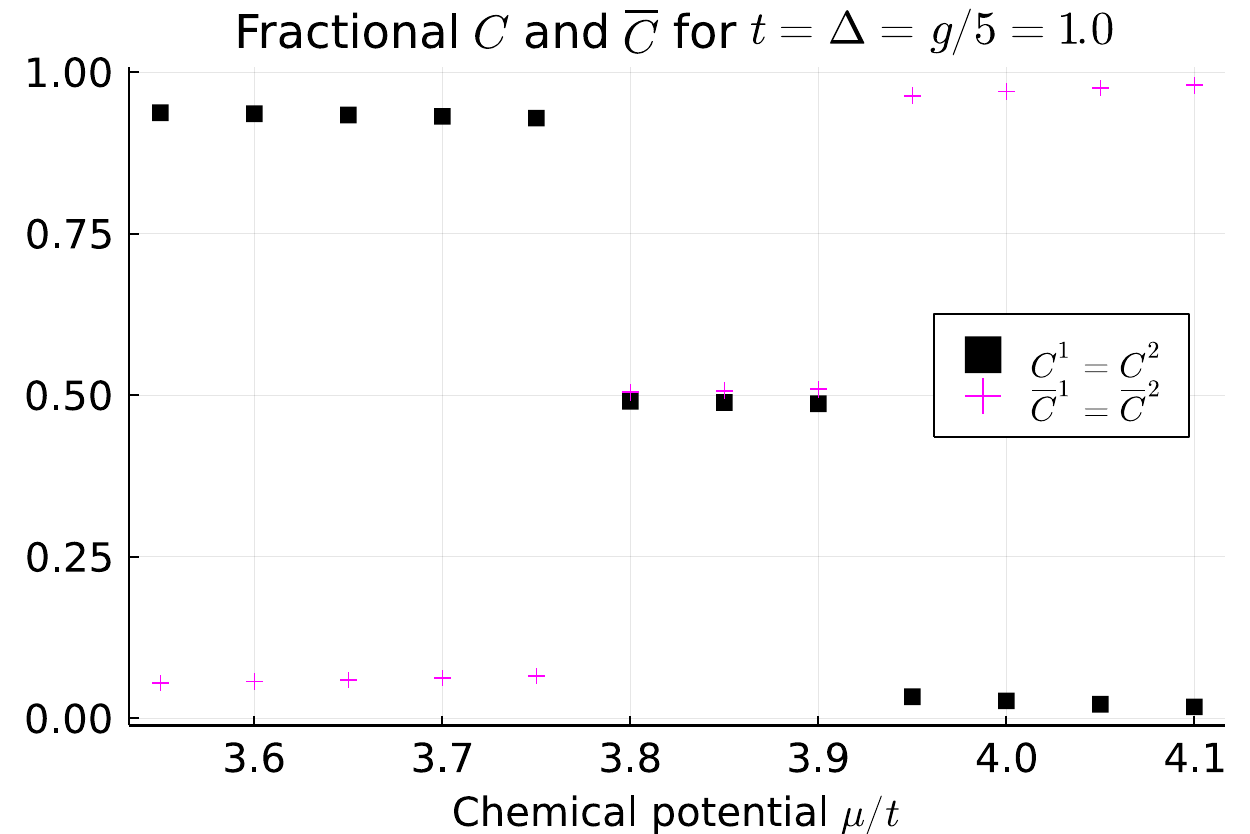}
    \caption{Chern numbers $C$ and $\overline{C}$ extracted from the GS for $L = 76a$ with PBCs. In the parameter range where the DCI phase was found \cite{Herviou_2016}, both topological numbers are found to be fractional $=1/2$. On either side of this extended region, we identify the \emph{4MF} and \emph{polarized} phases with $(C,\overline{C}) = (1,0)$ and $(C,\overline{C}) = (0,1)$ respectively.}
    \label{fig:ChalfPlatmu}
\end{figure}
\begin{figure}[h!]
    \centering
    \includegraphics[width = 0.49\textwidth]{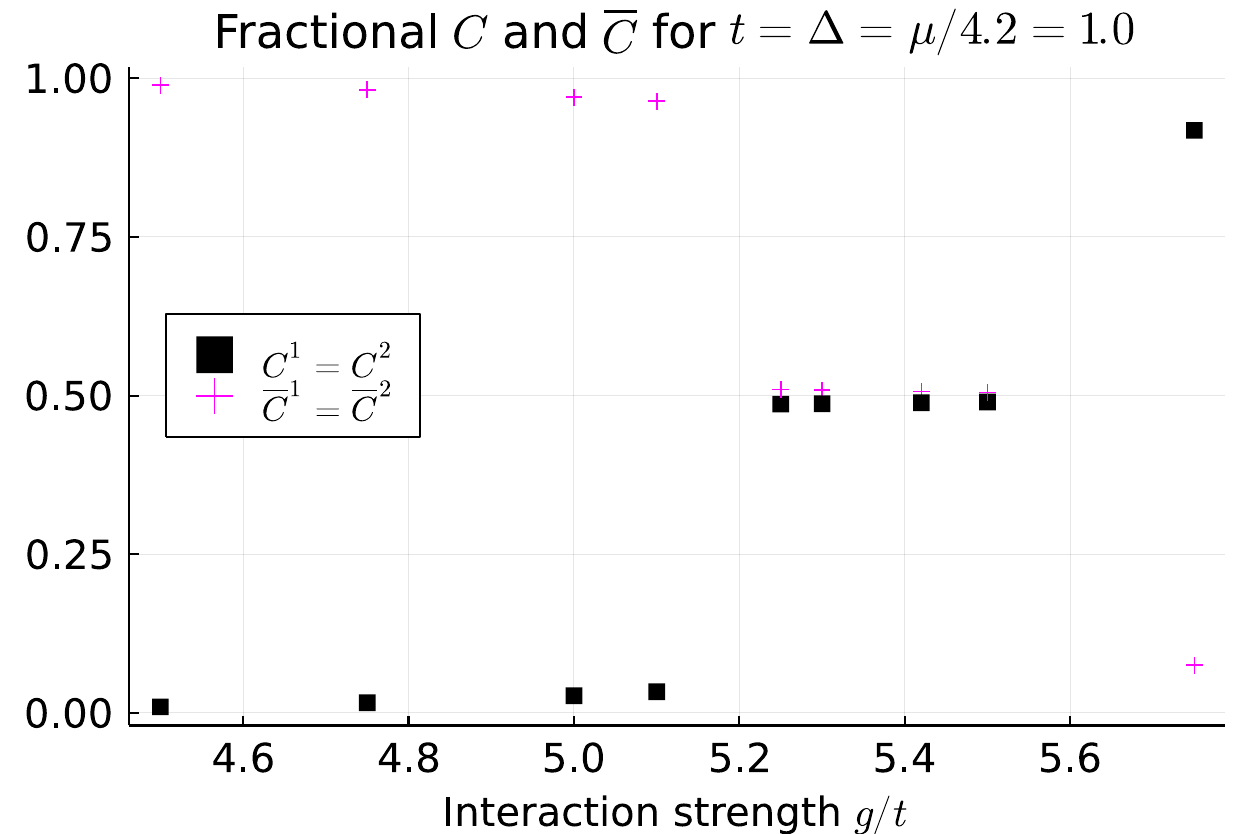}
    \caption{Chern numbers $C$ for the wires. We see again the fractional nature of the DCI phase around $g = 5t$ and $t = \Delta$.}
    \label{fig:ChalfPlatg}
\end{figure}

Evaluating the correlations we a priori expect that the exact values of the non-interacting $C = 1/2$ at $\delta\mu = 0$ will be smeared out by non-zero $g$. However, as can be seen in Figs.  \ref{fig:ChalfPlatmu} and \ref{fig:ChalfPlatg}, also for large $g$ the DCI phase is found as a constant $C = 1/2$ plateau. This can be explained within the Bloch sphere representation in Sec. \ref{Extending_correspondence}. Numerically, for a considerable range of interactions, we determined the expectation value $\langle S^{z,\sigma}_{k = 0}\rangle$ to be always unity. The effects of non-zero $g$ only manifest themselves in $C$ and $\overline{C}$ at the south pole, i.e. $ka = \pi$. For two-interacting spheres, the fractional $C =1/2$ phase stems from the entanglement of the two polarizations at the south pole \cite{Hutchinson_2021}. 
For two wires, the DCI phase should be then characterized by a GS which lies in the degenerate GS of the two parity sectors $|\text{GS}\rangle_{01}$ and $|\text{GS}\rangle_{10}$, \emph{cf.} discussion around equation \eqref{PhysGS_swQCP}. 
A-priori we thus expect the \emph{physical} GS to be given by 
\begin{equation}\label{physGS_coupled}
    |GS\rangle = \chi_{01}|\text{GS}\rangle_{01} + \chi_{10}|\text{GS}\rangle_{10},
\end{equation}
In this case we find (numerically) that $C^{1} =\overline{C}^{1} = \chi^{2}_{01}$ and $C^{2} = \overline{C}^{2} =  \chi^{2}_{10}$. Whilst from the single wire QCP at $g = 0$ we expect all parity sectors to be degenerate, an interaction $g>0$ introduces a gap between the energetically lower-lying states $|GS\rangle_{01}$ or $|GS\rangle_{10}$, and the $|GS\rangle_{00}$ and $|GS\rangle_{11}$ states. The inherent exchange symmetry between both wires fixes $\chi^{2}_{01/10} = 1/2$. In this case, it follows that $C^{1} = C^{2} = 1/2$. By expressing the z-spin expectation values by the Chern and dual markers we obtain the following relation
\begin{equation}\label{SzPiDCI}
    \langle GS| S^{z,1}_{ka = \pi}|GS\rangle = \langle GS| C^{1} - \overline{C}^{1}|GS\rangle
\end{equation}
This expectation value vanishes, which is the generalization of the entanglement property discussed prior in the QCP of a single wire, \emph{cf.} discussion following equation \eqref{entanglement_prop_wire}. In fact, similar considerations for the case of two (weakly) interacting spheres reveal the emergence of the two (complex) chiral fields $\psi_{R/L}$ around the entangled poles, making a link to the critical QFT of the DCI phase.
At $ka =0$ we write
\begin{equation}
\begin{aligned}
       \langle GS| S^{z,1}_{ka = 0}|GS\rangle = \langle GS| C^{1} + \overline{C}^{1}|GS\rangle.
\end{aligned}
\end{equation}
which is found numerically to always be $\approx 1$ in the ranges considered in figures \ref{fig:ChalfPlatmu} and \ref{fig:ChalfPlatg}.
Equivalent relations for the expectation values for wire $\sigma = 2$ exist. Thus, around $g = 0$ we find similar to the single wire case, that 
\begin{equation}\label{Csigma}
    C^{\sigma} = \frac{1}{2}\langle S^{z, \sigma}_{k = 0} \rangle \approx  \frac{1}{2}.
\end{equation}
At larger interaction strengths $g$ we no longer find that the topological numbers are quantized to $\pm 1$. However, we yet find the entanglement property in \eqref{SzPiDCI} to hold approximately in all parameter ranges considered, \emph{cf.} figures \ref{fig:ChalfPlatmu} and \ref{fig:ChalfPlatg}. The stability of this entanglement property explains the stability of the fractional $C=\overline{C} = 1/2$ in the DCI phase. 
% By continuity to the $g \approx 0$ critical line, the numbers $C^{\sigma}$ and $\overline{C}^{\sigma}$ defined in \eqref{Combined_pole_equations} reveal the fractional topology of the DCI phase. 
Thus, the DCI phase is characterized by both fractional central charge $c$ and topological markers $C$ and $\overline{C}$. As for the single wire, discussed in sec. \ref{FractionalTop_QCP}, we attribute this correspondence to the intrinsic duality with the model of two coupled and interacting spheres studied in \cite{Hutchinson_2021}.

\subsection{Outlook for measurement protocols}

Experimentally, the observation of Majorana fermions remains elusive, and measurement protocols do not always provide clear results: for example, in $dI/dV$ measurements Andreev bound states may also produce \emph{zero-bias-peaks} (ZBPs) \cite{Olesia_andreev}. By being defined globally, and in real space, the topological markers in \eqref{Combined_pole_equations} could provide a direct real-space technique to measure topology - also in the presence of disorder. \\

Numerically, the topological numbers $C$ and $\overline{C}$ in \eqref{Combined_pole_equations} are fairly simple to access, as they are fully determined by two-point correlation functions. Experimentally the non-locality $i$ of $\langle c^{\dagger}_{j+i}c_{j}\rangle$ may present potential hurdles. More recent advances in the cold-atoms community are establishing measurement techniques and protocols to obtain equal-time and spatially resolved correlation functions \cite{Zache_snapshots,ETcfs_1PI}. 
We can hope that in the near future, these techniques extend to non-local and higher-order correlation functions of quantum wires. In the following, we show that measuring the capacitance, charge, and band structure is already sufficient to probe the Chern numbers $C$. The capacitance is, essentially, the total charge density $Q/N$ of the wire. This, in Fourier space results in 
\begin{equation}
    \frac{Q}{N} = \frac{1}{N}\sum_{k \in BZ} \langle c^{\dagger}_{k}c_{k}\rangle.
\end{equation}

For the wire, the appropriate GS is the \emph{BCS} wave function, for which one finds
\begin{equation}
    \langle c^{\dagger}_{k}c_{k}\rangle =  \sin^{2}\left(\frac{\vartheta_{k}}{2}\right).
\end{equation}
We remind the reader that $\vartheta_{k}$ is given in by
\begin{equation}
    \cos\left(\vartheta_{k}\right) =  \frac{\mu+2t\cos\left(ka\right)}{\sqrt{\left(\mu + 2t\cos\left(ka\right)\right)^{2} + \Delta^{2}\sin^{2}\left(ka\right)}}.
\end{equation}
Making use of the trigonometric identity $\cos\left(2x\right) = \cos^{2}\left(x\right) - \sin^{2}\left(x\right)$, and going over into the continuum limit, we find alternatively the capacitance as
\begin{equation}
\label{onehalf}
    Q/N = \frac{1}{2} - \frac{1}{2\pi}\int_{0}^{+\pi/a} \cos\left(\vartheta_{k}\right) \  \text{d} k.
\end{equation}
At half-filling $\mu = 0$ the $\vartheta_{k}$ function is simply $k$ and hence $\frac{Q}{N} = 1/2 = C/2$. This identity comes from the fact that for one Bloch sphere, from geometry \cite{Hutchinson_2021} we have the identification $\cos^2\frac{\vartheta_{k}}{2} + \sin^2\frac{\vartheta_{k}}{2} = C$ which can be precisely used to identify the first term in Eq. (\ref{onehalf}) as $\frac{C}{2}$. As long as we remain in the topological phase, i.e. $|\mu| <2t$, the momentum integral above will be bound from above by the critical value $I_{crit}$ for $\mu = 2t$. On the other hand, $\vartheta_{k} = \pi$, thus the integral is strictly $1$. More generally, if $|\mu| > 2t$, it is bounded from below by $I_{crit}$. \\

For the DCI phase of the two wires' model, from Eq. (\ref{Csigma}), it could be suggestive to perform a direct charge measurement from a momentum-resolved probe \cite{Momentum,MomentumKLH} accessing then $\langle S_k^{z,\sigma} \rangle$ similarly as observing the motion of a spin-$\frac{1}{2}$
when driving from north to south pole such that for a wire $j$
\begin{equation}
C^{\sigma} = \frac{1}{2} = -\frac{1}{2}\int_{0}^{\pi/a} dk \frac{\partial \langle S_{k}^{z,\sigma} \rangle}{\partial k}.
\end{equation}

\section{The effects of an inter-wire hopping term $t_{\bot}$}\label{sec:TP}

Expanding on the phase diagram \ref{fig:trajectories_tp05} obtained in \cite{Herviou_2016}, we now consider the effects of an inter-wire hopping amplitude $t_{\bot}$. We use both QFT and numerical methods to study the properties at various ranges of $t_{\bot}$, and extract the Chern number $C$ and the dual number $\overline{C}$ to enhance the phase diagram. We emphasize here that the hopping term $t_{\bot}$ does not have a precise analogue on the two interacting Bloch spheres, therefore studying the effect of such a term within the wires is certainly justified and also physical as charges can leak
from one wire to another. In Appendix \ref{Appendix_bosonization}, we comment on the role of an inter-wire SC-pairing term. We also address the vicinity of the Mott phase(s) for strong interactions.\\

The non-interacting Hamiltonian $H_{0}$ of the combined two wires system could be written as a sum of two single-wire Hamiltonians $H_{K,\sigma}$  
\begin{equation}
H = H_{0} + H_{int} = H_{K,\sigma=1} + H_{K,\sigma = 2} + H_{int} .
\end{equation} 

In terms of a combined wires basis, the (free) Hamiltonian $H_{0}$ was given by a block diagonal matrix with $H_{K,\sigma=1,2}$ as diagonal entries. However, including an inter-wire hopping $t_{\bot}$ through the following addition to the Hamiltonian
\begin{equation}\label{Add_ham_tbot}
    H_{\bot} = -t_{\bot}\sum_{i} c_{i}^{1\dagger}c_{i}^{2} + \text{h.c.},
\end{equation}
introduces off-diagonal elements mixing both non-interacting wire Hamiltonians $H_{K, \sigma}$. 
The Hamiltonian is block diagonalized by the following transformation \cite{Ledermann_2000, KarynMaurice, Yang_2020} of the $c^{\sigma}$ basis
\begin{equation} \label{bonding_antibonding_basis}
c^{\pm}=\frac{1}{\sqrt{2}}\left(c^{1} \pm c^{2}\right) \Rightarrow  \begin{cases}    c^{1}=\frac{1}{\sqrt{2}}\left(c^{+}+c^{-}\right)\\  c^{2}=\frac{1}{\sqrt{2}}\left(c^{+}-c^{-}\right).     \end{cases}    
\end{equation} 
In this \emph{bonding/anti-bonding basis}, the $t_{\bot}$ term contributes as a shift in the chemical potential, however with an opposite sign for the respective bands: 
\begin{equation}\label{Bonding-anti-bonding_Ham}
\begin{aligned}
H_{0}= &-t \sum_{j, \kappa=\pm} \left(c_{j+1}^{\kappa\dagger} c_{j}^{\kappa} +\frac{\left(\mu +\left(-1\right)^{\kappa}t_{\bot}\right)}{2} c_{j}^{\kappa\dagger} c_{j}^{\kappa} \right) \\ &+\Delta \sum_{j, \kappa=\pm} \left(e^{i\varphi}c_{j}^{\kappa\dagger} c^{\kappa\dagger}_{j+1} + e^{-i\varphi}c_{j+1}^{\kappa} c^{\kappa}_{j}\right) \\ &+ \frac{g}{2}\sum_{j}\left(n_{j}^{+} - \frac{1}{2}\right)\left(n_{j}^{-} - \frac{1}{2}\right) + \text{h.c.},
\end{aligned}
\end{equation} 
where we defined $c^{+\dagger}c^{+} = n^{+}$ and so on. Since a priori two independent chemical potentials $\mu_{\sigma}$ could have been chosen, we work from now on with the effective potentials  
\begin{equation}    
\mu_{\pm} = \overline{\mu} \pm\frac{\Delta \mu}{2}  \equiv \frac{\mu_{+} + \mu_{-}}{2} \pm \frac{1}{2}\left(\mu_{+} - \mu_{-}\right).
\end{equation} 
To reproduce the above case \eqref{Bonding-anti-bonding_Ham} we thus simply set for the chemical potentials $\overline{\mu} = \mu$ and $\Delta \mu = t_{\bot}$. \\

The effects of adding an inter-wire hopping amplitude $t_{\bot}$ on the phase diagram in \cite{Herviou_2016} depend strongly on the phase. A first important property of the additional term to the full Hamiltonian \eqref{Add_ham_tbot} is that at zero interaction $g =0 $ it does not break \emph{time-reversal-symmetry} (TRS) for $t_{\bot} \in \mathbb{R}$. Therefore \eqref{Add_ham_tbot} cannot gap out the topological edge modes of the \emph{4MF} phase using the Chern number $C$ and $\overline{C}$ for the two bands. By continuity to the $g = 0$ limit, both the \emph{4MF} and Polarized phases \cite{Herviou_2016} will remain robust against a non-zero inter-wire hopping. At non-zero interactions, such symmetry protection is no longer a sufficient condition in general, to ensure the robustness of a topological phase \cite{Fidkowski_2010}. In the large $|g|$ limit we derive, by projecting onto the effective low-energy Hamiltonian, that in addition to the (anti-)ferromagnetic ordering, an additional paramagnetic \emph{Mott} order is induced due to a non-zero $t_{\bot}$. For sufficiently small $t_{\bot}$, the two \emph{Mott insulating} MI-AF and MI-F phases \cite{Herviou_2016} in figure \ref{fig:trajectories_tp05} are stable up to a critical interaction value $g^{*}$, where inevitably the paramagnetic order takes over. A key property of both the single QCP and DCI phase is the existence of gapless critical Ising modes. 

From the QFT away from half-filling we demonstrate that, in the thermodynamic limit, a  non-zero inter-wire hopping collapses the DCI phase and replaces the doubly critical line $g_{\phi} = 0$ by two \emph{single critical Ising} lines. The phase which emerges between both critical lines is a \emph{2MF} phase, where the wires host two dangling edge modes on the four possible ends in the case of OBCs, which is supported by numerical results for the Chern number $C$ and $\overline{C}$ for the two bands. However, we demonstrate numerically that at finite wire sizes, the DCI phase can be found to prevail below a critical scale $t_{\bot}^{*}$. For lengths of orders of a few hundred sites, we verify numerically a critical value $t_{\bot}^{*}/t \sim /L$ with $h/2\pi = 1$. Adjusting the length of the wires or the distance between them then allows for
the observation of various phases. Through the identification $T\sim v_F/L$ (with the Boltzmann constant $k_B=1$), varying the temperature $T$ also allows observation of the different phases.

\subsection{Phases at high $t_{\bot}$ and high $|g|$}\label{sec:high_g_param}

We first investigate the high $|g|$ limit for a $t_{\bot}$ term relevant at our length scales. Whilst we focus in particular here on cases at half-filling ($\mu = 0$), the results remain valid also when deviating from this. In the $t_{\bot} = 0$ phase diagram it was revealed by transforming into the low-energy subspace via a Schrieffer-Wolff transformation \cite{Herviou_2016}, that at sufficiently large $g>0$ a transition from the (doubly-topological) \emph{4MF} phase to a \emph{Mott insulating anti-ferromagnetic} (MI-AF) occurred. For $g<0$ conversely the transition was to a \emph{Mott insulating ferromagnetic} (MI-F) state. \\

To generalize the low-energy analysis of the Hamiltonian for large $|g|$ in the presence of a non-zero $\Delta\mu$, it is judicious to transform again into the bonding ($+$) and anti-bonding ($-$) basis \eqref{bonding_antibonding_basis} for which the Hamiltonian becomes block diagonal \eqref{Bonding-anti-bonding_Ham}. To obtain the spin-degrees of freedom characteristic of Mott physics, we perform additionally a Jordan-Wigner transform analogous to \cite{Herviou_2016}, defined by
\begin{equation}
\begin{aligned}
\sigma_{j}^{z} &=c_{j}^{+\dagger} c_{j}^{+}-c_{j}^{-\dagger} c_{j}^{-} \\
\sigma_{j}^{x} &=c_{j}^{+\dagger} c_{j}^{-}+c_{j}^{+\dagger} c_{j}^{+} \\
\sigma_{j}^{y} &=i\left(c_{j}^{-\dagger} c_{j}^{+}-c_{j}^{+\dagger} c_{j}^{-}\right).
\end{aligned}
\end{equation}
This transformation agrees with the one in \cite{Herviou_2016} in terms of $\sigma = 1$ and $\sigma = 2$ wire labels, as a Mott phase at half-filling implicitly means $n_1+n_2=1$ per rung. The change of basis in \eqref{bonding_antibonding_basis} leaves this invariant, i.e. $n^+ + n^-=1$. A subsequent Schrieffer-Wolff transformation results in
the low-energy effective Hamiltonian \cite{Herviou_2016}
\begin{equation}\label{EffHam_lowE_high_g}
  \begin{aligned}
&H_{\mathrm{eff} , g+} =\frac{t^{2}-\Delta^{2}}{2 g} \sum_{j} \sigma_{j}^{z} \sigma_{j+1}^{z} + \frac{\Delta\mu}{2}\sum_{x} \sigma^{x}_{j}
 \\ & +\frac{t^{2} + \Delta^{2}}{2 g} \sum_{j} \sigma_{j}^{y} \sigma_{j+1}^{y}+\frac{t^{2}-\Delta^{2}}{2 g} \sum_{j} \sigma_{j}^{x} \sigma_{j+1}^{x}.
\end{aligned}
\end{equation}
Taking $t = \Delta$ in the effective low-energy Hamiltonian in \eqref{EffHam_lowE_high_g} reduces the model to a   1$D$ quantum Ising model.  Compared to the low-energy Hamiltonian in  \cite{Herviou_2016}, the inter-wire hopping amplitudes map onto a transverse field contribution $\sim t_{\bot}$. As long as it is sub-dominant, i.e. $t^{2}/g > \Delta\mu$, we expect similar physics as in the $t_{\bot}$ limit, i.e. a (anti-)ferromagnetic ordering in $y-$direction:
\begin{equation}\label{MI_ordering_y}
    \sigma_{j}^{y} \sigma_{j+1}^{y} |GS\rangle = - |GS\rangle.
\end{equation}
We assumed here the limit $t=\Delta$ in Eq. \eqref{EffHam_lowE_high_g}, thus effectively dropping the $xx$ and $zz$ terms. As can be seen in figure \ref{fig:yyspins_notp} in Appendix \ref{Appendix:Figures} in the absence of a $t_{\bot}$ the $\sigma^{y}$ two-point function shows an (anti-)ferromagnetic ordering for large $|g|$. \\

\begin{figure}[h!]
    \centering
    \includegraphics[width = 0.5\textwidth]{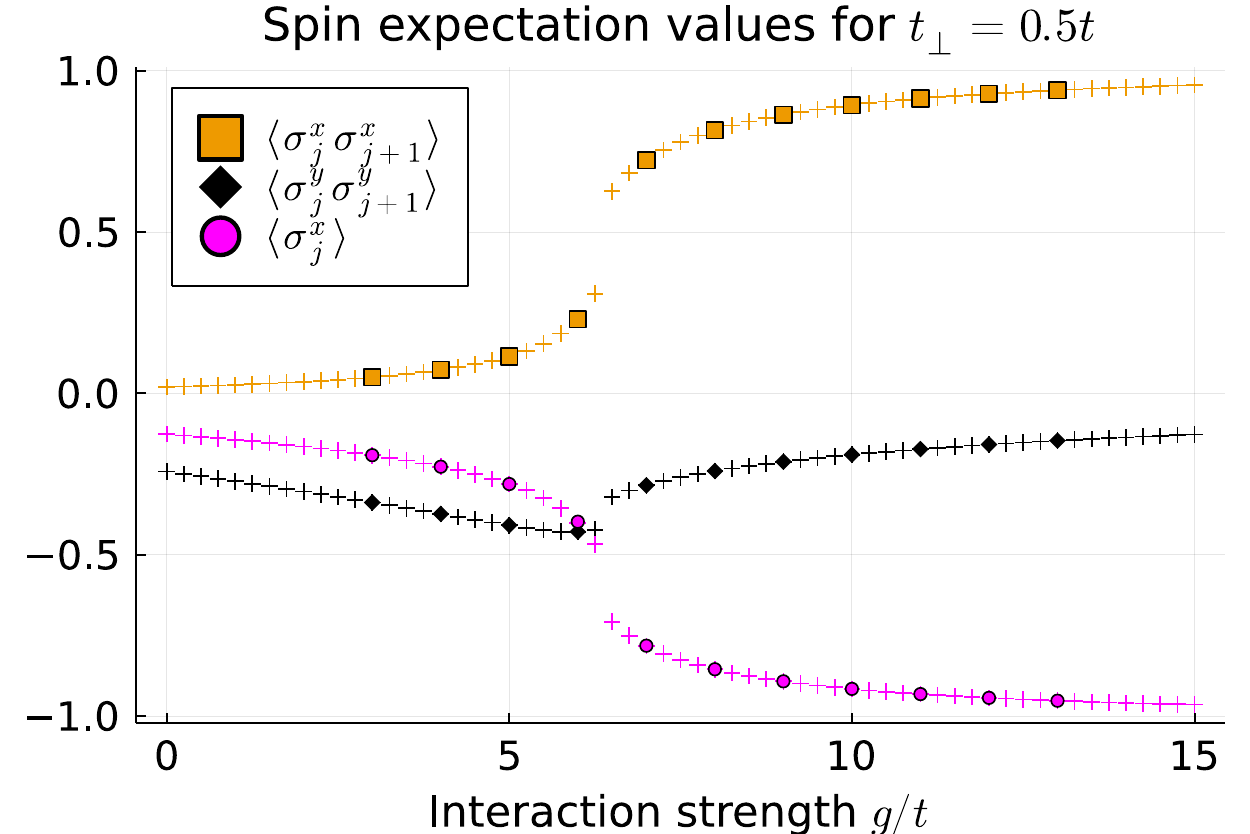}
    \caption{ One and two-point correlation functions for both $x$ and $y$ spin components extracted from the DMRG GS, in the presence of an inter-wire hopping $t_{\bot} = 0.5$. Crosses correspond to DMRG calculations with $L =20a$, whilst filled markers to $L = 30a$. The transverse field contribution $\Delta \mu$ drives the paramagnetic ordering, as can be seen by the $\langle \sigma^{x}\rangle$ and  $\langle \sigma^{x}_{j}\sigma^{x}_{j+1}\rangle$ correlation functions, which rapidly approach the values in \eqref{paramagnetic_ordering_x} for $g \approx 6t$.}
    \label{fig:xandyspins}
\end{figure}

At $\Delta\mu \sim t^{2}/g$, a transition to a paramagnetic ordered phase occurs, characterized by
\begin{equation}\label{paramagnetic_ordering_x}
\begin{aligned}
        \sigma_{j}^{x} |GS\rangle =& - |GS\rangle \\
    \Rightarrow    \sigma_{j}^{x} \sigma^{x}_{j+1}|GS\rangle =&  + |GS\rangle.
\end{aligned}
\end{equation}
The one and two-point functions of $\sigma^{y}$ are zero in this ground state. In the intermediate regime, i.e. finite $g$ or non-zero $t$, both the antiferromagnetic and paramagnetic ordering compete against each other. For large enough $\Delta \mu \sim t_{\bot}$, this intermediate phase can be tuned away completely. We test this numerically with DMRG, by extracting the spin and spin-spin correlation functions from the GS. We find that up to a threshold interaction $g_{crit}$ both a $\sigma^{y}_{j}\sigma^{y}_{j+1}$ and $\sigma^{x}$ ordering builds up, \emph{cf.} figure \ref{fig:xandyspins}. However, after crossing $g_{crit}$, the $\sigma^{y}\sigma^{y}$ correlations decay quickly to zero and only the ordering in $\sigma^{x}$ direction rapidly emerges. This is precisely the behaviour predicted from the low-energy effective Hamiltonian \eqref{EffHam_lowE_high_g}. \\

If we take $g > 0$, then we expect the transition from the Mott insulating MI-AF (y-ordered) to the \emph{paramagnetic} phase (x-ordered) to be a second-order phase transition. For this transition a power-law behaviour for the $\langle\sigma^{y}_{j}\sigma^{y}_{j+1}\rangle$ correlation function is expected, whilst $\langle\sigma^{x}\rangle$ should be continuous as a function of $g$ \cite{Pfeuty}. Similar arguments can be made for $g< 0$ and the MI-F phase. A more in-depth numerical analysis of this transition is necessary and may be done at a later stage, for example when investigating the role of Mott-physics for strongly interacting wire structures.

\subsection{Phase diagram in the high $t_{\bot}$, low $g$ limit}\label{sec:2MFphases}
We now turn to the phase diagram at interactions below the Mott scale, and focus on the stability of the remaining phase diagram in \ref{fig:trajectories_tp05} in the presence of an inter-wire hopping $t_{\bot}$. Recognizing the structural similarity of the Hamiltonians \eqref{Bonding-anti-bonding_Ham} and \eqref{Ham_coupled}, we generalize the definitions of the mixed wires fermions $\Gamma$, $\Theta$, and the chiral modes $\psi_{R/L}$ to the bonding/anti-bonding bands. 
Adapting the labels $\sigma =1,2$ with $\kappa = \pm$ in the previous definitions in Sec. \ref{sec:Fractional}, we define henceforth 
\begin{equation}\label{Mixed_wire_ferms_def_tperp}
\begin{aligned}
\Gamma_{j} \equiv&   \frac{1}{2}\left(\gamma_{j}^{A,+} + i \gamma_{j}^{A,-}\right) \\ 
\Theta_{j} \equiv& \frac{1}{2}\left(\gamma_{j}^{B,-} - i \gamma_{j}^{B,+}\right).
\end{aligned}
\end{equation}
The chiral modes are then defined exactly as in \eqref{chiral_modes_def}. { We now treat the wire-mixing component to the Hamiltonian $\sim t_{\bot}$ as a perturbation on top of the Hamiltonian in \eqref{Final_own_bosonized}, and we investigate its effects on the phase diagram using the same bosonization procedure as in Sec. \ref{sec:DCI}.\\

First, we rewrite the (band-) fermions $c^{\pm}$ in terms of the mixed-band fermions $\Gamma$ and $\Theta$. By comparing with previous expressions, we readily notice that  
\begin{equation}     \Gamma_{j}^{\dagger}\Theta^{\dagger}_{j} + \text{h.c} = n_{j}^{+} - n_{j}^{-}.
\end{equation} 
Therefore, the additional contribution to the total Hamiltonian takes the simple form $\sim \sum_{j} \Gamma_{j}^{\dagger}\Theta^{\dagger}_{j} + \text{h.c}$. Written in terms of the chiral modes $\psi_{R/L}$, this is equal to
\begin{equation} 
i\frac{\Delta\mu}{2}\int_{x} \text{d}x \left(\psi^{\dagger}_{R}\psi^{\dagger}_{L} - \psi_{L}\psi_{R}\right).
\end{equation}  
This is reminiscent of a superconducting pairing term \emph{cf.} \eqref{Ham_c}.} Indeed,  after performing the bosonization with analogous conventions as prior \emph{cf.} equation \eqref{bosonization_defs}, one finds to lowest order 
\begin{equation}
     i\left(\psi^{\dagger}_{R}\psi^{\dagger}_{L} - \psi_{L}\psi_{R}\right) = -\frac{1}{\pi \alpha} \cos\left(2\theta\right) + \mathcal{O}\left(\alpha^{0}\right).
\end{equation}
The short-distance cut-off $\alpha$ is again taken to be of the order of the lattice constant $a$. Thus, with a non-zero $\Delta\mu \sim t_{\bot}$ the full Hamiltonian is given by 
\begin{equation}\label{Boson_hamiltonian_final}
\begin{aligned}
   H =& \frac{v_{F}}{2\pi}\int_{x}\text{d}x \left[ K^{-1}\left(\partial_{x}\phi\right)^{2} + K\left(\partial_{x}\theta\right)^{2}\right] \\ &- \int_{x}\text{d}x \Big[g_{\phi} \cos\left(2\phi\right) - \gamma_{\theta}\cos\left(2\theta\right)\Big].
\end{aligned}
\end{equation}
The LL parameters are defined similarly to the $t_{\bot} = 0$ case as $v_{F}K^{-1} = a\left(4t  - \delta \overline{ \mu}  + \frac{2g}{\pi}\right)$, $g_{\phi} = \left(\frac{\delta \overline{\mu}}{2\pi \alpha} +  \frac{ga}{2\pi^{2}\alpha^{2}}\right) $. The effect of a non-zero $t_{\bot}$ enters as $\gamma_{\theta} = \frac{\Delta\overline{\mu}}{2\pi\alpha} \sim t_{\bot}$. In terms of the short-distance cut-off $a$ the flow equations {for both interaction } parameters $g_{\phi}$ and $\gamma_{\theta}$ are given by
\begin{equation}
\begin{aligned}
\frac{d g_{\phi}}{d l}=&\left(2- K\right) g_{\phi} \\ 
\frac{d \gamma_{\theta}}{d l}=&\left(2- K^{-1}\right) \gamma_{\theta},
\end{aligned}
\end{equation}
whilst for the LL parameter $K$ conversely the flow is described by 
\begin{equation}
    \frac{d K}{d l}=-\frac{4 \pi^{2}}{v_{F}^{2}}\left(g_{\phi}^{2} -\gamma_{\theta}^{2} \right)K^{2}.
\end{equation}
Therefore, a non-zero $t_{\bot}$ adds a gap to the critical chiral modes. Along the previously \emph{double critical Ising} line, i.e. for $g_{\phi} = 0$, the additional $\cos\left(2\theta\right)$ will keep a gap open. Thus, the single $c= 1$ critical line is replaced with two $c=1/2$ critical lines at $g_{\phi} = \pm 2\gamma_{\theta} \sim t_{\bot}$. \\

\begin{figure}[h!]
    \centering
    \includegraphics[width = 0.475\textwidth]{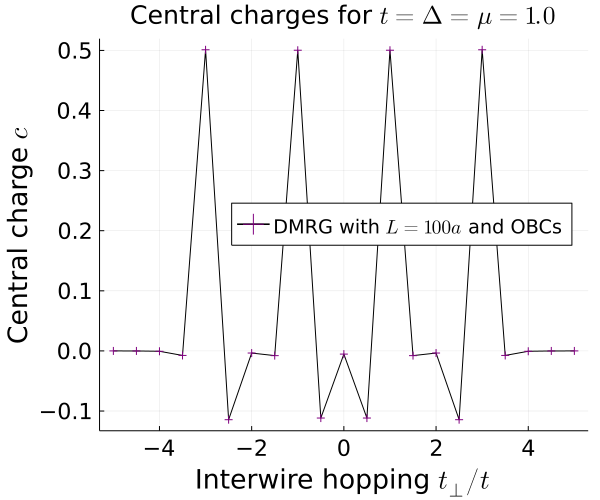}
    \caption{Central charges for $t = \Delta = 1.0$, $g = 0$ and $\mu = t$, obtained from the entanglement entropy for $L=100a$ and OBCs. The clear $c = 0.5$ peaks at the transitions $|t_{\bot}| = 2t\pm \mu \in \{-3,-1,+1,+3\}$ are visible. 
    % {\color{black}As DMRG is not necessary at $g = 0$, this offers a good benchmark for successive runs.}
    }
\label{fig:four_lifshitz_transitions}
\end{figure}

To understand the nature of these transition lines, it is judicial to re-fermionize the Hamiltonian and introduce Majorana fermions \cite{ Herviou_2016, Karyn1999}
\begin{equation}\label{re-fermionized_Hamiltonian_tp_relevant}
\begin{aligned}
H =& \sum_{\kappa=+,-} \int \text{d} x \frac{\left(-i v_{F}\right)}{2} (\gamma_{R}^{\sigma} \partial_{x} \gamma_{R}^{\sigma}-\gamma_{L}^{\sigma} \partial_{x} \gamma_{L}^{\sigma}) \\ 
- & i\pi\int \text{d} x \Big(\left(g_{\phi}+\gamma_{\theta}\right) \gamma_{R}^{+} \gamma_{L}^{+}+\left(g_{\phi}-\gamma_{\theta}\right) \gamma_{R}^{-} \gamma_{L}^{-}\Big).
\end{aligned}
\end{equation}
In this representation, it is visible that each line corresponds to a topological phase transition of one of both $\pm$ bands, independently. Along the fine-tuned line $g_{\phi} = \gamma_{\theta}$, the $-$ wire (band) is gapless and has a central charge of $c = 1/2$, whilst for $g_{\phi} = -\gamma_{\theta}$ the bonding band ($+$)  is gapless, i.e. massless. This corresponds to an Ising transition with a central charge $c=1/2$. We probed these transitions using DMRG for an open chain. Figure \ref{fig:four_lifshitz_transitions} shows the central charges, extracting from the entanglement entropy, as a function of $t_{\bot}$ for the non-interacting case $g = 0$ and $\mu = 1$. The four $c=1/2$ critical points at $|\mu| = 2t \pm t_{\bot}$, as well as their equidistant separations are visible.
The re-fermionized Majorana representation \eqref{re-fermionized_Hamiltonian_tp_relevant} also reveals that at least one of both pairs of chiral Majorana modes $\gamma^{\pm}_{R/L}$ will always have a non-zero mass-term $\sim i\gamma^{\pm}_{R}\gamma^{\pm}_{L}$. Adding an interaction $g$ only linearly shifts $g_{\phi}$ away from its non-interacting value. The phases for $g =0$ in \eqref{fig:PD_g0_tp} will therefore also extend to non-zero interaction strengths. The transition lines $g_{\phi} = \pm \gamma_{\theta}$ separate the phases and are given by straight lines $g\left(\delta\mu\right)$, for appropriate $g$ and $\delta\mu$ where \eqref{Boson_hamiltonian_final} is accurate.
\\

The central charge is a useful marker to identify critical regions of the phase diagram, since a gap results in $c = 0$. Thus, to investigate the phase diagram and characterize the topological properties in the presence of $t_{\bot} \sim t$, we make use again of the markers \eqref{Combined_pole_equations}. As we performed a change of basis to obtain the Hamiltonian \eqref{Bonding-anti-bonding_Ham}, it is also necessary to adapt the definition of $C$ and $\overline{C}$ in terms of the correlation functions, to obtain the correct markers for the bonding and anti-bonding bands, {labeled henceforth as $C^{\pm}$ and $\overline{C}^{\pm}$. They can be obtained from the wire basis by }adding to each $\mathfrak{S}^{z}_{i}$ the following mixed-wire correlation functions
\begin{equation}
  \mathfrak{S}^{z}_{i} \longrightarrow \mathfrak{S}^{z}_{i} + \frac{1}{2}\sum_{j}\left(c^{+\dagger}_{j}c^{-}_{j+i} - c^{-\dagger}_{j}c^{+}_{j+i} + \text{h.c.}\right).
\end{equation}
{ As predicted from the bosonized Hamiltonian \eqref{re-fermionized_Hamiltonian_tp_relevant} we expect two additional and distinct phases, next to the ``stable" \emph{4MF} and two polarized ones. This coincides with previous results for the Kitaev ladder, for example in \cite{ladder1} and \cite{ladder2}, in the non-interacting limit. 

We unravel the topological properties of these additional phases using the adapted topological numbers \eqref{Combined_pole_equations}, showing that the additional phases are \emph{2MF}, with a single Majorana edge mode per side for both wires.} We label these \emph{2MF-a} and \emph{2MF-b}, where the $a$ phase lies to the left of $\mu = 0$ and the $b$ phase to the right. { We performed several DMRG calculations, verifying the phase diagram in \ref{fig:PD_g0_tp} for sufficiently \footnote{ie. where no transition to \emph{Mott} physics has occurred.} low interaction $g$ and relevant inter-wire-hopping $t_{\bot}$, i.e. where no finite-size effects may occur. \\
\begin{figure}[h!]
    \centering
    \includegraphics[width = 0.475\textwidth]{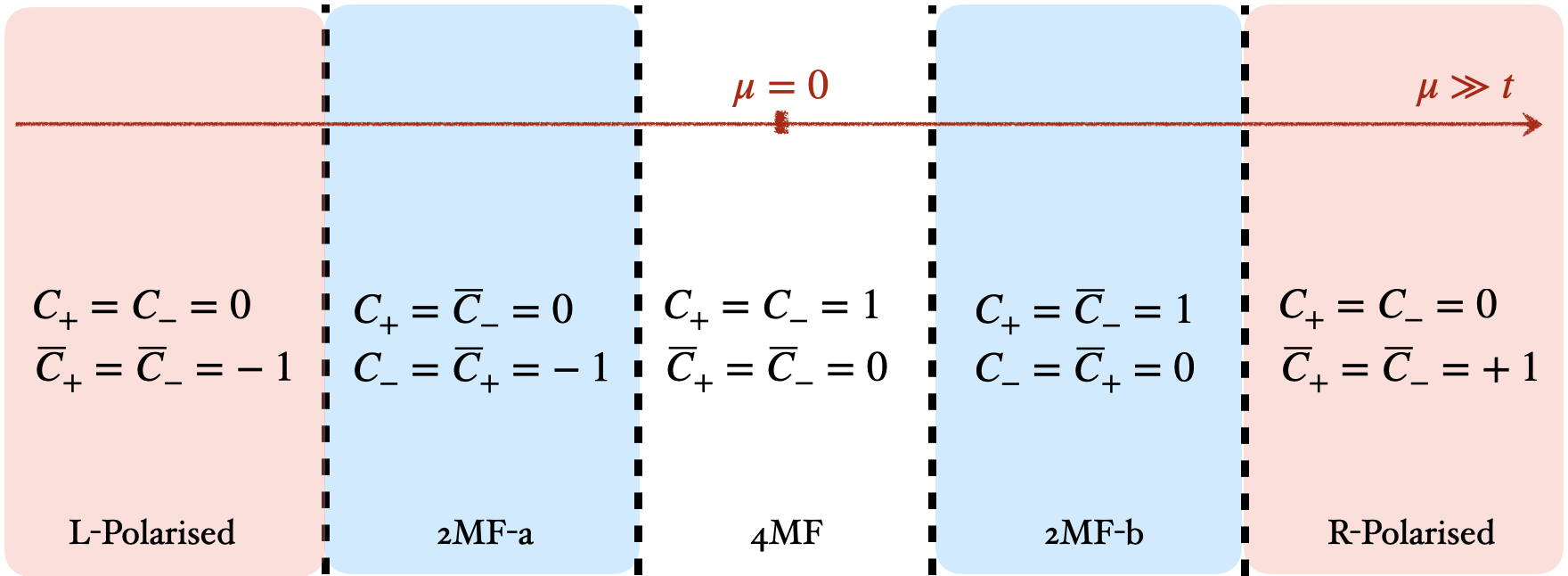}
    \caption{Phase diagram in the presence of $t_{\bot} \sim t$, which results in two additional phases \emph{2MF-a} and \emph{2MF-b}, with non-zero width $\sim t_{\bot}$ also at $g= 0$. We tested the diagram for up to relatively high $g\sim 5t$. Above this, a transition to \emph{Mott} physics occurs, as discussed prior in Sec. \ref{sec:high_g_param}. For non-zero interactions, the critical lines extend into sloped lines.}
    \label{fig:PD_g0_tp}
\end{figure}

The phase diagram was probed extensively, and we present the numerical results along three distinct ``trajectories" \emph{C2, C4} and \emph{B1} \emph{cf.} figure \ref{fig:trajectories_tp05}, in the $g-\mu$ plane. Without loss of generality, we present here the results for $t_{\bot}/t = 1/2$. The various phases are classified using the Chern $C^{\pm}$ and $\overline{C}^{\pm}$ numbers, which we extracted from the DMRG obtained GS. Beginning with the trajectory labeled as \emph{B1}, which starts deep in the polarized phase and then drives upward with interaction strength $g$, \emph{cf.}  figure \ref{fig:traj_B1}. The $C^{\pm}$ and $\overline{C}^{\pm}$ numbers are very stable against increasing $g$, compared to previous results for $t_{\bot} = 0$. The figure shows, that around $g \sim 3t$, the $+$band undergoes a transition from trivial to topological. The antibonding band ($-$)  remains trivial however, which signals the transition from the \emph{R-Polarized} to \emph{2MF-b} phases in figure \ref{fig:PD_g0_tp}.
\begin{figure}[h!]
    \centering
    \includegraphics[width = 0.475\textwidth]{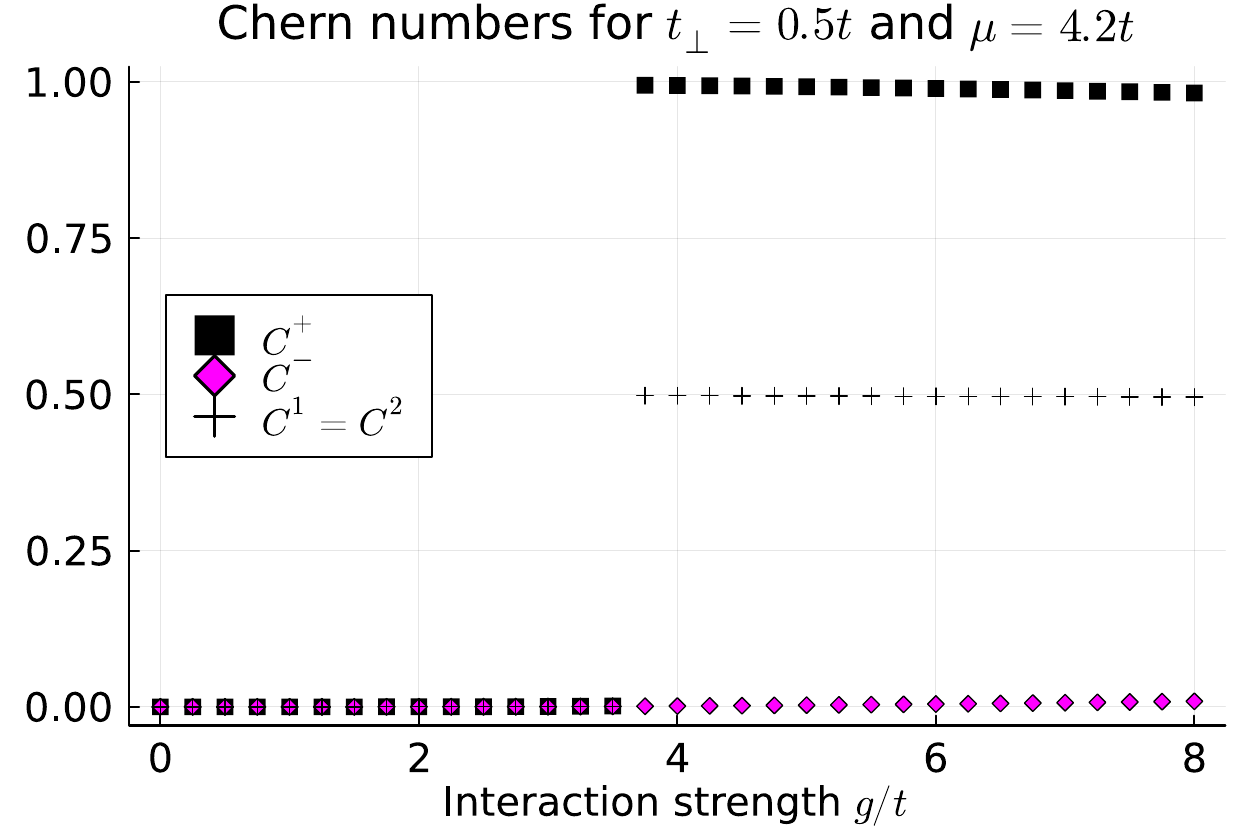}
    \caption{(Trajectory \emph{B1}) Transition from \emph{R-Polarized} to \emph{2MF-b}. DMRG for $t = \Delta = 1.0$ and $\mu = 4.2t$ and $L = 76a$ with PBCs. Inter-wire hopping $t_{\bot} = 0.5t$. The numbers $C^{\pm}$ are given in blue and black respectively, with (marginal) deviations from quantized $C^{+} = 1$ and $C^{-} = 0$ values.}
    \label{fig:traj_B1}
\end{figure}
As can be seen in figures \ref{fig:traj_C2} and \ref{fig:cbar_pd} in the Appendix, trajectory \emph{C2} drives with $\mu$ at constant $g = 0.5t$, and thus probes the transitions from \emph{4MF} to \emph{2MF-b} into the \emph{R-Rolarized} phase. The \emph{4MF} phase, signaled by $C^{\pm}= 1 $ and $\overline{C}^{\pm} = 0$, transitions into the \emph{2MF-b} phase around $\mu/t \sim 1.5$. Here $\overline{C}^{-}$ becomes one and $C^{-} = 0$, whilst the bonding band ($+$)  remains topological. Around $\mu/t = 2.75$ both bands become trivial, and the \emph{R-Polarized} phase is determined by $\overline{C}^{\pm} = 1$, corresponding to the r.h.s. of the phase diagram \ref{fig:PD_g0_tp}. This is revealed by the topological numbers in \ref{fig:traj_C2}, where both $\pm$ bands are in a topological $C^{\pm} = 1$ and $\overline{C}^{\pm} = 0$ phase up to $\mu/t \sim 1.5$ . Then, the $-$ Band becomes trivial, followed by bonding band ($+$) around $\mu/t \sim 2.8$.\\
\begin{figure}[h!]
    \centering
    \includegraphics[width = 0.475\textwidth]{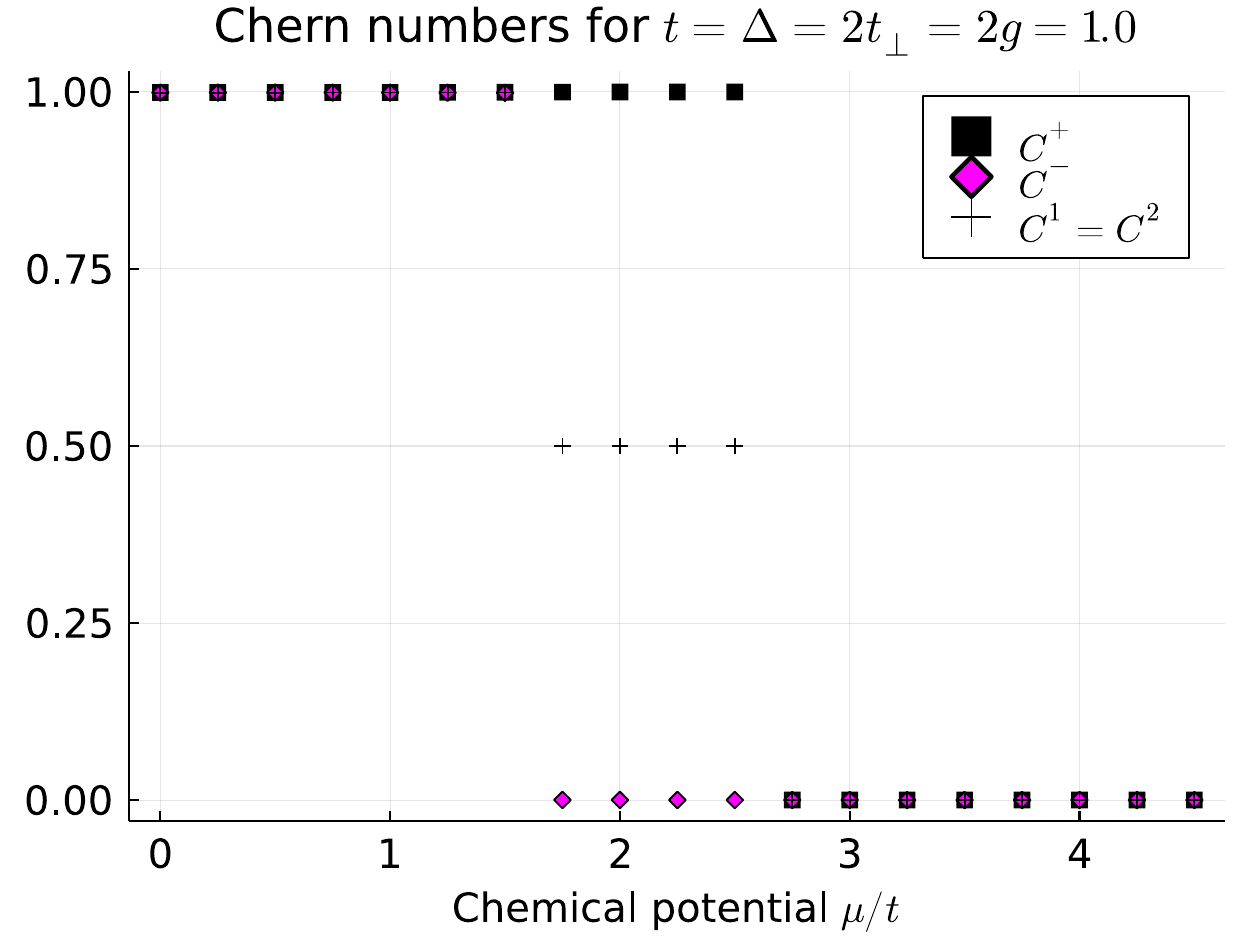}
   \caption{Trajectory \emph{C2}: Chern numbers ($C^{\pm}$) and $C^{1/2}$, with values almost quantized to $0,1/2,1$. See \ref{fig:cbar_pd} in Appendix \ref{Appendix:Figures} for duals $\overline{C}$.}
    \label{fig:traj_C2}
\end{figure}

The figures \ref{fig:C4_1} and \ref{fig:C4_2} show the trajectory \emph{C4} from DMRG with $L = 76a$ and PBCs, where the transition \emph{4MF} to \emph{2MF-b} is again hailed by $C^{+} = 0$ and $\overline{C}^{+} = 1$. It occurs around $\mu/t \sim 2.4$, whilst the subsequent transition to \emph{R-polarised} occurs at $\mu/t = 5$, and is signaled by the jump in the topological numbers. The phase remains stable also in the presence of strong interactions $g$, and comparison with figure  \ref{fig:traj_C2} reveals the shifting of the transition points $\mu_{crit}$ to higher values - i.e. the separatrix is sloped as in the $t_{\bot} = 0$ case in figure \ref{fig:trajectories_tp05}. Due to the inherent $\mathbb{Z}_{2}$ symmetry, the phase diagram is again symmetric wrt. $\mu \rightarrow -\mu$. However, the distinction between \emph{L-Polarized} and \emph{R-Polarized} is again possible with the dual numbers $\overline{C}^{\pm}$ which change signs.\\ \\
\begin{figure}[h!]
  \centering
  \includegraphics[width=.93\linewidth]{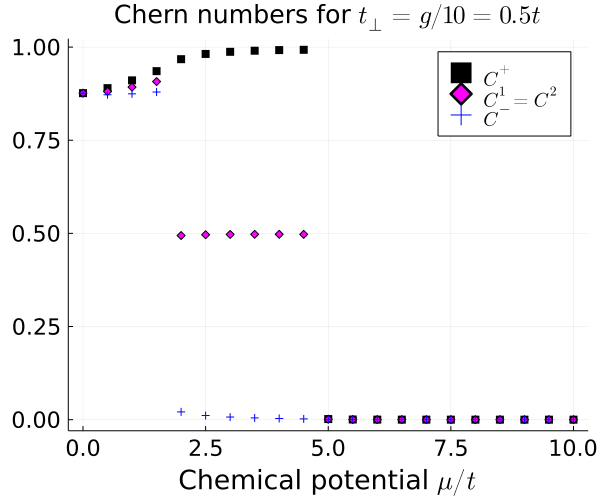}
  \caption{$C^{\pm}$ and $C^{1/2}$ for $t = \Delta = 1.0$ and $g = 5t$ and inter-wire hopping strength is $t_{\bot} = 0.5t$. Band Chern numbers $C^{\pm}$ are blue and black respectively, whilst $C^{1/2}$ is given in pink.}
  \label{fig:C4_1}
\end{figure}

The topological markers $C^{\pm}$ and $\overline{C}^{\pm}$ reveal the non-trivial topology of the \emph{2MF-a} and $\emph{-b}$. However, the bonding-/anti-bonding basis explicitly mixes both wires. To investigate the properties of a single wire it is necessary to consider $C^{1}$ and $C^{2}$ as well.  Extracting the relevant correlation functions reveals that, in the two \emph{2MF} phases each wire has effectively $C = 1/2$ as well, as can be seen in figure \ref{fig:traj_C2}. However, as known from the $c = 0$ central charges in \ref{fig:four_lifshitz_transitions}, this cannot be a DCI phase, { or any critical phase for that matter. Instead, since the $\pm$ basis mixes both wires equally, we infer that it corresponds to a cat-like superposition of the wires: let $T = 1$ label topological and $T = 0$ trivial, then the ground state of both wires is $\sim |T= 0, T=1\rangle + |T = 1, T =0\rangle$. Thus with OBCs, the combined system of two wires admits \emph{two} dangling edge modes for a total of \emph{four} edges, i.e. one Majorana on each side. Thus the Majorana has an equal probability to be localized on either of the two edges of the two wires. This results in an effective $C^{\sigma} = 1/2$, i.e. another instance of fractional topology fundamentally distinct from the DCI phase. To complement our current results and offer deeper insights into the two \emph{2MF} phases, a next study could focus with more detail on these shared edge modes and their, possibly, new applications. \\

\begin{figure}[h!]
  \centering
 \includegraphics[width=.93\linewidth]{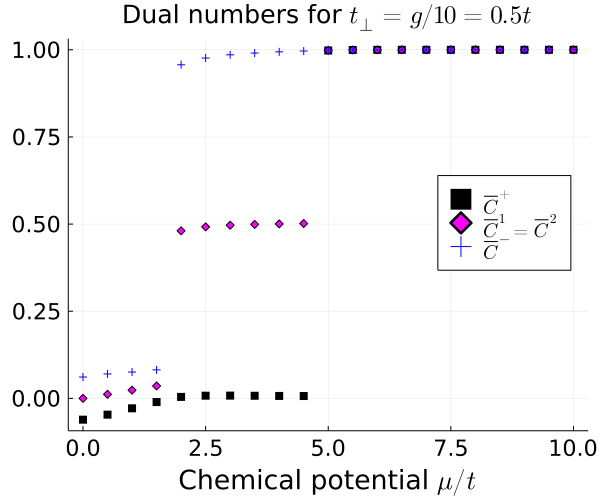}
   \caption{Dual numbers $\overline{C}^{\pm}$ are given in black and blue respectively, whilst $\overline{C}^{1/2}$ are presented in pink.}
   \label{fig:C4_2}
\end{figure}

The occurrence of the shared-edge modes may have interesting implications for the topological properties of coupled wire systems, and there are questions which remain to be addressed in future work. For example, do the shared edges have measurable effects on the topological zero-bias-peaks (ZBPs), which are the result of tunneling into the edge modes \cite{Pan_2021}? Can such shared Majorana states offer a potentially clear experimental signature for the existence of Majorana zero modes (MZMs)? The relevance of Majorana physics, both academically and in technological endeavours, warrants further investigation.

\subsection{Stability of the DCI phase for finite sizes and small $t_{\bot}$}

In the previous section we focused on the low-$g$ and high $t_{\bot}$ limit, yet the Hamiltonian in \eqref{re-fermionized_Hamiltonian_tp_relevant} is also valid at large $g$ and far away from half-filling. At $t_{\bot} = 0$, the existence of an extended DCI phase was predicted \cite{Herviou_2016} from RG arguments. Away from the critical line $g_{\phi} = 0$ the Hamiltonian in \eqref{Final_own_bosonized} is a priori gapped by the operator $\cos\left(2\phi\right)$, which scales as $\sim K$.  At large enough interactions the LL parameter $K$ can flow to values $K > 2$, rendering the $\cos\left(2\phi\right)$ operator irrelevant in the RG sense, with the effect that the critical line opens to an extended DCI phase. \\ 

In the present case, we showed that in the chiral $\psi_{R/L}$ basis a nonzero $t_{\bot}$ results in an additional $\sim \cos\left(2\theta\right)$ operator in \eqref{Boson_hamiltonian_final}. The dual field $\theta$ scales inversely to $\phi$ with dimension $K^{-1/2}$, which implies $\cos\left(2\theta\right) \sim K^{-1}$. The effect is clear: whilst $\cos\left(2\phi\right)$ becomes irrelevant for $K>2$, the $\cos\left(2\theta\right)$ operator is super-relevant. Thus, the gapless DCI phase cannot open in the continuum (thermodynamic) limit, and the critical lines are instead $c = 1/2$ seperated by $\sim t_{\bot}$, \emph{cf.} figure \ref{fig:four_lifshitz_transitions}. However, for} finite systems we may yet find some parameter range where the DCI phase remains stable against small $t_{\bot}$. 
% This is because, as long as the inter-wire hopping amplitude $t_{\bot}$ is not large enough to open a gap in the critical modes, the physics far below the length scales of $1/t_{\bot}$ should remain unperturbed. 
A way to understand this is by considering instead a temperature scale {$T^{*}$} set by the energy-gap $\sim \gamma_{\theta}$. Due to the linear spectrum of the chiral modes, their energy is simply set by $E_{G} \sim v_{F}k$, in units of $h/2\pi = 1$. Similarly, at finite temperature { T} we find $E_{gap} \sim T$, for $k_{B} = 1$. The criticality condition is thus when both scales are comparable, i.e. $E_{gap} \sim E_{G}$. Therefore, with the lowest momentum number $k$ at finite length being set by $k \sim 1/L$, we find simply $T^{*} \sim v_{F}/L^{*}$, i.e. when $L^{*} \sim 1/T^{*}$.} With lengths of the order of a few hundred sites, we thus expect the DCI phase to persist up to about $t_{\bot}/t \sim 10^{-3}$ or $t_{\bot}/t \sim 10^{-2}$.
However, due to the presence of interactions and superconducting pairings, the bare parameters \emph{flow} in the RG sense. It is therefore the \emph{effective}, renormalized $t_{\bot}$ which is relevant in this analysis. As we expect the system parameters to flow to larger values under an RG process using the definitions of \cite{Herviou_2016}, the critical value $t^{*}_{\bot}$ for which the DCI phase can be observed is expected to decrease.

\begin{figure}[h!]
    \centering
    \includegraphics[width = 0.5\textwidth]{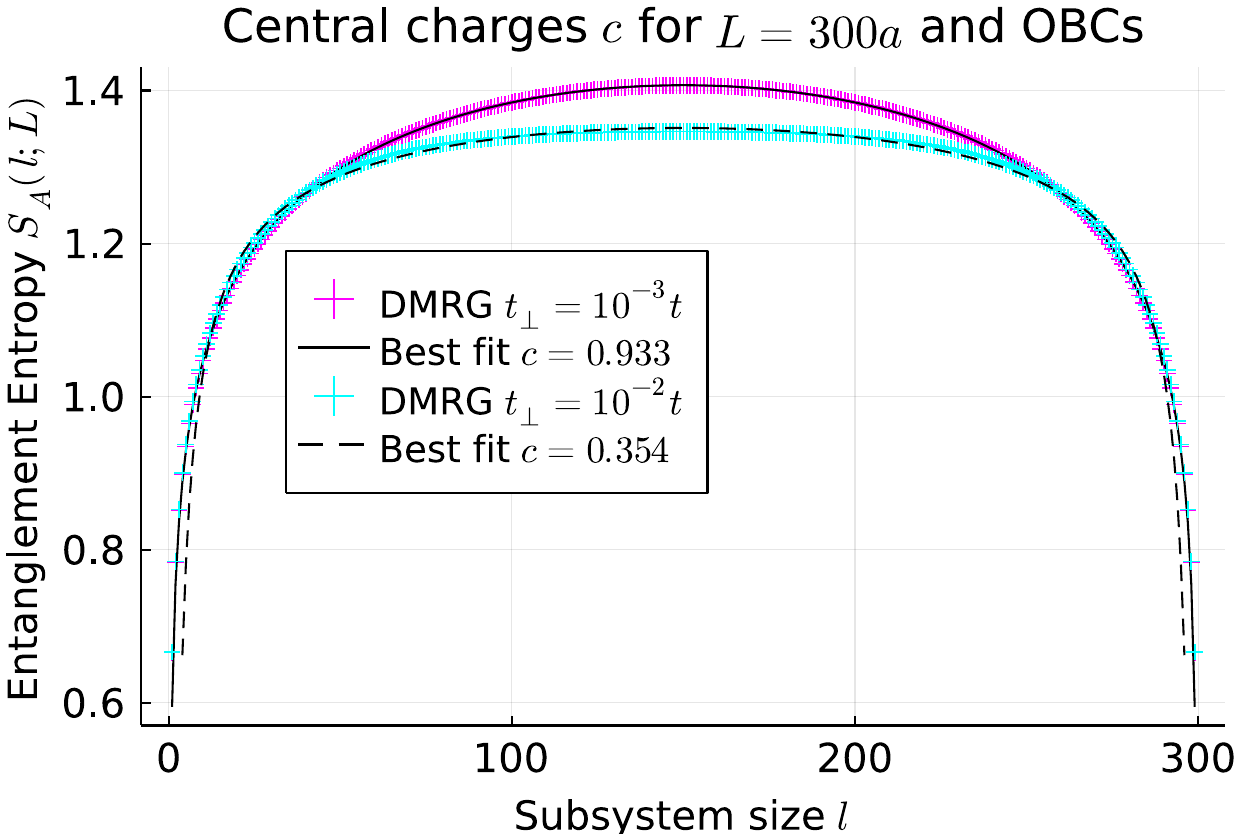}
    \caption{Entanglement entropies and central charges for $t_{\bot} = 10^{-3}t$ and $10^{-2}t$. Results extracted from OBC DMRG calculations for $L=300a$ per wire, and $t= \Delta = 1.0$. The interaction strength is $g = 5.0t$ and the chemical potential is set to $\mu = 3.85t$. A plateau, visible for $t_{\bot} = 10^{-2}t$, signals the non-criticality ($c = 0$) of the system. The central charge $c = 0.354$ is nonphysical and due to the finite system size, \emph{cf.} Appendix \ref{appendix:fitting}.}
    \label{fig:EntTPOBC_300}
\end{figure}
Numerically we analysed both the entanglement entropy and central charge, as well as the bipartite charge fluctuations \cite{Herviou_2016,Song_2010} to determine the stability of the DCI phase at different length scales. As seen for OBCs at $L = 300a$ in figure \ref{fig:EntTPOBC_300}, the entanglement entropy for $t_{\bot} = 10^{-3}t$ shows clear logarithmic (critical) behaviour. A central charge of $c \approx 1$ is found after fitting \emph{cf.} equation \eqref{ententropy_def}. Conversely, at $t_{\bot} = 10^{-2}t$, a visible plateau emerges, indicating non-criticality of the system. This is also supported by a central charge of (approximately) zero. Extracted from the logarithmic contribution of the entanglement entropy, \emph{cf.} equation \eqref{ententropy_def}, the central charge also depends on boundary effects for finite systems, as well as the fitting domain chosen for the analysis. Thus, a nonphysical central charge of $c > 0$ may emerge, however by modifying the fitting regime so that only the central, plateaued region is considered, yields $c \approx 0$. Conversely, for a critical phase, changing the fitting regime only impacts the resulting central charge slightly. Alternatively, we also considered larger systems sizes as well as open boundary conditions, both of which verified the non-criticality. Therefore, the phase becomes gapped beyond $t_{\bot} \sim 10^{-2}t$, and previous results in \ref{sec:2MFphases} as well as \cite{ladder1, ladder2} imply we are in the topological \emph{2MF-b} phase \emph{cf.} figure \ref{fig:PD_g0_tp}.\\ 

As a final test, we extracted as well the bipartite charge fluctuations, which are predicted to have a negative sub-dominant logarithmic contribution in the DCI phase \cite{Herviou_2016}. Due to the much more dominant linear contribution, it is more strategic to analyse instead the functional behaviour of 
 \begin{equation}
     F_{log}\left(l\right)  \equiv F_{Q}\left(2l\right) - 2F_{Q}\left(l\right),
 \end{equation}
for which any linear term is automatically removed. However, due to the convention of signs, the logarithmic contribution of $F_{log}$ is therefore expected to be \emph{positive} in the DCI phase (i.e. \emph{negative} when $c = 1$ per wire instead of $c= 1/2$ \cite{Herviou_2016}). As can be seen in figure \ref{fig:FlLogTPPBC_100}, and figure \ref{fig:FlLogTPPBC_76} in the appendix, this is indeed the case for $t_{\bot} \sim 10^{-3}t$. \\ 

Additional DMRG calculations are presented in figures \ref{fig:EntTPPBC_76} and \ref{fig:EntTPPBC_100} in the Appendix \ref{Appendix:Figures}. Similarly, in Appendix \ref{appendix:fitting}, results for smaller systems (order of tens of sites), and conversely larger $t_{\bot}$ (order of $0.1t$), offers more details and insight on the non-physical central charges of $c >0$ in the gapped, $2MF$ phases. We thus confirm a stability of the DCI phase against inter-wire hoppings up to values of order $10^{-3}t$, {for wires sizes of the order of a hundred sites. }

\begin{figure}[h!]
    \centering
    \includegraphics[width = 0.5\textwidth]{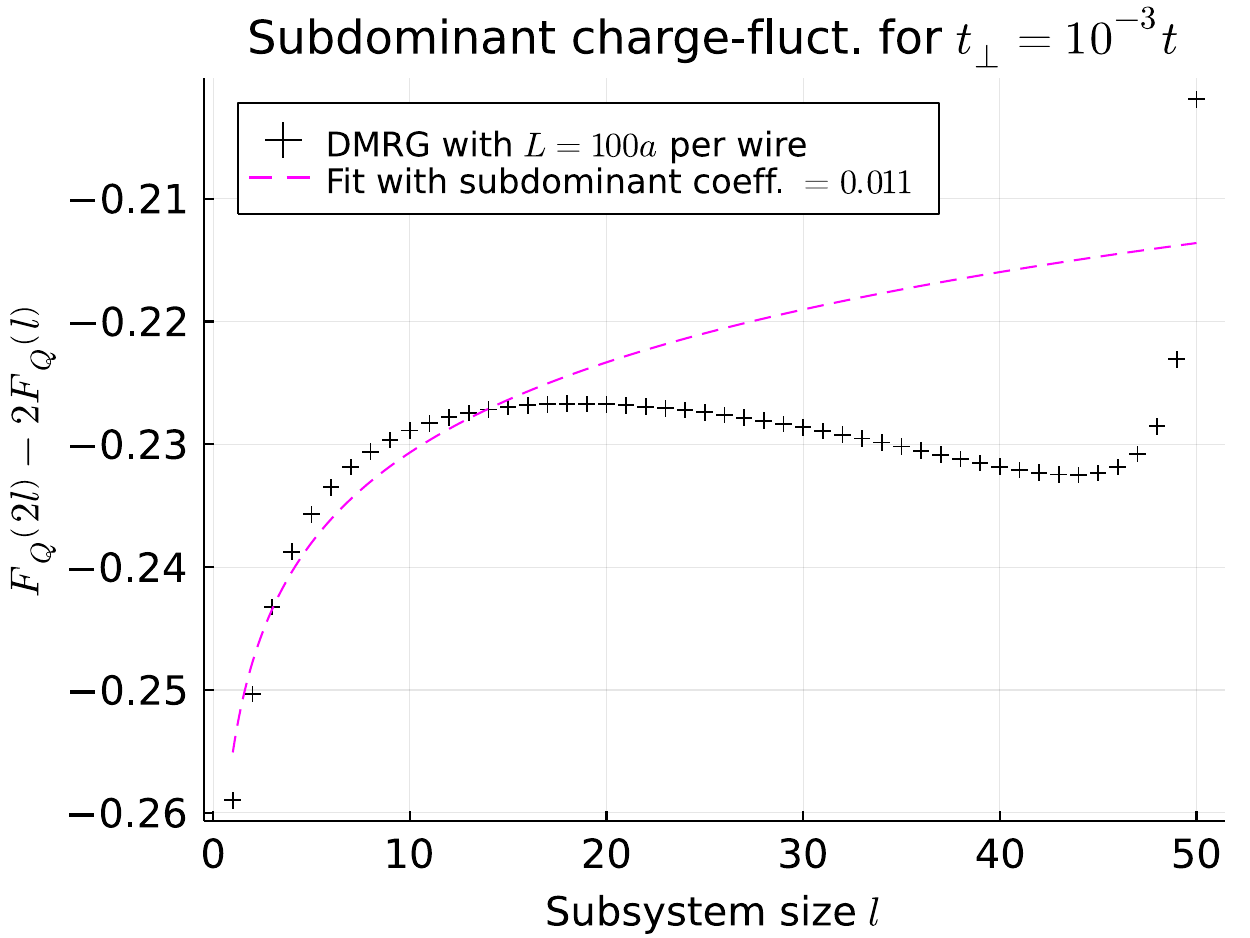}
   \caption{The subdominant logarithmic contributions are generally very fragile and susceptible to boundary-, and finite-size effects. A control run for $L = 100a$ per wire with $t_{\bot} = 10^{-3}t$  shows a clear \emph{positive} logarithmic cusp. The same holds true for smaller systems of $L = 76$, see figure \ref{fig:FlLogTPPBC_76}.}
    \label{fig:FlLogTPPBC_100}
\end{figure}

\section{Conclusion}\label{sec:Conclusion}
Different properties of two interacting, spinless p-wave superconducting wires were investigated to achieve a better understanding of the interacting physics of wire arrays or other heterostructures. Additionally, we also analyzed the effects an inter-wire hopping amplitude $t_{\bot}$ has on the phase diagram. Essential to this analysis were the two topological markers, the Chern number $C$ and its \emph{dual} $\overline{C}$ defined for superconducting wires, which can be expressed completely by two-point correlation functions in real-space. Our analysis presented perspectives and a deepened understanding of the \emph{double critical Ising} phase first discovered in \cite{Herviou_2016}, and revealed additional topological phases in the presence of inter-wire hopping amplitudes.

At the heart of our approach are the results in \cite{Hutchinson_2021}, which demonstrated how a Chern number can be defined purely from the poles for a Bloch sphere system. In momentum space the Kitaev wire can be mapped onto the Bloch sphere, such that topological numbers can be defined solely from expectation values of the particle-hole spin operator $S^{z}_{k}$ at $k = 0$ and $ka = \pi$. Since these expectation values can be expressed purely from real-space two-point correlation functions of the GS, they are particularly powerful numerical tools, with the potential also for experimental implementation. By studying the low-energy properties of two interacting wires we highlight, both theoretically and numerically with DMRG, that the topological markers remain a sensible tool to distinguish various topological phases of coupled wires and in the presence of relatively strong interactions.
These results demonstrate the usefulness of the topological numbers both as indicators for topological transitions and to distinguish the various phases. Since they are defined in real-space, the topological numbers in \eqref{Combined_pole_equations} could also provide a platform to investigate topology of Kitaev wires in the presence of disorder. Together with their expression as two-point correlation functions, we believe the topological numbers to be particularly interesting for both numerical and experimental applications. \\

The main focus of this paper was to deepen our understanding of the \emph{double critical Ising} phase (DCI). It appears as an extended gapless phase for two strongly interacting wires far from half-filling and is directly related to the quantum critical point (QCP) of the single Kitaev wire.  By revisiting the quantum field theory of the two wires in \ref{sec:DCI}, we showed how the model can be recast in terms of mixed-wire fermions $\Gamma, \Theta$ and subsequently two chiral Dirac fields $\psi_{R/L}$. The chiral Dirac modes are directly related to the chiral Majorana modes of each wire,  which provides further evidence that the extended critical region in \cite{Herviou_2016} is a \emph{double critical Ising} phase. Using DMRG to obtain the GS, we extracted the topological numbers $C$ and $\overline{C}$, which revealed that both are fractional $C = \overline{C} =1/2$ in the DCI phase. To the knowledge of the authors, this presents a first example of a fractional topological phase for interacting one-dimensional superconducting wires, possible due to the interplay of strong interactions and chemical potentials far from half-filling. These results further reinforce the correspondence between two interacting Bloch spheres \cite{Hutchinson_2021} and two Kitaev wires \cite{Herviou_2016}, beyond the perturbative limit of small interactions $g$. It also introduces a correspondence between one ``free" Majorana fermion at a pole on the Bloch sphere and
a gapless quantum fluid with central charge c=1/2 in real space that may deserve further analysis and applications.\\

Finally, using QFT methods, we demonstrate that in the thermodynamic limit, the inter-wire hopping $t_{\bot}$ will always gap out the DCI phase, resulting instead in two topological \emph{2MF} phases. However, at finite length scales for the system, the survival of the DCI phase is predicted and verified numerically for hopping amplitudes $t_{\bot}/t$ smaller than $\sim 1/L$. This may have important consequences for applications and modern technology, wherein components and constituents will inevitably have a finite length scale.
The numerical analysis in the presence of an inter-wire hopping $t_{\bot} > 0$ reproduced the results predicted from the QFT description, and offers a test of the phase diagram in figure \ref{fig:PD_g0_tp} up to large values of $g$ and $\mu$. Importantly, our investigation of the effects of an inter-wire hopping amplitude $t_{\bot}$ revealed the emergence of a topological phase characterized by two Majorana edge modes, shared between both ends of the wires on either side. Despite the relatively large interaction strength $g = 5t$, the topological numbers in \ref{fig:C4_1} and \ref{fig:C4_2} were found to only deviate marginally from the $g \approx 0$ values, \emph{cf.} figures \ref{fig:PD_g0_tp} and \ref{fig:traj_C2}. Together with the fractional value $C^{1} = C^{2} = 1/2$ in the wire basis, these results underline the power and usefulness of these topological numbers, not only as numerical markers to unravel the distinct topological phases of coupled super-conducting wires in the presence of (strong) interactions. For very strong interactions $|g| \gg t$ we found that an additional ordering in $x-$direction is introduced by a non-zero $t_{\bot}$, independent of $g$.  This result followed from a Schrieffer-Wolff transformation into the low-energy sub-sector. Since the \emph{(anti-)ferromagnetic} ordering in $y-$direction scales with $1/g$, a paramagnetic Mott phase opens up. This was also found numerically, with the transition seemingly being a smooth cross-over. However, a full numerical analysis of the Mott physics in the presence of an inter-wire coupling is still outstanding and could be an interesting next topic of study.
\\

The work presented in this paper not only revealed the \emph{doubly critical Ising} phase as one of fractional topology but also deepened the understanding of its physical properties. This was achieved both by the introduction of topological markers directly measurable from correlation functions and by the quantum field theoretic description of the gapless modes as chiral $R/L-$movers. With this enhanced ``theorists" toolbox, we are motivated to continue further research into interacting topological phases of superconducting wires. An immediate next step in the study of coupled-wire geometries is to consider the case of weakly-coupled ladder geometries, as an approximation to a quasi-two-dimensional model. Another interesting avenue of research is certainly the inclusion of light-matter interaction, and the DCI phase's response to external perturbations. Both could lead to further measurement protocols and topological probes. 

\begin{acknowledgments}
We thank colleagues and friends at the \emph{CPhT} of Ecole Polytechnique for their continued support and invaluable feedback. In particular, we thank E. Bernhardt and J. Schneider for their time and insight, and also thank J. Legendre, S. al Saati, J. Hutchinson, and C. Mora. F. del Pozo also thanks the \emph{IP Paris Ecole Doctorale} for its funding and support, and the organisers and participants of the TQM22, TMS22, and TOPDYN22 schools and workshops for the many fruitful and insightful discussions. L. Herviou acknowlesdges the support of the Swiss National Science Foundation (FM) grant 182179. This work was supported by the Deutsche Forschungs-gemeinschaft (DFG, German Research Foundation) under Project No. 277974659 via Research Unit FOR 2414 and via ANR BOCA. K.L.H also acknowledges interesting discussions related to this work at Aspen Center for Physics, which is supported by National Science Foundation grant PHY-1601671, at Dresden conferences TOPCOR22, TOPDYN22, and in the MANEP Workshop in Les Diablerets.
\end{acknowledgments}

\appendix
\section{Topology of the Bloch-sphere}\label{Appendix:Topo_BS}
An intuitive way to understand a non-zero Chern number (or TKNN invariant), is by considering a single-spin $\vec{S}$ in an external magnetic flux $\vec{\mathcal{B}}$. The Hamiltonian is determined from the interaction of the spin magnetic moment with the external field $H = -\mu_{B} \left(\vec{S}\cdot \vec{\mathcal{B}}\right)$. Assuming that $\vec{\mathcal{B}}$ is sourced by a monopole charge on the $z$-axis at position $M/B$, the Hamiltonian in terms of spherical-polar coordinates $r = B$ and angles $\left(\vartheta, \tilde{\varphi}\right)$ is
\begin{equation}\label{SpinBfield}
    H\left(B,M;\vartheta, \varphi\right) = -B \begin{pmatrix} {S}^{x} \\ {S}^{y} \\ {S}^{z} \end{pmatrix}^{T} \cdot  \begin{pmatrix} \cos\left(\tilde{\varphi}\right)\sin\left(\vartheta\right) \\
   \sin\left(\tilde{\varphi}\right)\sin\left(\vartheta\right)\\
   \cos\left(\vartheta\right) 
    \end{pmatrix}
   -MS^z
\end{equation}
where we chose $\mu_{B} = 1$ for convenience. When $M=0$, the \emph{ground state} (GS) of this Hamiltonian lies on the Bloch sphere and can be written as  
\begin{equation}\label{Bloch_GS}
 |GS; \vec{\mathcal{B}}\rangle = \cos\left(\vartheta/2\right) |\uparrow\rangle + e^{i\tilde{\varphi}}\sin\left(\vartheta/2\right)|\downarrow\rangle,
\end{equation}
where the \emph{up} and \emph{down} vectors $|\uparrow\rangle$ and $|\downarrow\rangle$ denote the spin-$1/2$ eigenvalues of the $S^{z}$ operator. When $M \neq 0$, we can still map the Hamiltonian \eqref{SpinBfield} onto the Bloch sphere, however only by re-defining the angular coordinate $\tilde{\vartheta}\left(\vartheta, M\right)$,  such that
\begin{equation}\label{BS_angle_semenoff}
    \cos\left(\tilde\vartheta\right) = \frac{M + B\cos\left(\vartheta\right)}{\sqrt{\left(B\cos\left(\vartheta\right) + M\right)^{2} + B^{2}\sin\left(\vartheta\right)^{2}}}.
\end{equation}

The effects of $M$ can be understood by evaluating the expectation value of $S^{z}$. Since the position of the monopole on the z-axis is determined by $M$, as long as $|M| < B$ the magnetic charge will remain within the Bloch sphere of radius $B$ centered at the origin. For this case one expects $\langle S^{z}\rangle$ to have opposite values at the North and South poles, i.e. $\pm 1$. Similarly, if $|M|>B$, the monopole is situated outside of the sphere, and therefore the spins will be aligned at both poles. The Chern number thus essentially counts the magnetic charges inside the Bloch sphere, in agreement with the Poincar\' e-Hopf theorem, which aligns or anti-aligns the spins at the poles. The formula \eqref{Cherns_JH} thus presents a (quasi) local measurement of a global topological invariant.

\section{DMRG}\label{Appendix_Numerics}
To find the GS of a Hamiltonian such as in \eqref{Ham_c} it is judicial to reformulate the problem as a matrix-diagonalization on the Hilbert space $\mathfrak{H}$, where the Hamiltonian is a $\text{dim}\left(\mathfrak{H}\right)\times \text{dim}\left(\mathfrak{H}\right)$ matrix. The dimension usually scales exponentially with system size $\text{dim}\left(\mathfrak{H}\right) \sim x^{N}$, and the precise value of $x$ depends on the number of internal degrees of freedom. \\

Whilst symmetries constrain the Matrix form and may reduce the total dimension considerably, even with that in mind the limits for ED lie in the $N = 20-30$ sites on current devices. Thus, approximation algorithms are necessary to determine the GS. \\
\begin{figure}[h!]
    \centering
    \includegraphics[width = 0.475\textwidth]{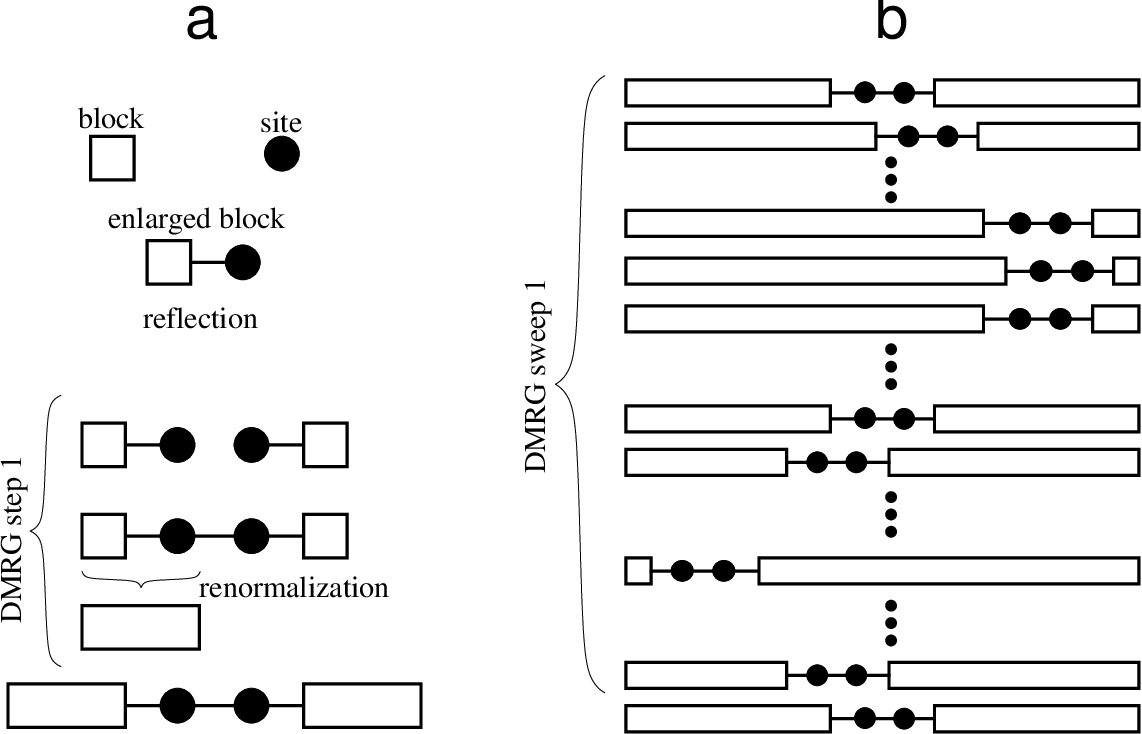}
    \caption{DMRG algorithm methodology, illustration taken from \cite{Chiara2006DensityMR}. \textbf{Left (a):} the infinite DMRG algorithm, where each step progressively ``grows" the chain. \\   \textbf{Right (b):} the finite size DMRG. Here the length $L = Na$ remains constant and the algorithm ``zips" through the system in \emph{sweeps}. For PBCs an additional link between both ends has to be made, resulting in a lower convergence and precision of the algorithm.}
    \label{fig:DMRG_method}
\end{figure}

A powerful and by now well-established iterative scheme to find the GS is the \emph{density matrix renormalization group} (DMRG) method, developed by White in the early nineties \cite{PhysRevLett.69.2863, PhysRevB.48.10345}. This algorithm computes the lowest energy eigenstates of quantum systems, by truncating the dimension of $\mathfrak{H}$ on each site at $D$ (bond dimension) at some cut-off value $\epsilon$. By \emph{sweeping} through the system back-and-forth this iteratively improves the initial guess $|\psi_{0}\rangle$ for a ground state, and in the language of \emph{matrix product states} (MPS) it is essentially a gradient descent \cite{Schollw_ck_2011}. See figure \ref{fig:DMRG_method} for a visual representation of the originally conceived DMRG algorithms for both infinite and finite systems. A standard marker for the convergence of the DMRG algorithm is the energy variance, which indicates how close the optimized state is to a true eigenstate. Another crucial marker when computing correlation functions or extracting central charges is the convergence of the entanglement entropy. \\

Whilst it has achieved unprecedented precision to approximate the GS for one-dimensional quantum systems \cite{Schollw_ck_2005}, the algorithm does have its limitations and subtleties. An intuitive way to understand this is, that truncating the Hilbert space is generally a very good approximation when the entanglement in the system is local i.e. decreases strongly with distance. For two- and higher-dimensional systems, or in the presence of long-range correlations, the number of \emph{relevant} eigenvalues to keep grows substantially. Whilst generally periodic boundary conditions (PBCs) are favoured analytically, the DMRG algorithm converges significantly slower and requires much higher bond dimensions and thus reduces the maximal size $N$ one can resolve. Due to the naturally occurring truncation errors, it is also not trivial to estimate the total error on highly non-local correlation functions, such as the ones involved for the topological numbers defined in \eqref{Combined_pole_equations} above. 

\subsection{DMRG for two coupled wires}
In the MPS language, it is clear that the natural formulation of the DMRG method is in one spatial dimension. For (quasi-)two-dimensional systems, such as the two coupled wires, this can be achieved by casting the (higher-dimensional) lattice into a $1$d chain. However, this may result in considerable changes, when previously nearest neighbours are now separated by two or more $1$d lattice sites. It is clear that this map is not unique, and can be chosen according to the microscopic details of the model. Considering both the interchain hopping $t_{\bot}$ and interaction $\sim g$, the following is a strategic choice to map the two coupled wires onto a single chain of length $2L$.  \\ 
\begin{figure}[h!]
    \centering
    \includegraphics[width = 0.5\textwidth]{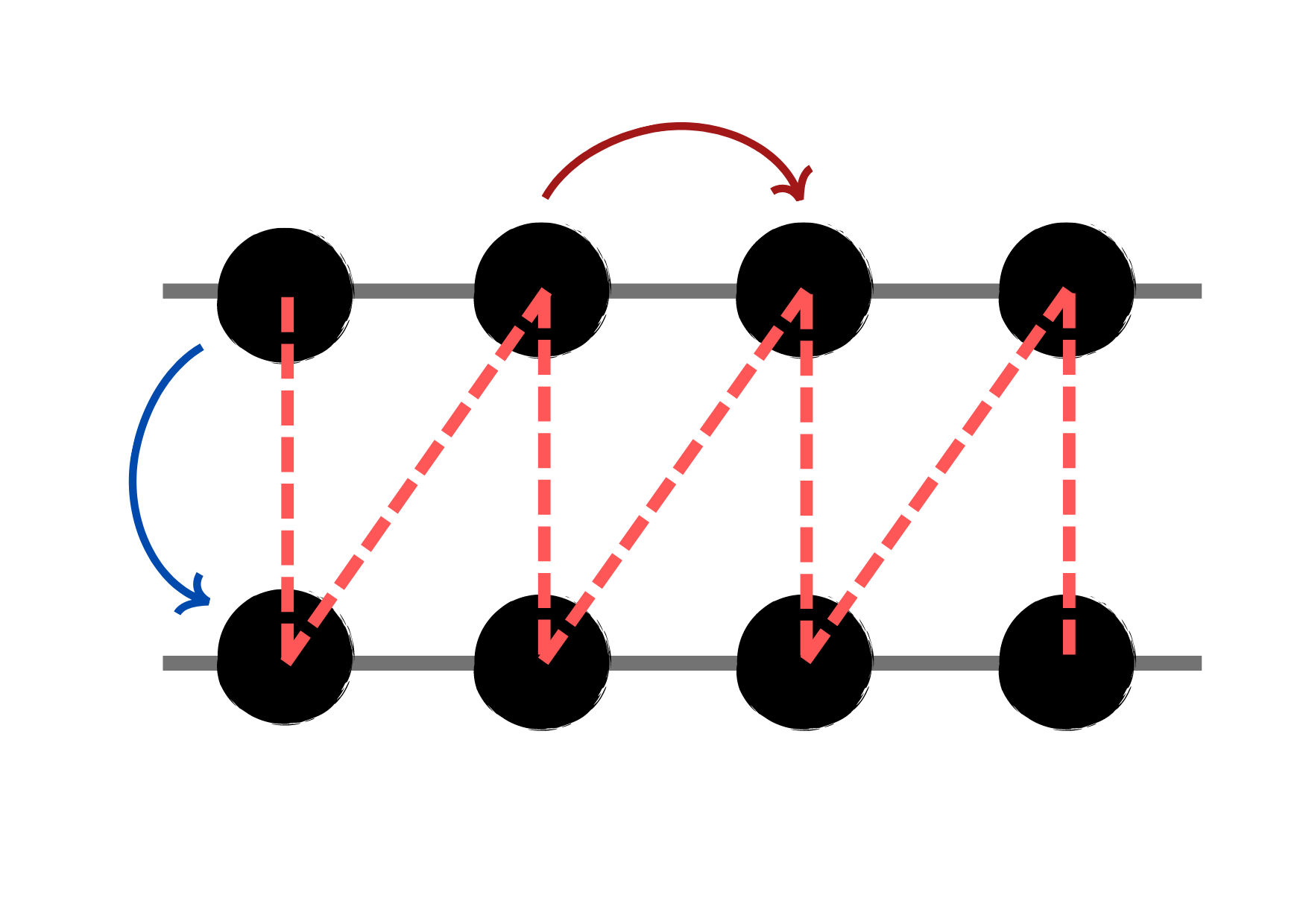}
    \caption{Zig-zag pattern, which was chosen to map two coupled wires onto a single chain. Previously nearest neighbour couplings are now next-nearest, and vice-versa. However, the choice of pattern is not unique.}
    \label{fig:DMRG_path}
\end{figure}

As can be seen in figure \ref{fig:DMRG_path}, previously nearest neighbours are now next-nearest (red), and the interchain terms are nearest-neighbours (blue). This is particularly efficient for studying the effects of both $g$ and $t_{\bot}$ on the coupled wire systems. In the case of PBCs, an additional link is added in the same way. Due to the doubling of the chain length, the convergence slows down and the time and maximum link dimension $D$ needed for a faithful approximation increases. This is exacerbated by adding PBCs, which limits our current investigation to $\sim 100$ sites per wire, whilst with OBCs we may achieve up to order $10^{3}$ sites. Particularly the critical phases, with their diverging correlation lengths, require both the longest run times and highest link dimensions, for a comparable GSE variance. \\ 

A subtlety that is easily overlooked is the issue of degenerate ground states, for example when symmetries are explicitly resolved. We found that the topological numbers $C$ and $\overline{C}$ are particularly sensitive to the explicit mixture $|GS\rangle = \alpha |GS,1\rangle + \beta |GS,2\rangle$ of both parity sectors $(P_{1},P_{2}) = (0,1)$ and $(1,0)$. It is thus important to resolve also the individual wires parities in DMRG.

\subsection{Convergence}
In general, gapped phases require lower bond dimensions to faithfully approximate the GS wave function. Critical systems with infinite correlation lengths in principle cannot be exactly described by MPS with finite bond dimension, though we can obtain good enough approximations. Naturally, the required dimensions also depend strongly on the system size and boundary conditions. A typical GSE variance that is considered sufficient lies below $10^{-5}$ or $10^{-6}$. 

\begin{figure}[h!]
    \centering
    \includegraphics[width = 0.5\textwidth]{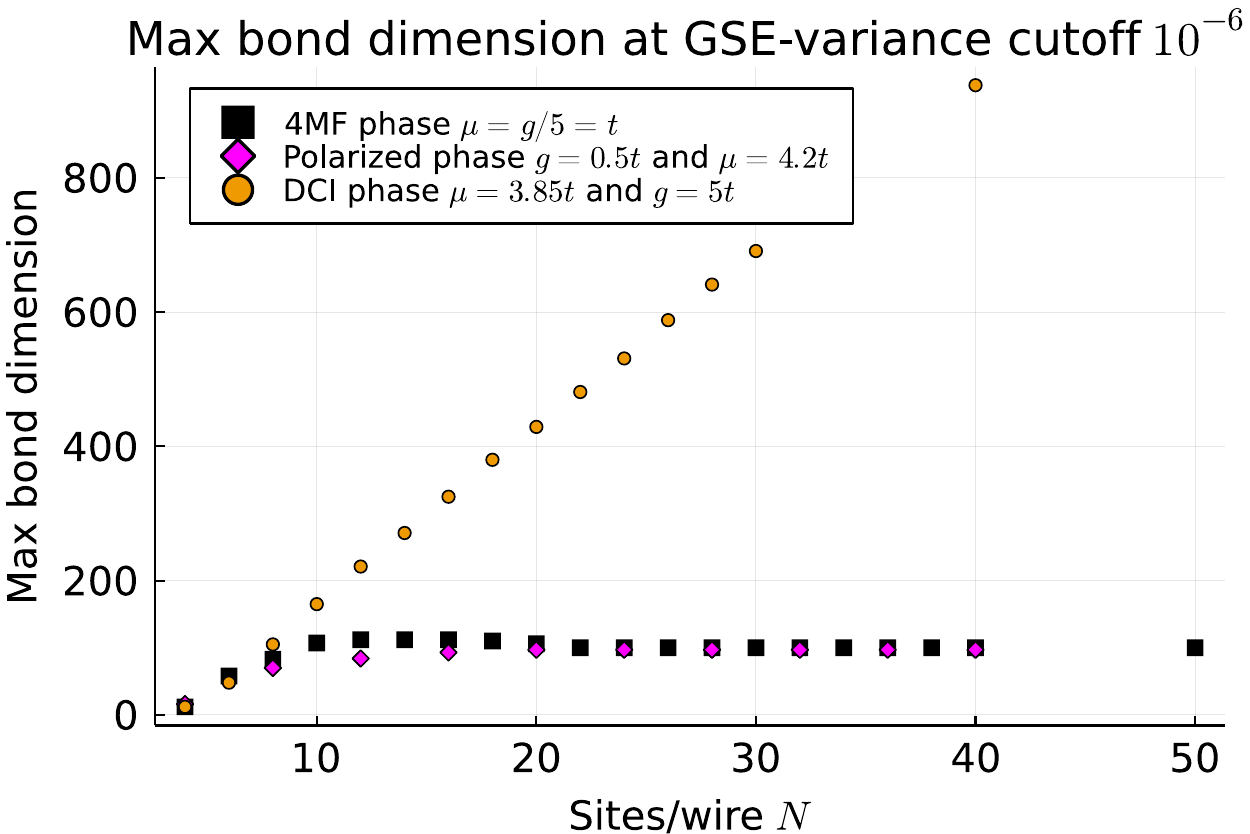}
    \caption{Maximum bond dimensions for both gapped and gapless phases, from small-system DMRG and PBCs.}
    \label{fig:Bonddims}
\end{figure}

\begin{figure}[h!]
    \centering
    \includegraphics[width = 0.5\textwidth]{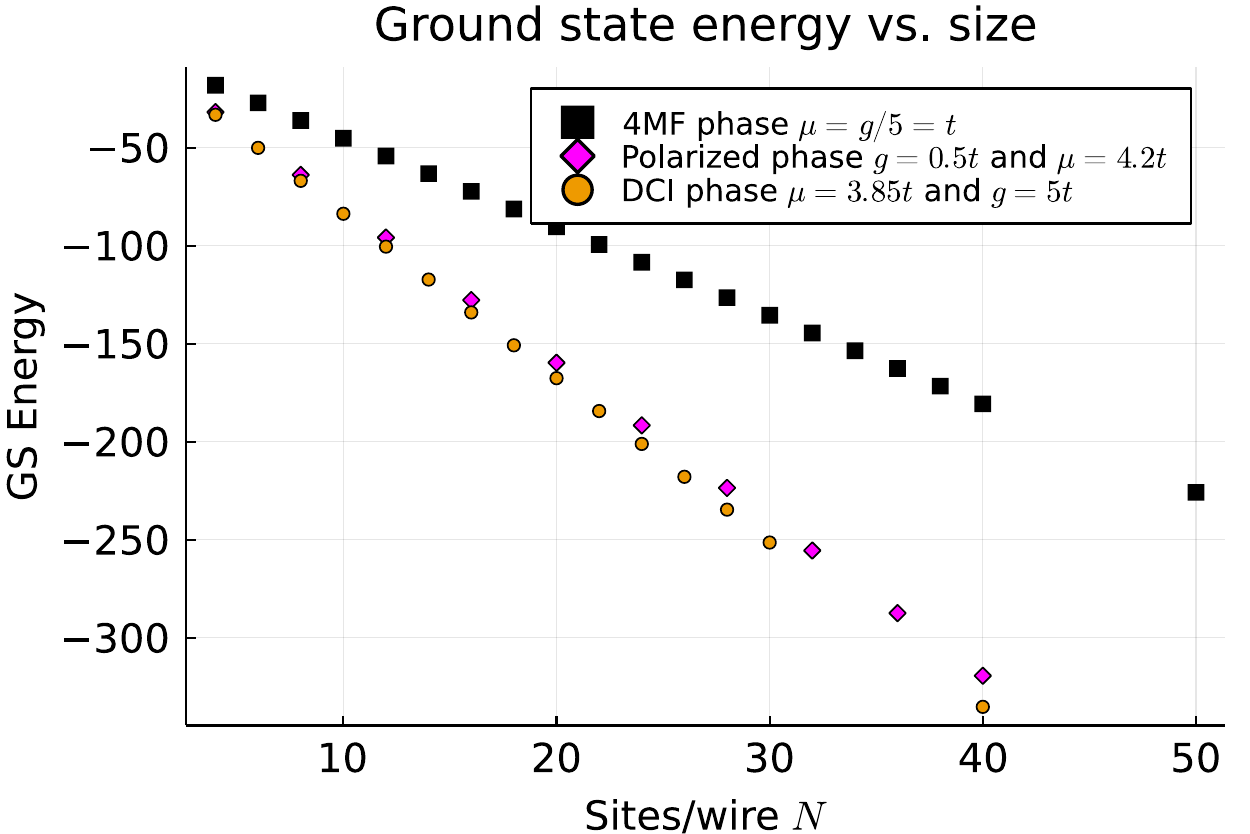}
    \caption{GS Energies as a function of system size for both gapped and gapless phases of two interacting superconducting wires. The DMRG was performed with PBCs.}
    \label{fig:energies}
\end{figure}

\begin{table}
\begin{center} 
\begin{tabular}{ |c|c|c|c| }
\hline
 Sites/wire  &  Polarized  & \emph{4MF}  &  DCI  \\ 
\hline
4 & $\sim 10^{-13}$ & $\sim 10^{-14}$ & $\leq 10^{-13}$  \\ 

6 & $10^{-7}$- $10^{-8}$ & $10^{-9}$ - $10^{-13}$ & $\sim 10^{-8}$ \\ 

30 & $\sim 10^{-6}$ & $10^{-6}$- $10^{-7}$ & $\sim 10^{-6}$ \\ 

50 & - & $\sim 10^{-6}$& $\sim 10^{-6}$ \\ 

76 & $\sim 10^{-6}$ & $\sim 10^{-6}$ & $\sim 10^{-6}$\\
\hline 
\end{tabular}
\end{center}
\caption{Variance of the groundstate energy of the two-wire models in the different phases we study in this paper. We work with periodic boundary conditions}
\label{tab:GSE}
\end{table}
As shown in Table~\ref{tab:GSE}, we achieved acceptable variances for both gapped and gapless phases up to $L = 76a$ per wire with PBCs. Generally with OBCs, we can achieve better convergences quicker, for lattices of up to a few thousand sites. For PBCs, while it is a priori possible to obtain good results for large systems, it is very costly both for computational resources as well as run-time. Therefore we stay with systems of up to a hundred sites per wire. The graphs in figures \ref{fig:Bonddims} and \ref{fig:energies}  showcase some typical values of GS energies and maximum bond dimension as a function of system size $L = Na$ for the various phases of two interacting topological wires.   \\

As shown in figure \ref{fig:Bonddims} above, the gapped phases have similar max. bond dimensions - irrespective of their topology. They plateau at some length scale, which is due to the fixed variance cut-off of $\approx 10^{-6}$. Since the \emph{4MF} and \emph{polarized} phases are gapped, the correlation length is finite and thus beyond a critical length the bond dimension is a function of the variance alone and becomes independent of system size. Conversely, the gapless modes in the DCI phase result in a diverging bond dimension, ie. a function of GSE energy variance \emph{and} system size. 
In figure \ref{fig:energies} we plot some representative GS energy values for the different phases.  
\subsection{Convergence of correlation functions}
\begin{figure}[h!]
    \centering
    \includegraphics[width = 0.5\textwidth]{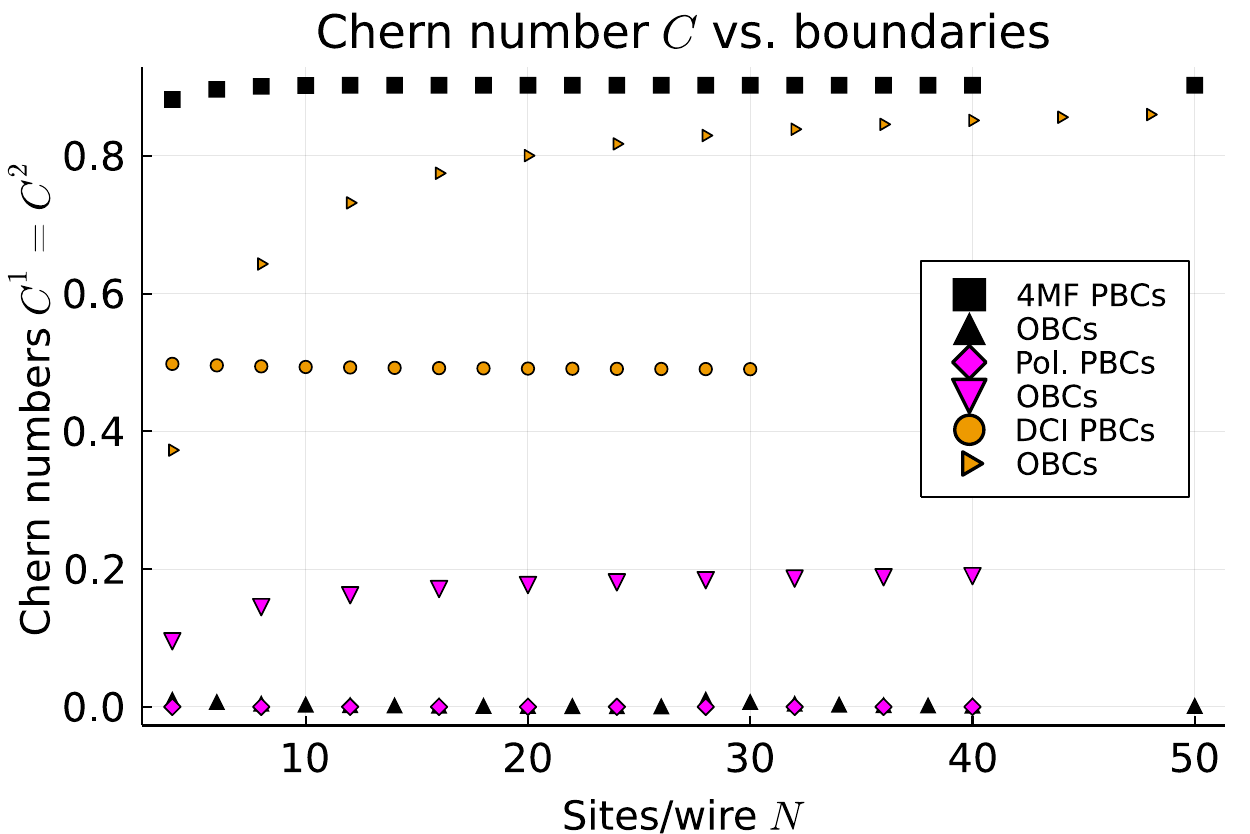}
    \caption{Chern numbers for PBCs vs. OBCs DMRG calculations. The \emph{4MF} phase here is chosen as $\mu = g/5 = t$, the DCI phase as $\mu = 3.85t$ and $g = 5t$ and the polarized phase as $\mu = 3.85t$ and $g = 0.5t$.}
    \label{fig:OBCvsPBC}
\end{figure}

A system near criticality will require a diverging bond dimension for $L \longrightarrow \infty$ to accurately compute non-local observables. The topological numbers \eqref{Combined_pole_equations} are determined by non-local correlation functions of all orders up to $L$, making them particularly challenging to simulate accurately. This is exacerbated by the periodic boundaries which are used in most instances throughout the paper.  A numerical comparison of the Chern numbers between OBCs and PBCs for a range of readily accessible system sizes is shown in figure \ref{fig:OBCvsPBC}. The defining relation for $C$ measured at the poles is, in fact, only valid in momentum space. However, momentum labels $k$ are only good quantum numbers when PBCs are imposed.\\

To summarise, the Chern numbers are thus highly dependent on the convergence of the DMRG, and also on the boundary conditions for accessible system sizes. Whilst convergence in the energy is a reliable marker for the accuracy of local observables, non-local correlation functions will naturally have errors. These will increase for longer-range correlators, with each approximation made at each bond contributing. The GSE variance cannot be used as a reliable quantifier for these errors. It is therefore paramount to consider a wide range of parameters, lengths, and GSE variances when investigating $C$ and $\overline{C}$. In some instances, we found that a poor quantization of the topological numbers was, in fact, due to convergence rather than physical reasons. 

\subsection{Zero central charge in gapped $2MF$ phases}\label{appendix:fitting}
The truncation error in DMRG can have significant impact on certain observables, as discussed above for the Chern and $\overline{C}$ markers. It is then sensible to wonder if the non-zero central charge presented in figure \ref{fig:EntTPOBC_300} is indeed due to the finite system sizes, and not an artefact of non-convergence of the DMRG algorithm. The truncation error may lead to a deformed, non-symmetric entanglement entropy, potentially leading to a non-zero central charge in gapped phases. 
We showcase below how system size and fitting domain have a considerable impact on the extracted central charge of the 2MF phases discussed in section \ref{sec:TP}, as opposed to the maximum bond-dimension. \\

\begin{figure}[h!]
    \centering
    \includegraphics[width = 0.5\textwidth]{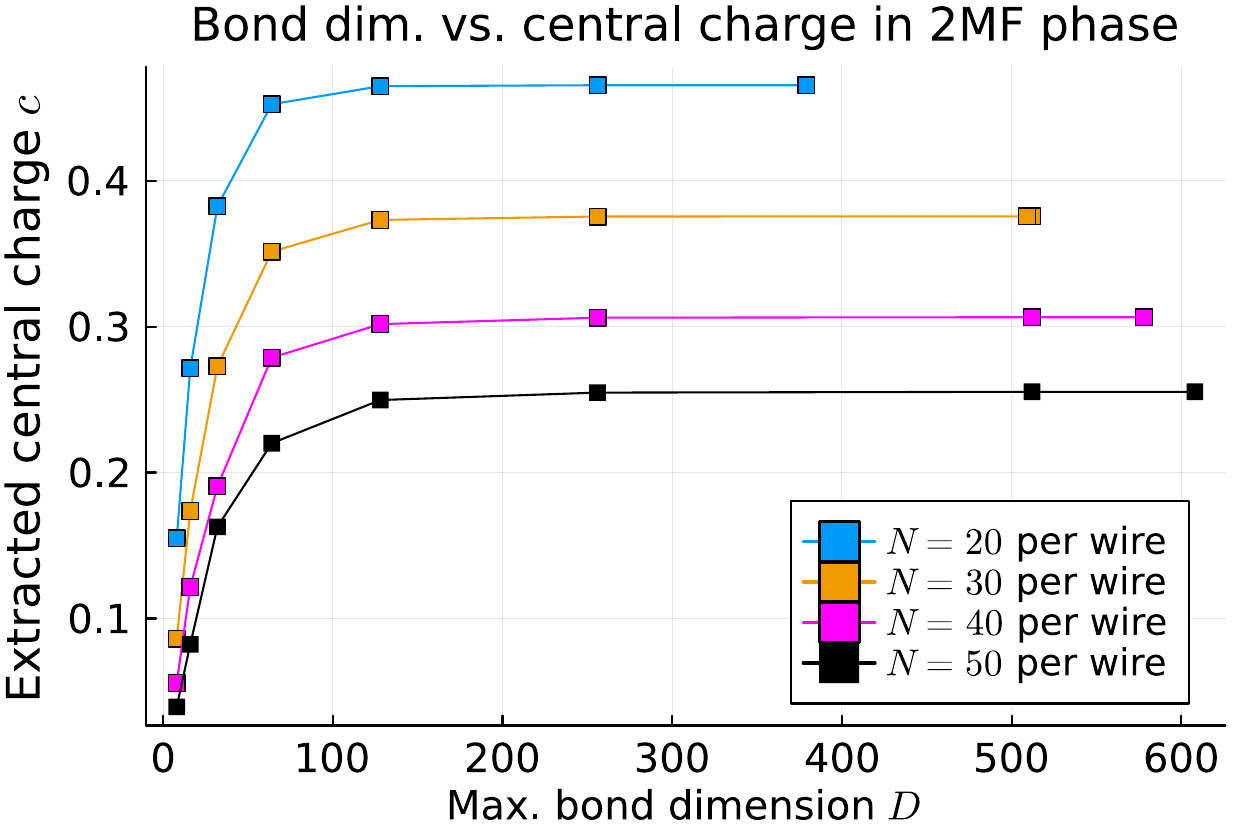}
    \caption{Central charge vs. bond dimension, with parameters $t =\Delta = g/5 = 1.0$, $\mu = 3.85t$ and $t_{\bot} = 0.1t$. }
    \label{fig:bondim}
\end{figure}

In figure \ref{fig:bondim} we show the central charge extracted at each step of the DMRG calculations. These were performed setting a fixed, maximum bond-dimension which the DMRG may exploit. The extracted maximum bond dimension of the (at this stage) optimized matrix product state (MPS) is plotted on the x-axis. Convergence is reached when the maximum bond-dimension is no longer saturated - and both energy and mid-chain entropy are then converged. \\

The parameters place us inside the DCI phase, with $t_{\bot} = 0.1$, a value comparable to the previously discussed ones given the smaller system size. The convergence of the entropy ensures that the extracted central charge no-longer depends on the bond dimension. The system sizes vary from $20$ to $50$ sites per wire. The initial increase of $c$ vs. $D$ can be explained from a non-converged entanglement entropy, which is manifest in fluctuations and the breakdown of the spatial symmetry, leading to a decreased effective central charge. More precisely, as long as the bond dimension is too small to fully capture the correlation length, the MPS appear nearly critical. If the total system size or the size of the subdomains are not significantly larger than the correlation length, we observe the same phenomenon. We showcase the latter for the $N = 50$ sites per wire in the figure \ref{fig:fittingdomain} below. As we can see, already changing the fitting domain slightly has significant effects on the extracted, numerical central charge, with a relatively quick convergence towards $c=0$.\\
\begin{figure}[h!]
    \centering
    \includegraphics[width = 0.5\textwidth]{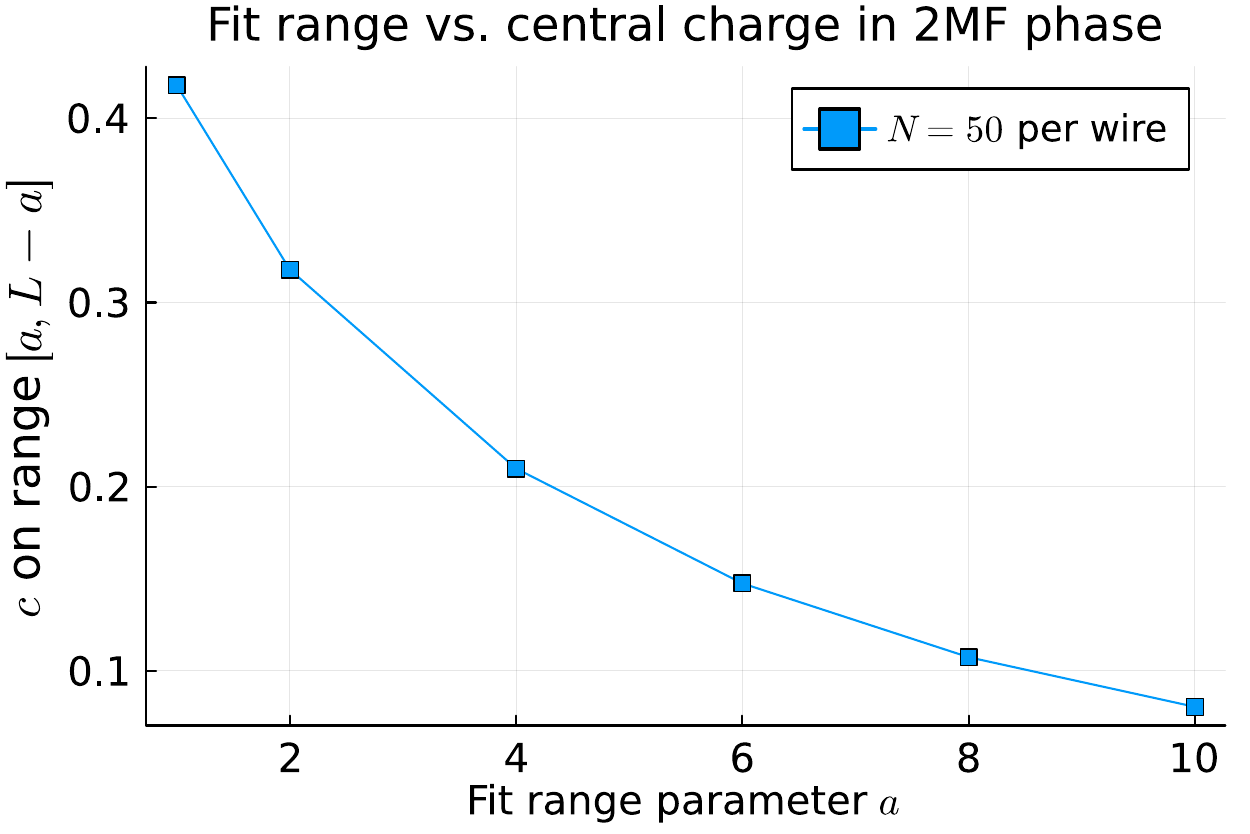}
    \caption{Central charge vs. fitting domain. Same parameters as in figure \ref{fig:bondim} above.}
    \label{fig:fittingdomain}
\end{figure}

Thus, we attribute the nonphysical value of $c \approx 0.35$ in figure \ref{fig:EntTPOBC_300} to the finite (and small-system) size of our numerical computations.

\section{QFT for coupled wires}
\label{Appendix_DCI_phase}
\subsection{Extending the Bloch sphere and wire correspondence}\label{Appendix:Spheres_Wires}

In the following, we develop the link between two interacting wires \cite{Herviou_2016} and Bloch spheres \cite{Hutchinson_2021} by rewriting the low-energy interaction Hamiltonian \eqref{Hint_wires} in momentum space. The on-site density-density interaction results in, \` a priori, a quartic coupling between modes of four independent momenta $k_{i}$. However, due to the sum over the site index $i$, one of these will be removed by the momentum-conservation  $k_{1} - k_{2} + k_{3} - k_{4} = 0$. The three remaining momenta can be written more suggestively way, by introducing 
\begin{equation}
\left\{\begin{array}{l}
k_{1} \rightarrow k+\Delta p \\
k_{2} \rightarrow k \\
q_{1} \rightarrow q-\Delta p \\
q_{2} \rightarrow q
\end{array}\right.
\end{equation}
where $\triangle p = k_{1} - k_{2} = q_{2} - q_{1}$ by momentum conservation. Then the interaction in momentum space is

\begin{equation}\label{FT_Coulomb}
    H_{int} = g\mathcal{A}^{-1}\sum_{\triangle p} \left[ \sum_{k} c^{1\dagger}_{k+\triangle p}c^{1}_{k}\right]\left[ \sum_{q} c^{2\dagger}_{q-\triangle p}c^{2}_{q}\right].
\end{equation}
The area factor $\mathcal{A}^{-1} = \left(2\pi/L\right)^{2}$ stems from the normalization factors of the Fourier transforms. The interaction in \eqref{FT_Coulomb} contains several physically distinct processes. Whilst $\triangle p > 0$ it is ``removed" from wire $2$ and ``added" to wire $1$. For $\triangle p < 0$ the situation is reversed, thus it is warranted to interpret $\triangle p$ as ``momentum transfer" between both wires. \\

For a wire, we have the identification
\begin{equation}
\sum_k c^{\dagger}_{k+\Delta p} c_k = \sum_i e^{i x \Delta p} c^{\dagger}_{j} c_{j},
\end{equation}
with $x = ja$. If we measure the expectation value of these two terms, we can approximate $\langle c^{\dagger}_i c_i\rangle$ by its mean density. In the continuum limit, then the dominant contribution comes from $\Delta p\rightarrow 0$. We can also rewrite $H_{int}$ 
as

\begin{equation}
H_{int} = g\mathcal{A}^{-1}\sum_{\Delta p}\sum_{i,j} e^{i(x_i - x_j)\Delta_p} c^{1\dagger}_i c_i^1 c^{2\dagger}_j c_j^2.
\label{Hint}
\end{equation}

In the sense of the Renormalization Group (RG) approach, terms with $\Delta p \neq 0$ would oscillate more rapidly with increasing the length of the system or decreasing the temperature such that we can safely take $\Delta p\rightarrow 0$ 
in Eq. (\ref{Hint}). A term that gives rise to Mott physics in the phase diagram is of the form ${\color{magenta} c^{1\dagger}_{R}c^1_{L}c^{2\dagger}_{L}c^2_{R}+h.c.}$ \cite{Herviou_2016} with $L$ and $R$ referring to left- and right movers or to the two Fermi points such that $c^{1\dagger}_{R}c^1_{L}$ involves a wavevector transfer $\Delta p=+2k_F$, $c^{2\dagger}_{L}c^2_{R}$ involves a momentum transfer of $-\Delta p=-2k_F$ and such that the total momentum transfer is effectively zero. This term can be therefore also relevant away from half-filling. The relevant term flowing to strong couplings is then local in real space with $x_i=x_j$
such that this can yet be described through the same form of interactions $H_{int} = gV \sum_{i,j} c^{1\dagger}_i c_i^1 c^{2\dagger}_j c_j^2$ where $i=j$. \\

Therefore in the low energy limit, the Hamiltonian is effectively similar to

\begin{equation}
    \begin{aligned}
     H_{int} \sim & \  g\mathcal{A}^{-1} \left[ \sum_{k} c^{1\dagger}_{k}c^{1}_{k}\right]\left[ \sum_{q} c^{2\dagger}_{q}c^{2}_{q}\right] \\  =& \ \frac{g\mathcal{A}^{-1}}{4} \left[ \sum_{k} c^{1\dagger}_{k}c^{1}_{k} + c^{1\dagger}_{-k}c^{1}_{-k} \right]\\ \ &\times \left[ \sum_{q} c^{2\dagger}_{q}c^{2}_{q} + c^{2\dagger}_{-q} c^{2}_{-q}\right].
\end{aligned}
\end{equation}

In the second equality, we symmetrised as both sums run over the entire Brillouin zone. Using the anti-commutation relations each bracket can be written in terms of the $S^{z}_{k}$ pseudo-spins introduced in \ref{subsec_cherns_for_single_wire}.
Finally, the full interaction Hamiltonian together with the linear shifts becomes

\begin{equation}\label{Hint_wire_sphere}
\begin{aligned}
          H_{int} =& \ \frac{g\mathcal{A}^{-1}}{4}\sum_{q,k} S^{z,1}_{k}S^{z,2}_{q} \\ =& \ \frac{g\mathcal{A}^{-1}}{4} \left(\sum_{q,k} \delta_{k,q}S^{z,1}_{k}S^{z,2}_{q} + \sum_{q \neq k} S^{z,1}_{k}S^{z,2}_{q} \right).
\end{aligned}
\end{equation}

The first of these terms corresponds to the same interaction as in the two-spheres model studied in \cite{Hutchinson_2021}, with a $k$-independent constant $g$ interaction. The second part can be absorbed into the two chemical potentials $\mu^{1/2}$ of each wire, by noting that the sums $\sum_{k\neq q} = \frac{1}{2} \sum_{k}\sum_{q \neq k} + \frac{1}{2}\sum_{q}\sum_{k\neq q}$. \\ 

\subsection{Mixed-wire fermion basis}
In the following Appendix, we aim to support the discussions in Sections \ref{sec:Fractional} and \ref{sec:TP}, in particular, the derivations of Hamiltonians \eqref{Final_own_bosonized_text} and \eqref{Boson_hamiltonian_final}.\\ 

 In what follows we work immediately in the $\Gamma$ and $\Theta$ basis introduced in \eqref{Mixed_wire_ferms_def}. Thus we first recast the coupled-wire Hamiltonian in this basis. Beginning with $g = 0 = \delta \mu$, we essentially have two copies of the QCP discussed in \ref{sec:Fractional}. In this basis, the free Hamiltonian is given by
\begin{equation}\label{Ham_crit_twowires}
\begin{aligned}
    H =&  \ 2t\sum_{j} \left(\Gamma^{\dagger}_{j+1}\Theta_{j} - \Gamma^{\dagger}_{j}\Theta_{j} + \text{h.c.}\right) \\ 
    &\stackrel{a\longrightarrow 0}{\longrightarrow} \ 2ta \int_{x}\text{d}x \left(\partial_{x}\Gamma\left(x\right)\right)\Theta\left(x\right) + \text{h.c}.
\end{aligned}
\end{equation}

In the last step, a continuum limit was made. This Hamiltonian has superficial (structural) similarity with the single wire critical theory \eqref{Hamcritp}, however now with Dirac Fermions $\Gamma$ and $\Theta$. Moving forward we can therefore analogously define chiral Fermions $\gamma^{R/L}$.  \\ 

Adding a non-zero shift from the critical points, i.e. $\delta \mu \neq 0$, results in an additional term proportional to 
\begin{equation}
    \sim -\sum_{j} \left(\Gamma^{\dagger}_{j+1}\Theta_{j} + \Gamma^{\dagger}_{j}\Theta_{j} + \text{h.c}\right).
\end{equation}
In the continuum this term is dominated by the $\sim \Gamma^{\dagger}_{j}\Theta_{j} + \text{h.c.}$ contribution to order $1/a$, such that we may write
\begin{equation}
\begin{aligned}
   \stackrel{a\longrightarrow 0}{\longrightarrow} \frac{2}{a}\int_{x}\text{d}x \left(\Gamma^{\dagger}\left(x\right)\Theta\left(x\right) + \text{h.c} \right).
\end{aligned}
\end{equation}
Finally, decomposing the interaction contribution 
\begin{equation}
    \sim \sum_{j} \left( n^{1}_{i} -\frac{1}{2} \right)\left( n^{2}_{i} -\frac{1}{2} \right),
\end{equation}
in the $\Gamma$ and $\Theta$ basis results in both linear and quadratic terms in $n$. The linear ones simply shift the chemical potential and are found to be given by 
\begin{equation}
\sim  -\sum_{j} \left(\Gamma^{\dagger}_{j}\Gamma_{j} + \Theta^{\dagger}_{j}\Theta_{j}\right). 
\end{equation}
More interesting are the quadratic ones, which by analogous computations are found as 
\begin{equation}
  \sim  -2\sum_{j} \Gamma^{\dagger}_{j}\Gamma_{j}\Theta^{\dagger}_{j}\Theta_{j}. 
\end{equation}
Grouping all contributions together, we find the final (continuum) Hamiltonian in the $\Gamma/\Theta$ basis as
\begin{equation}\label{Hamiltonian_GammaTheta}
    \begin{aligned}
        H = \ &-\int_{x}\text{d}x  \Big(v_{F} \left(\Theta\partial_{x}\Gamma\right)\left(x\right) - \frac{\delta\mu}{2}\left(\Gamma^{\dagger}\Theta\right)\left(x\right) + \text{h.c.}\Big) \\ &-\frac{g}{2}\int_{x}\text{d}x \Big(\left(\Gamma^{\dagger}\Gamma + \Theta^{\dagger}\Theta\right)\left(x\right) -2 \left(\Gamma^{\dagger}\Gamma\Theta^{\dagger}\Theta\right)\left(x\right)\Big).
    \end{aligned}
\end{equation}

\subsection{Chiral basis}\label{Appendix_chiral_basis}
Just as in the single wire at criticality we introduce left and right movers $\psi_{L/R}$ in terms of the complex $\Gamma$ and $\Theta$ Fermions
\begin{equation}
   \sqrt{2} \psi_{R/L}\left(x\right) = \Theta\left(x\right)\pm i \Gamma\left(x\right).
\end{equation}
These chiral modes $\psi_{R/L}$ are related directly to the chiral Majoranas $\gamma_{R/L}$ of each wire. Decomposing $\Gamma$ and $\Theta$ into their Majorana constituents, we find
\begin{equation}
    \sqrt{2}\psi_{R} = \frac{1}{2}\left(\gamma^{2}_{B} -i \gamma^{1}_{B} + i\gamma^{1}_{A} - \gamma^{2}_{A}\right) \sim \gamma^{2}_{R} -i\gamma^{1}_{R}.
\end{equation}
Similarly, for the other chirality, we find
\begin{equation}
      \sqrt{2}\psi_{L} = \frac{1}{2}\left(\gamma^{2}_{B} -i \gamma^{1}_{B} - i\gamma^{1}_{A} + \gamma^{2}_{A}\right) \sim \gamma^{2}_{L} -i\gamma^{1}_{L}.
\end{equation}
Thus these Fermions can be decomposed into the two chiral Majoranas on each wire. Since $g= 0$ we expected such a decomposition, since then the two-wire system is simply two copies of the single wire. \\ 

We now proceed to rewrite the Hamiltonian in \eqref{Hamiltonian_GammaTheta} in terms of the new chiral modes. The kinetic term can be written as
\begin{equation}
 \left(  \psi^{\dagger}_{R} \partial_{x} \psi_{R} - \psi^{\dagger}_{L} \partial_{x} \psi_{L}\right) = i\Theta^{\dagger}\partial_{x} \Gamma  - i\Gamma^{\dagger}\partial_{x}\Theta .
\end{equation}
Unsurprisingly, the left- and right-movers have a relative minus sign, reflecting the opposite directions of propagation with the effective velocities $v_{F} = \pm ita$. The two remaining quadratic terms yield additionally 
\begin{equation}
\begin{aligned}
   \delta\mu \left(\Gamma^{\dagger}\Theta + \text{h.c.}\right) \sim & \ \psi^{\dagger}_{R}\psi_{L} - \psi^{\dagger}_{L}\psi_{R} \\ 
     g \left(\Gamma^{\dagger}\Gamma  + \Theta^{\dagger}\Theta\right) \sim & \    \psi^{\dagger}_{R}\psi_{R} + \psi^{\dagger}_{L}\psi_{L} .
\end{aligned}
\end{equation}
Finally, the quartic interaction contributions in \eqref{Hamiltonian_GammaTheta} result in a myriad of terms. There are grouped into three classes of terms, preempting the subsequent bosonization and resulting Hamiltonian. 
\begin{equation}\label{quartic_contribs}
\begin{aligned}
       \text{I: \ }& \psi^{\dagger}_{R}\psi_{R}\psi^{\dagger}_{R}\psi_{R} + \psi^{\dagger}_{R}\psi_{R}\psi^{\dagger}_{L}\psi_{L} + \text{h.c.} \\ 
       \text{II: \ }&
\left(\psi_{R}^{\dagger} \psi_{R}+\psi_{L}^{\dagger} \psi_{L}\right)\left(\psi_{R}^{\dagger} \psi_{L}+\psi_{L}^{\dagger} \psi_{R}\right)
- \text{h.c.} \\ 
\text{III: \ }& \left(\psi_{R}^{\dagger} \psi_{L}+\psi_{L}^{\dagger} \psi_{R}\right)\left(\psi_{R}^{\dagger} \psi_{L}+\psi_{L}^{\dagger} \psi_{R}\right).
\end{aligned}
\end{equation}
With these, the full Hamiltonian written in terms of the $R/L$ movers is given by
\begin{equation}\label{RL_Ham}
    \begin{aligned}
           H = \ &i\frac{\tilde{v}_{F}}{2} \int_{x}\text{d}x \left(\psi^{\dagger}_{L} \partial_{x} \psi_{L} - \psi^{\dagger}_{R}\partial_{x} \psi_{R}\right) \\ &- i \frac{\delta \mu }{2a} \int_{x}\text{d}x \left( \psi^{\dagger}_{R}\psi_{L} - \psi^{\dagger}_{L}\psi_{R} \right)\\ &- \frac{g}{2a}\int_{x}\text{d}x \left(\psi^{\dagger}_{R}\psi_{R} + \psi^{\dagger}_{L}\psi_{L} \right)\\ 
           &-\frac{g}{4a}\int_{x}\text{d}x \left((\text{I}.) - (\text{II}.) - (\text{III}.)\right).
    \end{aligned}
\end{equation}
Here $\tilde{v}_{F} = a(4t - \delta\mu)$ and we introduced a phase factor
\begin{equation}
\psi_{R/L} \longrightarrow e^{\pm i\frac{\pi}{4}}\psi_{R/L},
\end{equation}
which slightly modifies some signs and factors of $i$ in the interaction contributions \eqref{quartic_contribs}. This phase factor is judicious when bosonizing the Hamiltonian.

\subsection{Bosonization}\label{Appendix_bosonization}

A powerful tool to investigate the low-energy physics of a Fermion system in one-dimension is \emph{bosonization} \cite{https://doi.org/10.1002/andp.19985100401, Senechal}. Chiral Fermion fields $\psi_{R/L}$ are rewritten in terms of bosonic degrees of freedom.
\begin{equation}
\psi_{R/L}\left(x\right)=\frac{U_{R/L}}{\sqrt{2 \pi \alpha}} e^{-i\left(\pm \phi\left(x\right)-\theta\left(x\right)\right)}.
\end{equation}
Here we defined the boson field $\phi$ and its dual $\theta$, and a low-energy cut-off $\alpha$ of the order of $a$. The Fermion statistics are imposed by introducing the two \emph{Klein factors} $U_{R/L}$ respectively, defined over the relations $U^{\dagger}_{R/L}U_{R/L} = 1$ and $U^{\dagger}_{R}U_{L} = i$. In what follows we sketch the main steps in deriving the final bosonized Hamiltonians in \eqref{Final_own_bosonized} and \eqref{Boson_hamiltonian_final}. \\

The bosonization for the kinetic term is straightforward and results in \cite{Senechal, giamarchi2004quantum}
\begin{equation}
H_{kin} =  \frac{\tilde{v_{F}}}{2\pi}\int_{x}\text{d}x \Big(\left(\partial_{x}\phi\right)^{2} + \left(\partial_{x}\theta\right)^{2}\Big)  .
\end{equation}
Next, consider the quadratic Fermion terms in \eqref{RL_Ham}. The $\sim g$ term simply reduces to $\partial_{x}\phi$, which constitutes a linear shift in the chemical potential and will be henceforth discarded. The bosonization of the second quadratic term $\sim \delta\mu$ on the other hand is not so trivial.
In fact, upon introducing the boson expressions, we find 
\begin{equation}
\begin{aligned}
&-\frac{\delta \mu}{2}\int_{x}\text{d}x \Big(\psi^{\dagger}_{R}\psi_{L} + \psi^{\dagger}_{L}\psi_{R}\Big) \\ & \sim -\frac{\delta \mu}{4\pi  a}\int_{x}\text{d}x \Big(U^{\dagger}_{R}U_{L} e^{2\phi i} + \text{h.c}. \Big)
\end{aligned}
\end{equation}
Using the relation $U_{R}U_{L} = i$ the above term thus simply becomes
\begin{equation}\label{cos_term_1}
\frac{\delta \mu}{2\pi \alpha}\int_{x}\text{d}x \sin\left(2\phi\right) .
\end{equation}
Finally, we consider the quartic (interaction) terms. For bosonization it is important to consider normal ordered expressions \cite{giamarchi2004quantum}, therefore we must first apply Wicks theorem to write the four-point functions in normal ordered form, denoted by ``$: \text{...} :$" \cite{WeaklydisorderedSpinLadders}. We thus consider the following limits of, for example the $\left(\psi^{\dagger}_{R}\psi_{R}\right)^{2}$ contribution, which after applying Wicks theorem results in
\begin{widetext}
    \begin{equation}\label{Fourpoint_kinetic_ren1}
\begin{aligned}
\psi^{\dagger}_{R}\psi_{R} \psi^{\dagger}_{R}\psi_{R}  =& \lim_{\epsilon \longrightarrow 0}\left(  \psi^{\dagger}_{R}\left(x+\epsilon\right)\psi_{R}\left(x+\epsilon\right)\psi^{\dagger}_{R}\left(x\right)\psi_{R}\left(x\right)\right) \\ =& \lim_{\epsilon \longrightarrow 0}\left(  :\psi^{\dagger}_{R}\left(x+\epsilon\right)\psi_{R}\left(x+\epsilon\right)\psi^{\dagger}_{R}\left(x\right)\psi_{R}\left(x\right): + 
:\psi^{\dagger}_{R}\left(x+\epsilon\right)\psi_{R}\left(x\right):\left(C_{R}\left(x+\epsilon,x\right) + C_{R}\left(x, x+\epsilon\right)\right)\right).
\end{aligned}
\end{equation}
\end{widetext}
The point-splitting distance $\epsilon$ is of the order of $\alpha$ and $a$. Using the fact that the contractions of $R/L$-movers is given by \cite{giamarchi2004quantum}
\begin{equation}
    C_{R/L}\left(x,y\right) = \pm \frac{i}{2\pi \left(x-y\right)}.
\end{equation}
The factor of $i$ comes from the Baker-Campbell-Hausdorff formula (BCH) when calculating the leading order bosonized expression.
The bracketed expression at the end of \eqref{Fourpoint_kinetic_ren1} is identically zero. The same will hold for the equivalent $L$ term, as well as for the mixed $RL$ contribution. Therefore we only bosonize the normal ordered expression of $\text{(I. )}$, resulting in \cite{Senechal, giamarchi2004quantum}
\begin{widetext}
\begin{equation}\label{ren_kinetic_1}
    -\frac{ga}{2\cdot4\pi^{2}}\left[\left(\partial_{x}\phi- \partial_{x}\theta\right)^{2} + \left(\partial_{x}\phi +  \partial_{x}\theta\right)^{2} + 2\left(\partial_{x}\phi- \partial_{x}\theta\right) \left(\partial_{x}\phi+ \partial_{x}\theta\right)  \right] = -\frac{ga}{2\pi^{2}}\left(\partial_{x}\phi\right)^{2}.
\end{equation}
\end{widetext}

By a similar approach, we now calculate the leading order terms for the remaining $\text{II}.$ and $\text{III}.$ contributions in \eqref{RL_Ham}. Here it is important to carefully keep track of all factors of $\pm i$. With respect to the renormalization of the kinetic term by \eqref{ren_kinetic_1} we note that $+\frac{ga}{4}\left(i\psi_{R}^{\dagger} \psi_{L}-i\psi_{L}^{\dagger} \psi_{R}\right)\left(i\psi_{R}^{\dagger} \psi_{L}-i\psi_{L}^{\dagger} \psi_{R}\right)$. Non-zero are only the two mixed terms, and by considerations as in \eqref{Fourpoint_kinetic_ren1} one can compute
\begin{widetext}
\begin{equation}\label{Fourpoint_kinetic_ren2}
\begin{aligned}
        \frac{ga}{4}\left(\psi^{\dagger}_{R}\psi_{L}\psi^{\dagger}_{L}\psi_{R} + \psi^{\dagger}_{R}\psi_{L}\psi^{\dagger}_{L}\psi_{R}\right)  = &   \frac{ga}{4} \lim_{\epsilon \longrightarrow 0}\left(  \psi^{\dagger}_{R}\left(x+\epsilon\right)\psi_{L}\left(x+\epsilon\right)\psi^{\dagger}_{L}\left(x\right)\psi_{R}\left(x\right)\right) \\ +& \frac{ga}{4}   \lim_{\epsilon \longrightarrow 0}\left(  \psi^{\dagger}_{L}\left(x+\epsilon\right)\psi_{R}\left(x+\epsilon\right)\psi^{\dagger}_{R}\left(x\right)\psi_{L}\left(x\right)\right).
\end{aligned}
\end{equation}
\end{widetext}
The normal ordered two-point functions result in constant terms which can be discarded, and thus we consider only the normal ordered four-point functions:
\begin{widetext}
\begin{equation}
    : \psi^{\dagger}_{R}\left(x+\epsilon\right)\psi^{\dagger}_{L}\left(x\right)\psi_{L}\left(x+\epsilon\right)\psi_{R}\left(x\right): = \frac{1}{\left(2\pi \epsilon\right)^{2}}\left( e^{i\left(\phi_{x+\epsilon} - \theta_{x+\epsilon} - \phi_{x} - \theta_{x}\right)}\cdot e^{i\left(\phi_{x+\epsilon} + \theta_{x+\epsilon} - \phi_{x} + \theta_{x}\right)} \right) .
\end{equation}
\end{widetext}
Merging the product of exponentials using the BCH formula $e^{A}e^{B} = e^{A+B + 1/2\left[A,B\right]}$ we find that the commutator expressions cancel due to the fact that $\left[\phi_{x},\theta_{y}\right] = - \left[\phi_{y},\theta_{x}\right]$, \emph{cf.} \cite{Herviou_2016}.
Then, writing $\phi_{x+\epsilon} - \phi_{x} = \epsilon \partial_{x}\phi_{x}$ and $\theta_{x+\epsilon} + \theta_{x} = 2\theta_{x} + \epsilon\partial_{x}\theta_{x}$, one finds the leading terms to be 
\begin{equation}
\begin{aligned}
           : \psi^{\dagger}_{R}\left(y\right)\psi^{\dagger}_{L}&\left(x\right)\psi_{L}\left(y\right)\psi_{R}\left(x\right): \ =  \frac{1}{\left(2\pi \epsilon\right)^{2}}e^{2i\epsilon\partial_{x}\phi_{x}} \\  =& -\frac{4}{2\left(2\pi\right)^{2}}\left(\partial_{x}\phi\right)^{2} + i\frac{2}{\epsilon}\partial_{x}\phi_{x} \\ & +\text{subleading terms} ,
\end{aligned}
\end{equation}
with $y = x+\epsilon$. The total derivative term can be neglected, and so together with the second contribution to $III.$ this results in a second renormalization to the kinetic part, such that finally 
\begin{equation}
v_{F}\left(\partial_{x}\phi\right)^{2} \longrightarrow \left( v_{F} - \frac{g a}{\pi^{2}}\right)\left(\partial_{x}\phi\right)^{2}
\end{equation}
This corresponds to the effective $\phi$ velocity $v_{F}K^{-1}$, \emph{cf.} \eqref{Final_own_bosonized}. \\ 

Lastly consider the II. term, which follows directly the discussion in \cite{giamarchi2004quantum} and the previous calculation resulting in \eqref{cos_term_1}. Firstly, again introducing the phase of $\pi/4$ in the field definition, one has
    \begin{equation}
    \begin{aligned}
    \frac{ga}{4}  (II.) =& \
\frac{ga}{4}\left(\psi_{R}^{\dagger} \psi_{R}+\psi_{L}^{\dagger} \psi_{L}\right)\\ & \ \times\left(i\psi_{R}^{\dagger} \psi_{L}-i\psi_{L}^{\dagger} \psi_{R}\right) 
- \text{h.c}.
    \end{aligned}
\end{equation}
The normal ordered four-point function yield subdominant terms \cite{WeaklydisorderedSpinLadders}, such that only the non-trivial contractions of $\psi^{\dagger}_{R}\psi_{R}\psi^{\dagger}_{R}\psi_{L}$ etc. remain relevant. This results in 
\begin{equation}
-\frac{ga}{4\left(2\pi\epsilon\right)^{2}}U^{\dagger}_{R}U_{L}e^{2i\phi} -\frac{ga}{4\left(2\pi\epsilon\right)^{2}}U^{\dagger}_{L}U_{R}e^{-2i\phi} - h.c.
\end{equation}
Again with conventions $U_{R}U_{L} = i$ this results in 
\begin{equation}
  \frac{ga}{4}  (II.) = \frac{ga}{\left(4\pi\epsilon\right)^{2}}\sin\left(2\phi\right).
\end{equation}
Finally, by shifting $2\phi$ to $2\phi - \pi/2$ and turning the $\sim \sin\left(\phi\right)$ into a $\sim \cos\left(2\phi\right)$, we result in the following boson Hamiltonian
\begin{equation}\label{Final_own_bosonized}
    H = \int_{x}\text{d}x \frac{\tilde{v}_{F}}{2\pi}\left(\frac{1}{K}\left(\partial_{x}\phi\right)^{2} + K\left(\partial_{x}\theta\right)^{2}\right) - g_{\phi}\cos\left(2\phi\right).
\end{equation}
with the following parameters $\tilde{v}_{F}K^{-1} = a\left(4t - \delta \mu + \frac{g}{\pi}\right)$ and $\tilde{v}_{F}K = a\left(4t - \delta \mu\right) $, as well as $g_{\phi} = \left(\frac{\delta\mu}{2\pi \epsilon} +  \frac{ga}{4\pi^{2}\epsilon^{2}}\right)$. \\ 

Last but not least we note that for $t_{\bot} \neq 0$ an additional term $\sim i\left(\psi^{\dagger}_{R}\psi^{\dagger}_{L} - \psi_{L}\psi_{R}\right)$ appears in the Hamiltonian. Bosonizing this contribution in an anaologous fashion results in a \emph{SC-pairing-like} terms \cite{Herviou_2016}, given by 
\begin{equation}
    \sim t_{\bot}\cos\left(2\theta\right).
\end{equation}

\subsection{Inter-wire superconducting pairing term $\Delta_{\bot}$}

Similarly, we investigate the effects on the bosonized Hamiltonian \eqref{Boson_hamiltonian_final} when additionally introducing an inter-wire SC-pairing $\Delta_{\bot}$. In the bonding/anti-bonding basis such a pairing enters as 
\begin{equation}
  H_{\Delta_{\bot}} =  -  \sum_{j} \left(\Delta_{\bot} c^{-\dagger}_{j}c^{+\dagger}_{j} + \Delta^{*}_{\bot} c^+_{j}c^-_{j}\right).
\end{equation}
Due to the gauge choice of the SC-phase $\phi_{\bot}$, the pairing potential is a priori a complex number.
When $\Delta_{\bot}$ is a real number, the Majorana representation of $H_{\Delta_{\bot}}$ is given by{
\begin{equation}
    H_{Re(\Delta_{\bot})} = i\frac{Re(\Delta_{\bot})}{2}\sum_{j} \left(\gamma^{+}_{A,j}\gamma^{-}_{B,j} - \gamma^{-}_{A,j}\gamma^{+}_{B,j}\right),
\end{equation}
resulting in $H_{Re(\Delta_{\bot})} \sim \sum_{j}\Gamma^{\dagger}_{j}\Theta_{j} + \text{h.c.}$.  In terms of the $R/L$ chiral basis this is $\sim \psi^{\dagger}_{R}\psi_{R} - \psi^{\dagger}_{L}\psi_{L}$, and subsequent bosonization produces an \emph{irrelevant} contribution. \\ 

On the other hand, if $\Delta_{\bot}$ is purely imaginary, one obtains analogously the terms
\begin{equation}
    H_{Im(\Delta_{\bot})} = i\frac{Im(\Delta_{\bot})}{2}\sum_{j} \left(\gamma^{+}_{A,j}\gamma^{-}_{A,j} - \gamma^{+}_{B,j}\gamma^{-}_{B,j}\right).
\end{equation}
Introducing the $\Gamma$ and $\Theta$ fermions yields
\begin{equation}
    H_{Im(\Delta_{\bot})} = Im(\Delta_{\bot})\sum_{j} \left(\Gamma^{\dagger}_{j}\Gamma{j}- \Theta^{\dagger}_{j}\Theta_{j}\right),
\end{equation}
which in the chiral basis is proportional to $\psi^{\dagger}_{R}\psi_{L} + \text{h.c.}$. Upon bosonization this contributes as $\sim \sin\left(2\phi\right)$ term, before shifting the $\phi$ fields by $\pi/4$. The imaginary part of $\Delta_{\bot}$ thus renormalizes the interaction $g_{\phi}$. This confirms that only an imaginary pairing $\Delta_{\bot}$ may gap out the topological modes, \emph{cf.} \cite{Yang_2020}. }

\section{Additional figures}\label{Appendix:Figures}
\begin{figure}[h!]
    \centering
    \includegraphics[width = 0.5\textwidth]{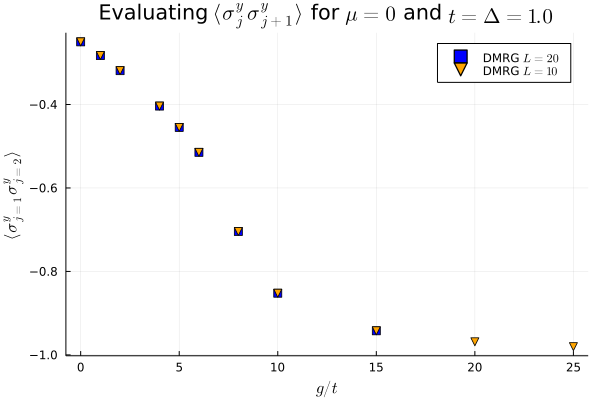}
    \caption{ Two-point correlation functions for both $y$ spin components extracted from the DMRG GS without inter-wire hopping $t_{\bot}$. An MI-AF emerges for $g \gg t$, seen by the ordering $\langle \sigma^{y}_{j}\sigma^{y}_{j+1}\rangle = -1$, as predicted, \emph{cf.} \eqref{MI_ordering_y}.    }
    \label{fig:yyspins_notp}
\end{figure}

{\color{magenta}
\begin{figure}[h!]
    \centering
    \includegraphics[width = 0.515\textwidth]{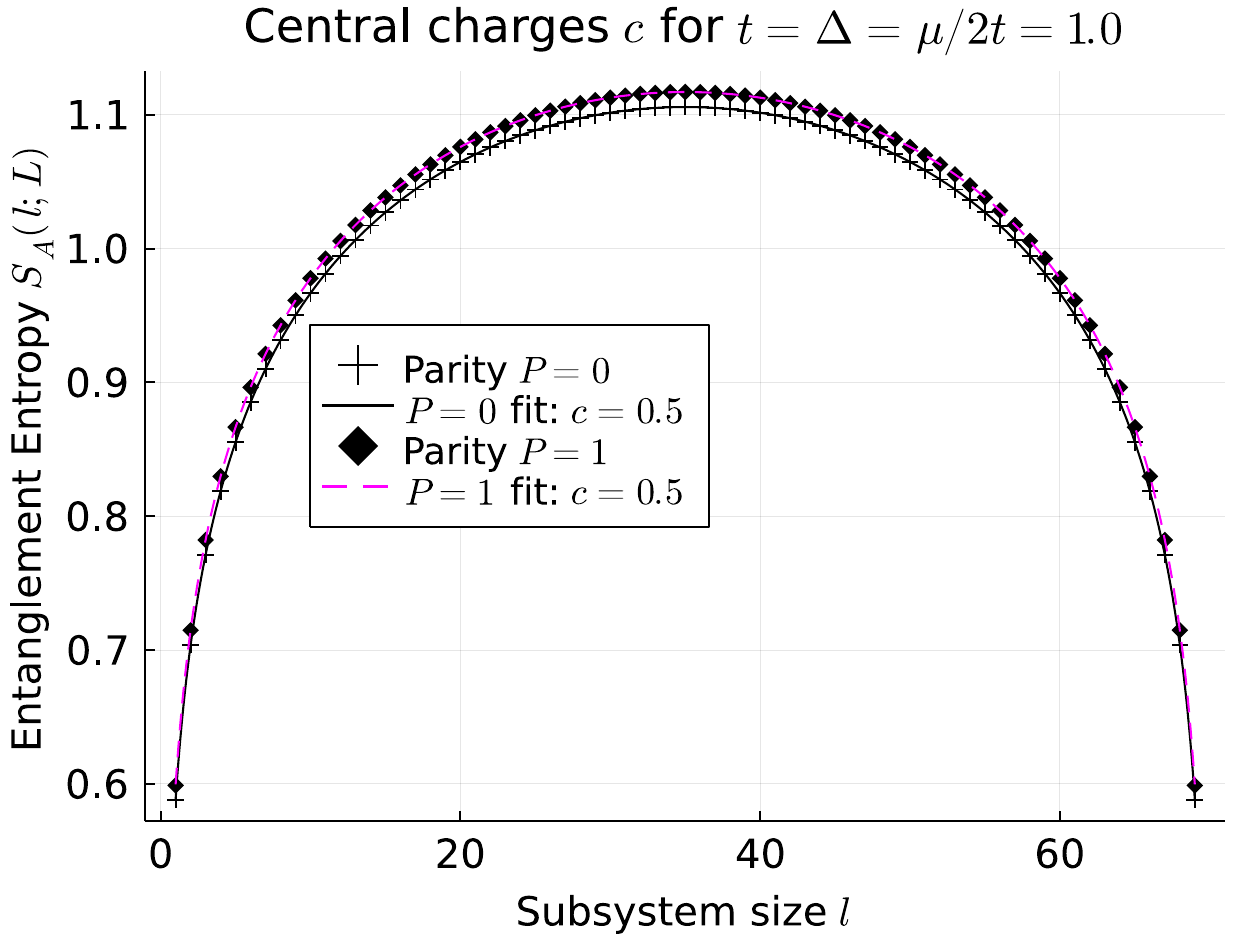}
    \caption{Entanglement entropy for a single Kitaev wire at the \emph{quantum critical point} (QCP), defined by $\mu = 2t$. Fitting the extracted entanglement entropy, we find the logarithmic contribution \cite{Song_2010}, i.e. central charge, which in both parity sectors yields $c= 1/2$. This corresponds to a free Majorana CFT, i.e. a single \emph{critical Ising phase}.}
    \label{fig:fractional_sw}
\end{figure}
}
Figure \ref{fig:degeneracy_twowires} below shows the energy gaps between the parity sectors $(P_{1},P_{2}) = (0,0)$ and $(1,1)$ wrt. to the sector $(P_{1},P_{2}) =(0,1)$ or $(1,0)$ equivalently.
\begin{figure}[h!]
    \centering
    \includegraphics[width = 0.5\textwidth]{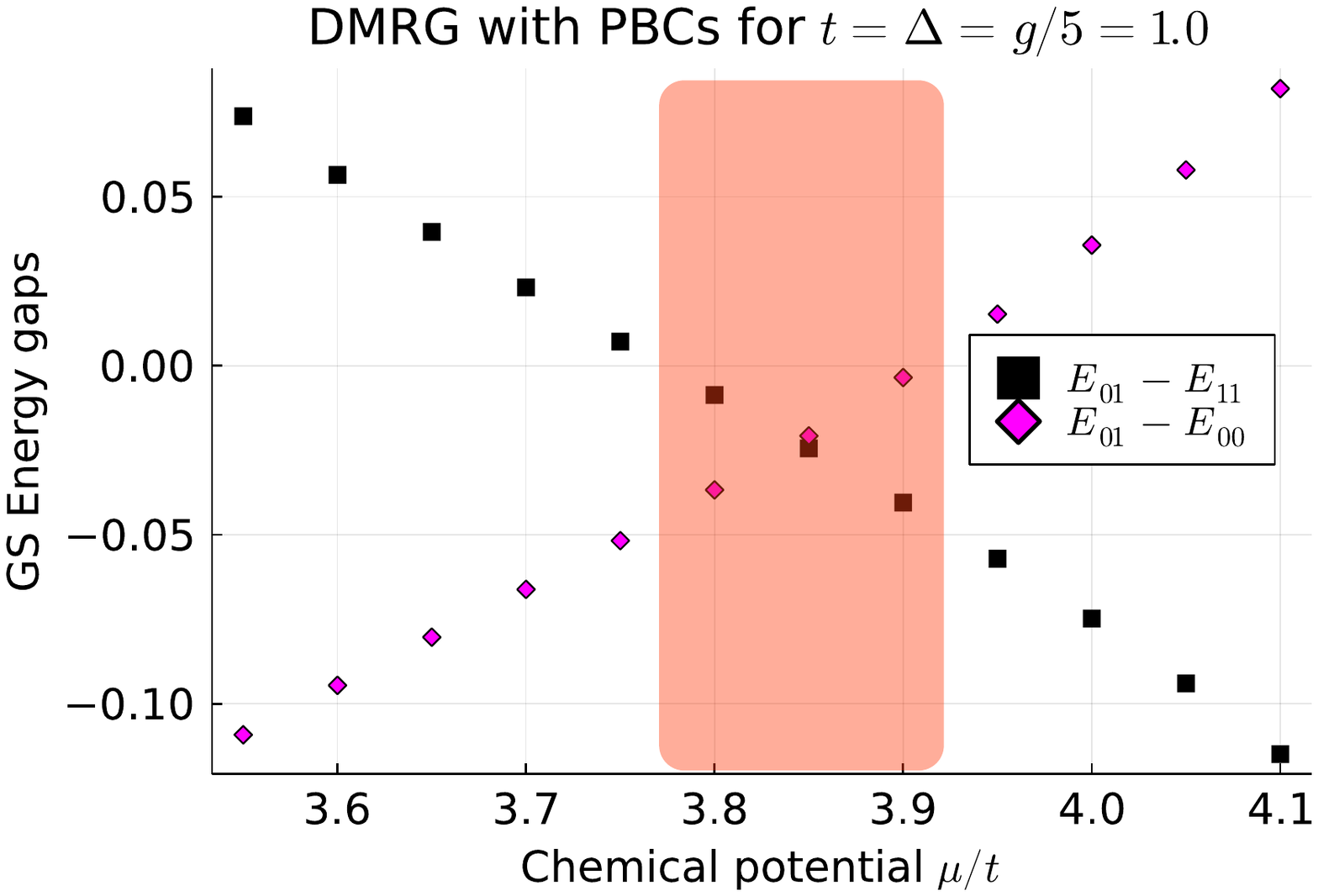}
    \caption{Energy gaps for $t = \Delta = 1.0$ and $g = 5t$ and $L = 76a$, wrt. $E_{01} = E_{10}$. In the central (shaded) region $E_{01}$ is the lowest eigenvalue, whilst the \emph{polarized} and \emph{4MF} phases are determined by the parities $(P_{1},P_{2}) = (0,0)$ and $(1,1)$ respectively.}
    \label{fig:degeneracy_twowires}
\end{figure}
Figure \ref{fig:Integer_12} below shows the dual topological numbers $\overline{C}^{1} = \overline{C}^{2}$ for various interaction streengths $g = -2.5t,0t, 2.5t$. They distinguish the two \emph{polarized} phases by $\overline{C}^{1/2} = \pm1$.
\begin{figure}[h!]
    \centering
    \includegraphics[width = 0.48\textwidth]{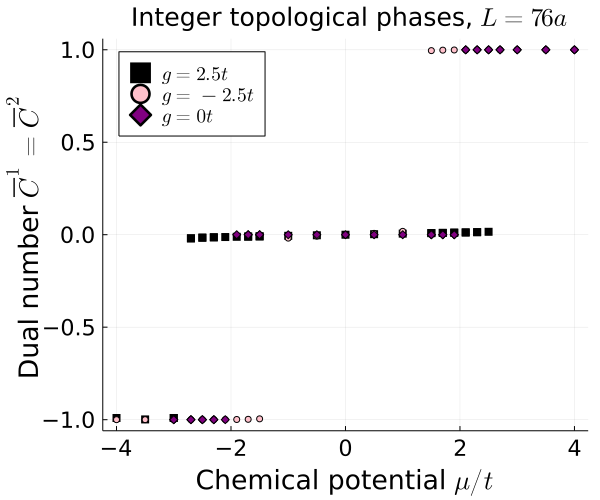}
    \caption{Dual numbers $\overline{C}^{\sigma}$ for $g = 0$ and $t = \Delta=1.0$, for $L = 76a$ per wire with PBCs. Dual numbers are thus able to distinguish between both polarized phases here as well.} 
    \label{fig:Integer_12}
\end{figure}
Figure \ref{fig:yyspins_notp} shows the $yy$-spin correlation functions in the limit of $t_{\bot} = 0$ for large interaction strengths $g$ and $\mu$. As is visible, the values quickly approach $-1$, verifying the GS property of the \emph{Mott-insulating anti-ferromagnetic} phase. This confirms the results previously obtained in \cite{Herviou_2016} equivalently.
{Figure \ref{fig:cbar_pd} again shows the dual topological numbers $\overline{C}^{\pm}$ and $\overline{C}^{1/2}$, which again reveal the transition from \emph{4MF} to \emph{2MF-b}, and then to the \emph{R-polarised} phase respectively. The \emph{2MF-b} phase is again characterized by $C^{1} = C^{2} = 1/2$, implying the existence of shared edge modes.
\begin{figure}[h!]
    \centering
    \includegraphics[width = 0.48\textwidth]{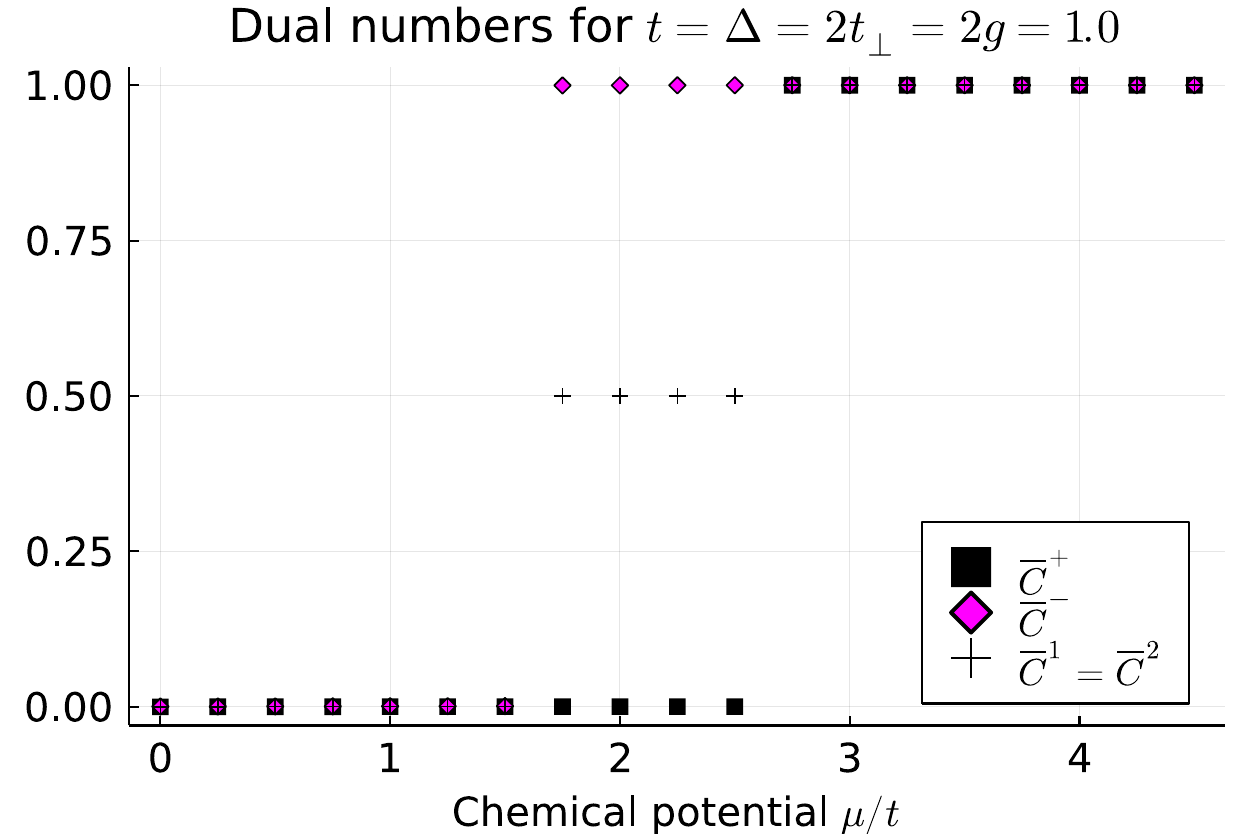}
   \caption{Dual numbers $\overline{C}$ for the transition from \emph{4MF} to \emph{2MF-b} and subsequently \emph{R-Polarized} phases. The fractional $\overline{C}^{1} = \overline{C}^{2} = 1/2$ value in the \emph{2MF-b} phase distinguishes this topological phase from the others.}
    \label{fig:cbar_pd}
\end{figure}
The subsequent figures \ref{fig:FlLogTPPBC_76}, \ref{fig:EntTPPBC_76} and \ref{fig:EntTPPBC_100} provide numerical evidence for the stability of the DCI against an inter-wire hopping amplitude $t_{\bot}$ at length scales below $\sim 1/t_{\bot}$. Both the central charges $c$ and the negative sub-dominant logarithmic contributions support the existence of the DCI phase.
\begin{figure}[h!]
    \centering
    \includegraphics[width = 0.5\textwidth]{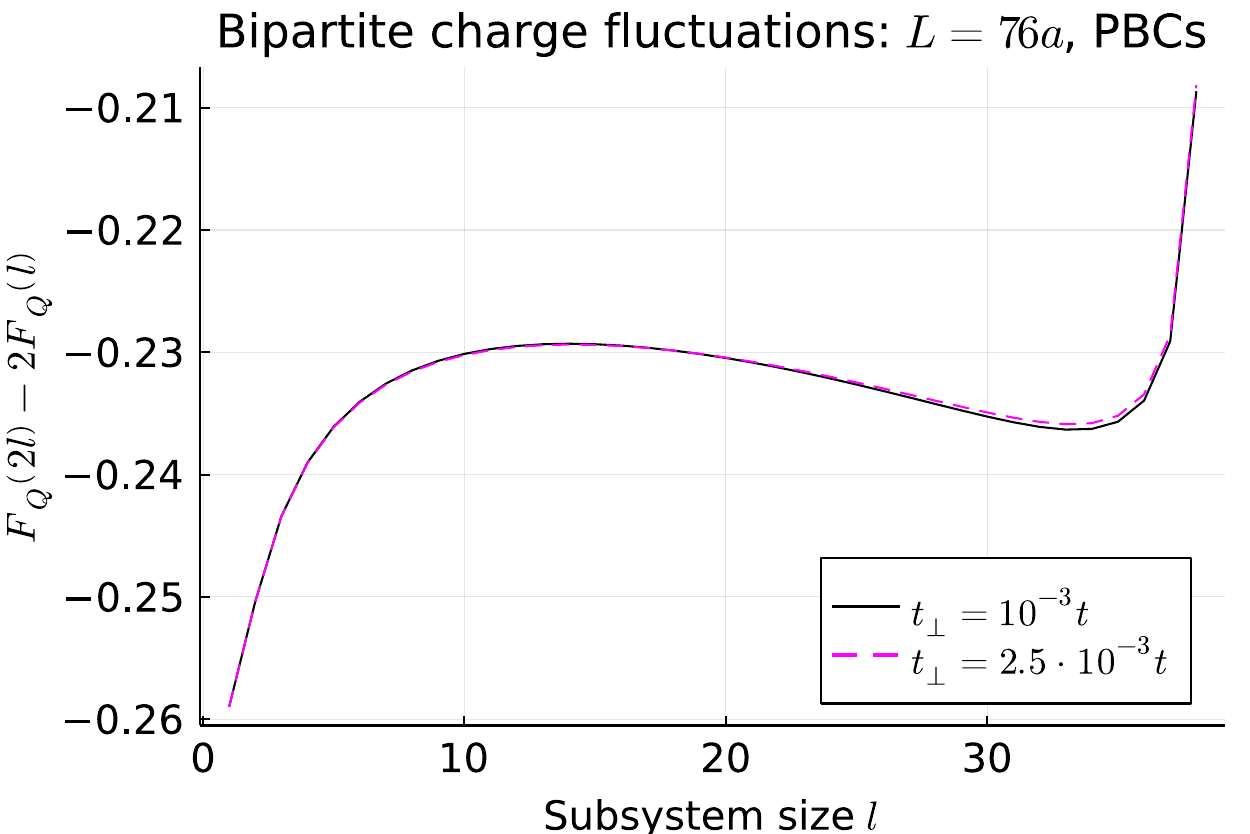}
   \caption{Sub-dominant logarithmic contributions of the bi-partite charge fluctuations \cite{Herviou_2017} for $t_{\bot} = 10^{-3}t$ and $t_{\bot} = 10^{-2}t$. Results from the same DMRG runs as in figure \ref{fig:EntTPPBC_76}. A clear positive logarithmic cusp initially is a defining feature of the DCI phase. }
    \label{fig:FlLogTPPBC_76}
\end{figure}

\begin{figure}[h!]
    \centering
    \includegraphics[width = 0.5\textwidth]{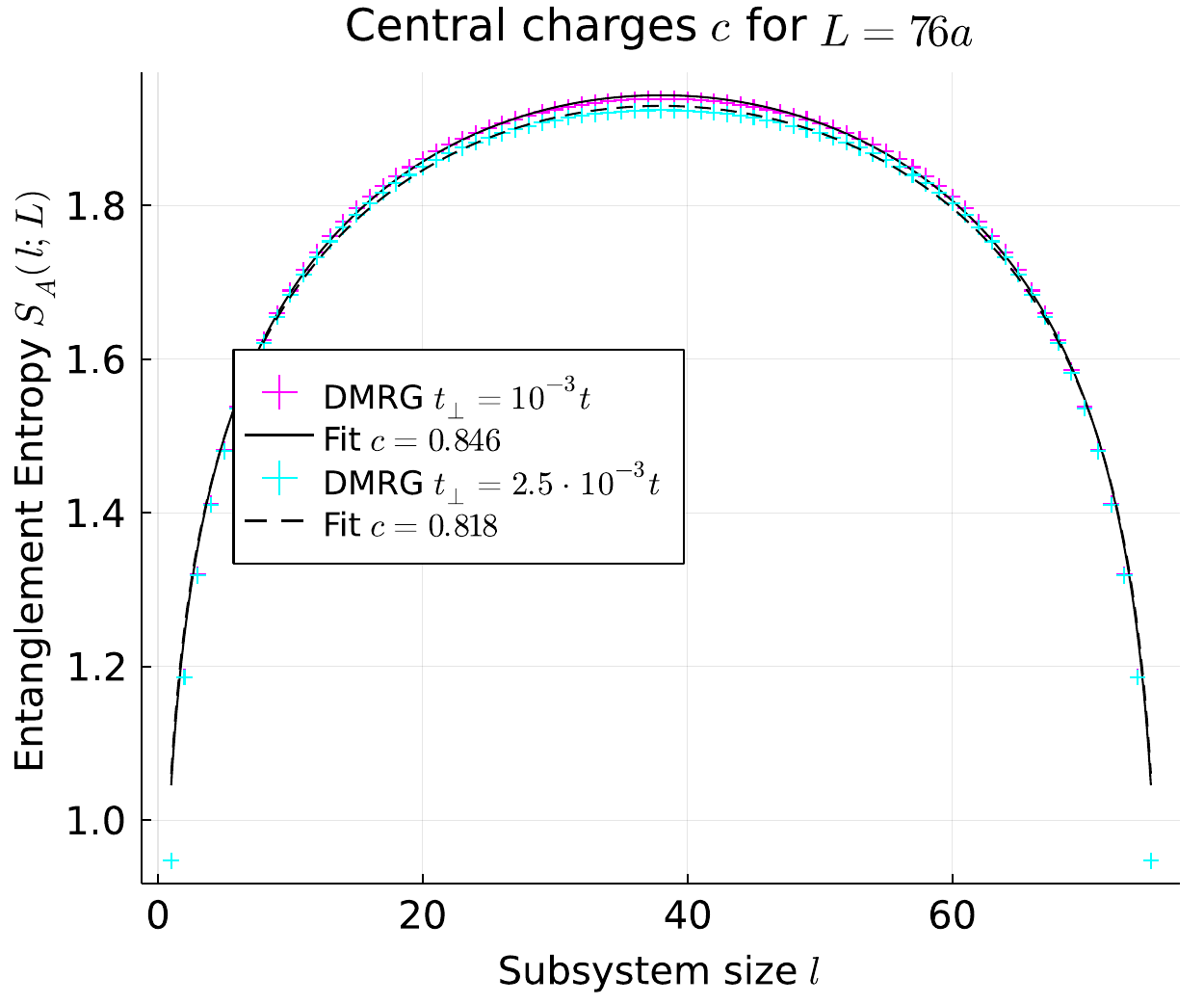}
   \caption{Central charges extracted for $t_{\bot} = 10^{-2}t$ and $t_{\bot} = 2.5\cdot 10^{-3}t$. Results for PBCs and same model parameters as in figure \ref{fig:EntTPOBC_300}. For both cases, the central charge is close to unity, with deviations possible due to finite size effects.}
    \label{fig:EntTPPBC_76}
\end{figure}

\begin{figure}[h!]
    \centering
    \includegraphics[width = 0.5\textwidth]{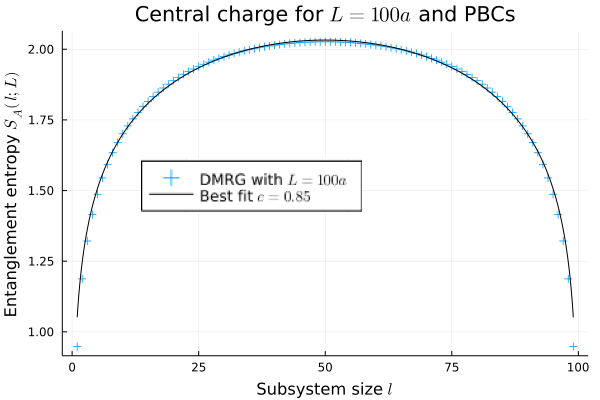}
   \caption{ To account for finite size effects for a phase in proximity to a critical region, we performed another DMRG calculation for $L = 100a$ per wire, and $t_{\bot} =10^{-3}t$. } 
    \label{fig:EntTPPBC_100}
\end{figure}
\clearpage
\bibliography{File}% Produces the bibliography via BibTeX.

%apsrev4-2.bst 2019-01-14 (MD) hand-edited version of apsrev4-1.bst
%Control: key (0)
%Control: author (8) initials jnrlst
%Control: editor formatted (1) identically to author
%Control: production of article title (0) allowed
%Control: page (0) single
%Control: year (1) truncated
%Control: production of eprint (0) enabled
\providecommand{\noopsort}[1]{}\providecommand{\singleletter}[1]{#1}%
\begin{thebibliography}{67}%
\makeatletter
\providecommand \@ifxundefined [1]{%
 \@ifx{#1\undefined}
}%
\providecommand \@ifnum [1]{%
 \ifnum #1\expandafter \@firstoftwo
 \else \expandafter \@secondoftwo
 \fi
}%
\providecommand \@ifx [1]{%
 \ifx #1\expandafter \@firstoftwo
 \else \expandafter \@secondoftwo
 \fi
}%
\providecommand \natexlab [1]{#1}%
\providecommand \enquote  [1]{``#1''}%
\providecommand \bibnamefont  [1]{#1}%
\providecommand \bibfnamefont [1]{#1}%
\providecommand \citenamefont [1]{#1}%
\providecommand \href@noop [0]{\@secondoftwo}%
\providecommand \href [0]{\begingroup \@sanitize@url \@href}%
\providecommand \@href[1]{\@@startlink{#1}\@@href}%
\providecommand \@@href[1]{\endgroup#1\@@endlink}%
\providecommand \@sanitize@url [0]{\catcode `\\12\catcode `\$12\catcode
  `\&12\catcode `\#12\catcode `\^12\catcode `\_12\catcode `\%12\relax}%
\providecommand \@@startlink[1]{}%
\providecommand \@@endlink[0]{}%
\providecommand \url  [0]{\begingroup\@sanitize@url \@url }%
\providecommand \@url [1]{\endgroup\@href {#1}{\urlprefix }}%
\providecommand \urlprefix  [0]{URL }%
\providecommand \Eprint [0]{\href }%
\providecommand \doibase [0]{https://doi.org/}%
\providecommand \selectlanguage [0]{\@gobble}%
\providecommand \bibinfo  [0]{\@secondoftwo}%
\providecommand \bibfield  [0]{\@secondoftwo}%
\providecommand \translation [1]{[#1]}%
\providecommand \BibitemOpen [0]{}%
\providecommand \bibitemStop [0]{}%
\providecommand \bibitemNoStop [0]{.\EOS\space}%
\providecommand \EOS [0]{\spacefactor3000\relax}%
\providecommand \BibitemShut  [1]{\csname bibitem#1\endcsname}%
\let\auto@bib@innerbib\@empty
%</preamble>
\bibitem [{\citenamefont {Herviou}\ \emph {et~al.}(2016)\citenamefont
  {Herviou}, \citenamefont {Mora},\ and\ \citenamefont
  {Le~Hur}}]{Herviou_2016}%
  \BibitemOpen
  \bibfield  {author} {\bibinfo {author} {\bibfnamefont {L.}~\bibnamefont
  {Herviou}}, \bibinfo {author} {\bibfnamefont {C.}~\bibnamefont {Mora}},\ and\
  \bibinfo {author} {\bibfnamefont {K.}~\bibnamefont {Le~Hur}},\ }\bibfield
  {title} {\bibinfo {title} {Phase diagram and entanglement of two interacting
  topological kitaev chains \\},\ }\bibfield  {journal} {\bibinfo  {journal}
  {Physical Review B}\ }\textbf {\bibinfo {volume} {93}},\ \href
  {https://doi.org/165142} {165142} (\bibinfo {year} {2016})\BibitemShut
  {NoStop}%
\bibitem [{\citenamefont {Kosterlitz}\ and\ \citenamefont
  {Thouless}(1973)}]{KT}%
  \BibitemOpen
  \bibfield  {author} {\bibinfo {author} {\bibfnamefont {J.~M.}\ \bibnamefont
  {Kosterlitz}}\ and\ \bibinfo {author} {\bibfnamefont {D.~J.}\ \bibnamefont
  {Thouless}},\ }\bibfield  {title} {\bibinfo {title} {Ordering, metastability
  and phase transitions in two-dimensional systems},\ }\href
  {https://doi.org/10.1088/0022-3719/6/7/010} {\bibfield  {journal} {\bibinfo
  {journal} {Journal of Physics C: Solid State Physics}\ }\textbf {\bibinfo
  {volume} {6}},\ \bibinfo {pages} {1181} (\bibinfo {year} {1973})}\BibitemShut
  {NoStop}%
\bibitem [{\citenamefont {Thouless}\ \emph {et~al.}(1982)\citenamefont
  {Thouless}, \citenamefont {Kohmoto}, \citenamefont {Nightingale},\ and\
  \citenamefont {den Nijs}}]{TKNN}%
  \BibitemOpen
  \bibfield  {author} {\bibinfo {author} {\bibfnamefont {D.~J.}\ \bibnamefont
  {Thouless}}, \bibinfo {author} {\bibfnamefont {M.}~\bibnamefont {Kohmoto}},
  \bibinfo {author} {\bibfnamefont {M.~P.}\ \bibnamefont {Nightingale}},\ and\
  \bibinfo {author} {\bibfnamefont {M.}~\bibnamefont {den Nijs}},\ }\bibfield
  {title} {\bibinfo {title} {Quantized hall conductance in a two-dimensional
  periodic potential},\ }\href {https://doi.org/10.1103/PhysRevLett.49.405}
  {\bibfield  {journal} {\bibinfo  {journal} {Phys. Rev. Lett.}\ }\textbf
  {\bibinfo {volume} {49}},\ \bibinfo {pages} {405} (\bibinfo {year}
  {1982})}\BibitemShut {NoStop}%
\bibitem [{\citenamefont {von Klitzing}\ \emph {et~al.}(1980)\citenamefont {von
  Klitzing}, \citenamefont {Dorda},\ and\ \citenamefont {Pepper}}]{QHE}%
  \BibitemOpen
  \bibfield  {author} {\bibinfo {author} {\bibfnamefont {K.}~\bibnamefont {von
  Klitzing}}, \bibinfo {author} {\bibfnamefont {G.}~\bibnamefont {Dorda}},\
  and\ \bibinfo {author} {\bibfnamefont {M.}~\bibnamefont {Pepper}},\
  }\bibfield  {title} {\bibinfo {title} {New method for high-accuracy
  determination of the fine-structure constant based on quantized hall
  resistance},\ }\href
  {https://doi.org/https://doi.org/10.1103/PhysRevLett.45.494} {\bibfield
  {journal} {\bibinfo  {journal} {Phys. Rev. Lett.}\ }\textbf {\bibinfo
  {volume} {45}},\ \bibinfo {pages} {494} (\bibinfo {year} {1980})}\BibitemShut
  {NoStop}%
\bibitem [{\citenamefont {Laughlin}(1983)}]{Laughlin}%
  \BibitemOpen
  \bibfield  {author} {\bibinfo {author} {\bibfnamefont {R.~B.}\ \bibnamefont
  {Laughlin}},\ }\bibfield  {title} {\bibinfo {title} {Anomalous quantum hall
  effect: An incompressible quantum fluid with fractionally charged
  excitations},\ }\href {https://doi.org/10.1103/PhysRevLett.50.1395}
  {\bibfield  {journal} {\bibinfo  {journal} {Phys. Rev. Lett.}\ }\textbf
  {\bibinfo {volume} {50}},\ \bibinfo {pages} {1395} (\bibinfo {year}
  {1983})}\BibitemShut {NoStop}%
\bibitem [{\citenamefont {Stormer}\ \emph {et~al.}(1999)\citenamefont
  {Stormer}, \citenamefont {Tsui},\ and\ \citenamefont {Gossar}}]{Stormer}%
  \BibitemOpen
  \bibfield  {author} {\bibinfo {author} {\bibfnamefont {H.~L.}\ \bibnamefont
  {Stormer}}, \bibinfo {author} {\bibfnamefont {D.~C.}\ \bibnamefont {Tsui}},\
  and\ \bibinfo {author} {\bibfnamefont {A.~C.}\ \bibnamefont {Gossar}},\
  }\bibfield  {title} {\bibinfo {title} {The fractional quantum hall effect},\
  }\href {https://doi.org/10.1103/RevModPhys.71.S298} {\bibfield  {journal}
  {\bibinfo  {journal} {Rev. Mod. Phys.}\ }\textbf {\bibinfo {volume} {71}},\
  \bibinfo {pages} {S298} (\bibinfo {year} {1999})}\BibitemShut {NoStop}%
\bibitem [{\citenamefont {Kapfer}\ \emph {et~al.}(2019)\citenamefont {Kapfer},
  \citenamefont {Roulleau}, \citenamefont {Santin}, \citenamefont {Farrer},
  \citenamefont {Ritchie},\ and\ \citenamefont {Glattli}}]{Kapfer_2019}%
  \BibitemOpen
  \bibfield  {author} {\bibinfo {author} {\bibfnamefont {M.}~\bibnamefont
  {Kapfer}}, \bibinfo {author} {\bibfnamefont {P.}~\bibnamefont {Roulleau}},
  \bibinfo {author} {\bibfnamefont {M.}~\bibnamefont {Santin}}, \bibinfo
  {author} {\bibfnamefont {I.}~\bibnamefont {Farrer}}, \bibinfo {author}
  {\bibfnamefont {D.~A.}\ \bibnamefont {Ritchie}},\ and\ \bibinfo {author}
  {\bibfnamefont {D.~C.}\ \bibnamefont {Glattli}},\ }\bibfield  {title}
  {\bibinfo {title} {A josephson relation for fractionally charged anyons},\
  }\href {https://doi.org/10.1126/science.aau3539} {\bibfield  {journal}
  {\bibinfo  {journal} {Science}\ }\textbf {\bibinfo {volume} {363}},\ \bibinfo
  {pages} {846} (\bibinfo {year} {2019})}\BibitemShut {NoStop}%
\bibitem [{\citenamefont {Alicea}(2012)}]{Alicea_2012}%
  \BibitemOpen
  \bibfield  {author} {\bibinfo {author} {\bibfnamefont {J.}~\bibnamefont
  {Alicea}},\ }\bibfield  {title} {\bibinfo {title} {New directions in the
  pursuit of majorana fermions in solid state systems},\ }\href
  {https://doi.org/10.1088/0034-4885/75/7/076501} {\bibfield  {journal}
  {\bibinfo  {journal} {Reports on Progress in Physics}\ }\textbf {\bibinfo
  {volume} {75}},\ \bibinfo {pages} {076501} (\bibinfo {year}
  {2012})}\BibitemShut {NoStop}%
\bibitem [{\citenamefont {Alicea}\ \emph {et~al.}(2011)\citenamefont {Alicea},
  \citenamefont {Oreg}, \citenamefont {Refael}, \citenamefont {von Oppen},\
  and\ \citenamefont {Fisher}}]{Alicea_2011}%
  \BibitemOpen
  \bibfield  {author} {\bibinfo {author} {\bibfnamefont {J.}~\bibnamefont
  {Alicea}}, \bibinfo {author} {\bibfnamefont {Y.}~\bibnamefont {Oreg}},
  \bibinfo {author} {\bibfnamefont {G.}~\bibnamefont {Refael}}, \bibinfo
  {author} {\bibfnamefont {F.}~\bibnamefont {von Oppen}},\ and\ \bibinfo
  {author} {\bibfnamefont {M.~P.~A.}\ \bibnamefont {Fisher}},\ }\bibfield
  {title} {\bibinfo {title} {Non-abelian statistics and topological quantum
  information processing in 1d wire networks},\ }\href
  {https://doi.org/10.1038/nphys1915} {\bibfield  {journal} {\bibinfo
  {journal} {Nature Physics}\ }\textbf {\bibinfo {volume} {7}},\ \bibinfo
  {pages} {412} (\bibinfo {year} {2011})}\BibitemShut {NoStop}%
\bibitem [{\citenamefont {Kitaev}(2001)}]{Kitaev_2001}%
  \BibitemOpen
  \bibfield  {author} {\bibinfo {author} {\bibfnamefont {A.~Y.}\ \bibnamefont
  {Kitaev}},\ }\bibfield  {title} {\bibinfo {title} {Unpaired majorana fermions
  in quantum wires},\ }\href {https://doi.org/10.1070/1063-7869/44/10s/s29}
  {\bibfield  {journal} {\bibinfo  {journal} {Physics-Uspekhi}\ }\textbf
  {\bibinfo {volume} {44}},\ \bibinfo {pages} {131} (\bibinfo {year}
  {2001})}\BibitemShut {NoStop}%
\bibitem [{\citenamefont {Majorana}(1937)}]{Majorana_OG}%
  \BibitemOpen
  \bibfield  {author} {\bibinfo {author} {\bibfnamefont {E.}~\bibnamefont
  {Majorana}},\ }\bibfield  {title} {\bibinfo {title} {Teoria simmetrica
  dell’elettrone e del positrone},\ }\bibfield  {journal} {\bibinfo
  {journal} {Il Nuovo Cimento (1924-1942), 14, Article number: 171 (1937)}\
  }\href {https://doi.org/10.1007/BF02961314} {10.1007/BF02961314} (\bibinfo
  {year} {1937})\BibitemShut {NoStop}%
\bibitem [{\citenamefont {Mi}\ \emph {et~al.}(2022)\citenamefont {Mi},
  \citenamefont {Sonner}, \citenamefont {Niu}, \citenamefont {Lee},
  \citenamefont {Foxen}, \citenamefont {Acharya}, \citenamefont {Aleiner},
  \citenamefont {Andersen}, \citenamefont {Arute}, \citenamefont {Arya},
  \citenamefont {Asfaw}, \citenamefont {Atalaya}, \citenamefont {Babbush},
  \citenamefont {Bacon}, \citenamefont {Bardin}, \citenamefont {Basso},
  \citenamefont {Bengtsson}, \citenamefont {Bortoli}, \citenamefont {Bourassa},
  \citenamefont {Brill}, \citenamefont {Broughton}, \citenamefont {Buckley},
  \citenamefont {Buell}, \citenamefont {Burkett}, \citenamefont {Bushnell},
  \citenamefont {Chen}, \citenamefont {Chiaro}, \citenamefont {Collins},
  \citenamefont {Conner}, \citenamefont {Courtney}, \citenamefont {Crook},
  \citenamefont {Debroy}, \citenamefont {Demura}, \citenamefont {Dunsworth},
  \citenamefont {Eppens}, \citenamefont {Erickson}, \citenamefont {Faoro},
  \citenamefont {Farhi}, \citenamefont {Fatemi}, \citenamefont {Flores},
  \citenamefont {Forati}, \citenamefont {Fowler}, \citenamefont {Giang},
  \citenamefont {Gidney}, \citenamefont {Gilboa}, \citenamefont {Giustina},
  \citenamefont {Dau}, \citenamefont {Gross}, \citenamefont {Habegger},
  \citenamefont {Harrigan}, \citenamefont {Hilton}, \citenamefont {Hoffmann},
  \citenamefont {Hong}, \citenamefont {Huang}, \citenamefont {Huff},
  \citenamefont {Huggins}, \citenamefont {Ioffe}, \citenamefont {Isakov},
  \citenamefont {Iveland}, \citenamefont {Jeffrey}, \citenamefont {Jiang},
  \citenamefont {Jones}, \citenamefont {Kafri}, \citenamefont {Kechedzhi},
  \citenamefont {Khattar}, \citenamefont {Kim}, \citenamefont {Kitaev},
  \citenamefont {Klimov}, \citenamefont {Klots}, \citenamefont {Korotkov},
  \citenamefont {Kostritsa}, \citenamefont {Kreikebaum}, \citenamefont
  {Landhuis}, \citenamefont {Laptev}, \citenamefont {Lau}, \citenamefont {Lee},
  \citenamefont {Laws}, \citenamefont {Liu}, \citenamefont {Locharla},
  \citenamefont {Lucero}, \citenamefont {Martin}, \citenamefont {McClean},
  \citenamefont {McEwen}, \citenamefont {Costa}, \citenamefont {Miao},
  \citenamefont {Mohseni}, \citenamefont {Montazeri}, \citenamefont {Morvan},
  \citenamefont {Mount}, \citenamefont {Mruczkiewicz}, \citenamefont {Naaman},
  \citenamefont {Neeley}, \citenamefont {Neill}, \citenamefont {Newman},
  \citenamefont {O'Brien}, \citenamefont {Opremcak}, \citenamefont {Petukhov},
  \citenamefont {Potter}, \citenamefont {Quintana}, \citenamefont {Rubin},
  \citenamefont {Saei}, \citenamefont {Sank}, \citenamefont {Sankaragomathi},
  \citenamefont {Satzinger}, \citenamefont {Schuster}, \citenamefont {Shearn},
  \citenamefont {Shvarts}, \citenamefont {Strain}, \citenamefont {Su},
  \citenamefont {Szalay}, \citenamefont {Vidal}, \citenamefont {Villalonga},
  \citenamefont {Vollgraff-Heidweiller}, \citenamefont {White}, \citenamefont
  {Yao}, \citenamefont {Yeh}, \citenamefont {Yoo}, \citenamefont {Zalcman},
  \citenamefont {Zhang}, \citenamefont {Zhu}, \citenamefont {Neven},
  \citenamefont {Boixo}, \citenamefont {Megrant}, \citenamefont {Chen},
  \citenamefont {Kelly}, \citenamefont {Smelyanskiy}, \citenamefont {Abanin},\
  and\ \citenamefont {Roushan}}]{Google_Maj}%
  \BibitemOpen
  \bibfield  {author} {\bibinfo {author} {\bibfnamefont {X.}~\bibnamefont
  {Mi}}, \bibinfo {author} {\bibfnamefont {M.}~\bibnamefont {Sonner}}, \bibinfo
  {author} {\bibfnamefont {M.~Y.}\ \bibnamefont {Niu}}, \bibinfo {author}
  {\bibfnamefont {K.~W.}\ \bibnamefont {Lee}}, \bibinfo {author} {\bibfnamefont
  {B.}~\bibnamefont {Foxen}}, \bibinfo {author} {\bibfnamefont
  {R.}~\bibnamefont {Acharya}}, \bibinfo {author} {\bibfnamefont
  {I.}~\bibnamefont {Aleiner}}, \bibinfo {author} {\bibfnamefont {T.~I.}\
  \bibnamefont {Andersen}}, \bibinfo {author} {\bibfnamefont {F.}~\bibnamefont
  {Arute}}, \bibinfo {author} {\bibfnamefont {K.}~\bibnamefont {Arya}},
  \bibinfo {author} {\bibfnamefont {A.}~\bibnamefont {Asfaw}}, \bibinfo
  {author} {\bibfnamefont {J.}~\bibnamefont {Atalaya}}, \bibinfo {author}
  {\bibfnamefont {R.}~\bibnamefont {Babbush}}, \bibinfo {author} {\bibfnamefont
  {D.}~\bibnamefont {Bacon}}, \bibinfo {author} {\bibfnamefont {J.~C.}\
  \bibnamefont {Bardin}}, \bibinfo {author} {\bibfnamefont {J.}~\bibnamefont
  {Basso}}, \bibinfo {author} {\bibfnamefont {A.}~\bibnamefont {Bengtsson}},
  \bibinfo {author} {\bibfnamefont {G.}~\bibnamefont {Bortoli}}, \bibinfo
  {author} {\bibfnamefont {A.}~\bibnamefont {Bourassa}}, \bibinfo {author}
  {\bibfnamefont {L.}~\bibnamefont {Brill}}, \bibinfo {author} {\bibfnamefont
  {M.}~\bibnamefont {Broughton}}, \bibinfo {author} {\bibfnamefont {B.~B.}\
  \bibnamefont {Buckley}}, \bibinfo {author} {\bibfnamefont {D.~A.}\
  \bibnamefont {Buell}}, \bibinfo {author} {\bibfnamefont {B.}~\bibnamefont
  {Burkett}}, \bibinfo {author} {\bibfnamefont {N.}~\bibnamefont {Bushnell}},
  \bibinfo {author} {\bibfnamefont {Z.}~\bibnamefont {Chen}}, \bibinfo {author}
  {\bibfnamefont {B.}~\bibnamefont {Chiaro}}, \bibinfo {author} {\bibfnamefont
  {R.}~\bibnamefont {Collins}}, \bibinfo {author} {\bibfnamefont
  {P.}~\bibnamefont {Conner}}, \bibinfo {author} {\bibfnamefont
  {W.}~\bibnamefont {Courtney}}, \bibinfo {author} {\bibfnamefont {A.~L.}\
  \bibnamefont {Crook}}, \bibinfo {author} {\bibfnamefont {D.~M.}\ \bibnamefont
  {Debroy}}, \bibinfo {author} {\bibfnamefont {S.}~\bibnamefont {Demura}},
  \bibinfo {author} {\bibfnamefont {A.}~\bibnamefont {Dunsworth}}, \bibinfo
  {author} {\bibfnamefont {D.}~\bibnamefont {Eppens}}, \bibinfo {author}
  {\bibfnamefont {C.}~\bibnamefont {Erickson}}, \bibinfo {author}
  {\bibfnamefont {L.}~\bibnamefont {Faoro}}, \bibinfo {author} {\bibfnamefont
  {E.}~\bibnamefont {Farhi}}, \bibinfo {author} {\bibfnamefont
  {R.}~\bibnamefont {Fatemi}}, \bibinfo {author} {\bibfnamefont
  {L.}~\bibnamefont {Flores}}, \bibinfo {author} {\bibfnamefont
  {E.}~\bibnamefont {Forati}}, \bibinfo {author} {\bibfnamefont {A.~G.}\
  \bibnamefont {Fowler}}, \bibinfo {author} {\bibfnamefont {W.}~\bibnamefont
  {Giang}}, \bibinfo {author} {\bibfnamefont {C.}~\bibnamefont {Gidney}},
  \bibinfo {author} {\bibfnamefont {D.}~\bibnamefont {Gilboa}}, \bibinfo
  {author} {\bibfnamefont {M.}~\bibnamefont {Giustina}}, \bibinfo {author}
  {\bibfnamefont {A.~G.}\ \bibnamefont {Dau}}, \bibinfo {author} {\bibfnamefont
  {J.~A.}\ \bibnamefont {Gross}}, \bibinfo {author} {\bibfnamefont
  {S.}~\bibnamefont {Habegger}}, \bibinfo {author} {\bibfnamefont {M.~P.}\
  \bibnamefont {Harrigan}}, \bibinfo {author} {\bibfnamefont {J.}~\bibnamefont
  {Hilton}}, \bibinfo {author} {\bibfnamefont {M.}~\bibnamefont {Hoffmann}},
  \bibinfo {author} {\bibfnamefont {S.}~\bibnamefont {Hong}}, \bibinfo {author}
  {\bibfnamefont {T.}~\bibnamefont {Huang}}, \bibinfo {author} {\bibfnamefont
  {A.}~\bibnamefont {Huff}}, \bibinfo {author} {\bibfnamefont {W.~J.}\
  \bibnamefont {Huggins}}, \bibinfo {author} {\bibfnamefont {L.~B.}\
  \bibnamefont {Ioffe}}, \bibinfo {author} {\bibfnamefont {S.~V.}\ \bibnamefont
  {Isakov}}, \bibinfo {author} {\bibfnamefont {J.}~\bibnamefont {Iveland}},
  \bibinfo {author} {\bibfnamefont {E.}~\bibnamefont {Jeffrey}}, \bibinfo
  {author} {\bibfnamefont {Z.}~\bibnamefont {Jiang}}, \bibinfo {author}
  {\bibfnamefont {C.}~\bibnamefont {Jones}}, \bibinfo {author} {\bibfnamefont
  {D.}~\bibnamefont {Kafri}}, \bibinfo {author} {\bibfnamefont
  {K.}~\bibnamefont {Kechedzhi}}, \bibinfo {author} {\bibfnamefont
  {T.}~\bibnamefont {Khattar}}, \bibinfo {author} {\bibfnamefont
  {S.}~\bibnamefont {Kim}}, \bibinfo {author} {\bibfnamefont {A.}~\bibnamefont
  {Kitaev}}, \bibinfo {author} {\bibfnamefont {P.~V.}\ \bibnamefont {Klimov}},
  \bibinfo {author} {\bibfnamefont {A.~R.}\ \bibnamefont {Klots}}, \bibinfo
  {author} {\bibfnamefont {A.~N.}\ \bibnamefont {Korotkov}}, \bibinfo {author}
  {\bibfnamefont {F.}~\bibnamefont {Kostritsa}}, \bibinfo {author}
  {\bibfnamefont {J.~M.}\ \bibnamefont {Kreikebaum}}, \bibinfo {author}
  {\bibfnamefont {D.}~\bibnamefont {Landhuis}}, \bibinfo {author}
  {\bibfnamefont {P.}~\bibnamefont {Laptev}}, \bibinfo {author} {\bibfnamefont
  {K.-M.}\ \bibnamefont {Lau}}, \bibinfo {author} {\bibfnamefont
  {J.}~\bibnamefont {Lee}}, \bibinfo {author} {\bibfnamefont {L.}~\bibnamefont
  {Laws}}, \bibinfo {author} {\bibfnamefont {W.}~\bibnamefont {Liu}}, \bibinfo
  {author} {\bibfnamefont {A.}~\bibnamefont {Locharla}}, \bibinfo {author}
  {\bibfnamefont {E.}~\bibnamefont {Lucero}}, \bibinfo {author} {\bibfnamefont
  {O.}~\bibnamefont {Martin}}, \bibinfo {author} {\bibfnamefont {J.~R.}\
  \bibnamefont {McClean}}, \bibinfo {author} {\bibfnamefont {M.}~\bibnamefont
  {McEwen}}, \bibinfo {author} {\bibfnamefont {B.~M.}\ \bibnamefont {Costa}},
  \bibinfo {author} {\bibfnamefont {K.~C.}\ \bibnamefont {Miao}}, \bibinfo
  {author} {\bibfnamefont {M.}~\bibnamefont {Mohseni}}, \bibinfo {author}
  {\bibfnamefont {S.}~\bibnamefont {Montazeri}}, \bibinfo {author}
  {\bibfnamefont {A.}~\bibnamefont {Morvan}}, \bibinfo {author} {\bibfnamefont
  {E.}~\bibnamefont {Mount}}, \bibinfo {author} {\bibfnamefont
  {W.}~\bibnamefont {Mruczkiewicz}}, \bibinfo {author} {\bibfnamefont
  {O.}~\bibnamefont {Naaman}}, \bibinfo {author} {\bibfnamefont
  {M.}~\bibnamefont {Neeley}}, \bibinfo {author} {\bibfnamefont
  {C.}~\bibnamefont {Neill}}, \bibinfo {author} {\bibfnamefont
  {M.}~\bibnamefont {Newman}}, \bibinfo {author} {\bibfnamefont {T.~E.}\
  \bibnamefont {O'Brien}}, \bibinfo {author} {\bibfnamefont {A.}~\bibnamefont
  {Opremcak}}, \bibinfo {author} {\bibfnamefont {A.}~\bibnamefont {Petukhov}},
  \bibinfo {author} {\bibfnamefont {R.}~\bibnamefont {Potter}}, \bibinfo
  {author} {\bibfnamefont {C.}~\bibnamefont {Quintana}}, \bibinfo {author}
  {\bibfnamefont {N.~C.}\ \bibnamefont {Rubin}}, \bibinfo {author}
  {\bibfnamefont {N.}~\bibnamefont {Saei}}, \bibinfo {author} {\bibfnamefont
  {D.}~\bibnamefont {Sank}}, \bibinfo {author} {\bibfnamefont {K.}~\bibnamefont
  {Sankaragomathi}}, \bibinfo {author} {\bibfnamefont {K.~J.}\ \bibnamefont
  {Satzinger}}, \bibinfo {author} {\bibfnamefont {C.}~\bibnamefont {Schuster}},
  \bibinfo {author} {\bibfnamefont {M.~J.}\ \bibnamefont {Shearn}}, \bibinfo
  {author} {\bibfnamefont {V.}~\bibnamefont {Shvarts}}, \bibinfo {author}
  {\bibfnamefont {D.}~\bibnamefont {Strain}}, \bibinfo {author} {\bibfnamefont
  {Y.}~\bibnamefont {Su}}, \bibinfo {author} {\bibfnamefont {M.}~\bibnamefont
  {Szalay}}, \bibinfo {author} {\bibfnamefont {G.}~\bibnamefont {Vidal}},
  \bibinfo {author} {\bibfnamefont {B.}~\bibnamefont {Villalonga}}, \bibinfo
  {author} {\bibfnamefont {C.}~\bibnamefont {Vollgraff-Heidweiller}}, \bibinfo
  {author} {\bibfnamefont {T.}~\bibnamefont {White}}, \bibinfo {author}
  {\bibfnamefont {Z.~J.}\ \bibnamefont {Yao}}, \bibinfo {author} {\bibfnamefont
  {P.}~\bibnamefont {Yeh}}, \bibinfo {author} {\bibfnamefont {J.}~\bibnamefont
  {Yoo}}, \bibinfo {author} {\bibfnamefont {A.}~\bibnamefont {Zalcman}},
  \bibinfo {author} {\bibfnamefont {Y.}~\bibnamefont {Zhang}}, \bibinfo
  {author} {\bibfnamefont {N.}~\bibnamefont {Zhu}}, \bibinfo {author}
  {\bibfnamefont {H.}~\bibnamefont {Neven}}, \bibinfo {author} {\bibfnamefont
  {S.}~\bibnamefont {Boixo}}, \bibinfo {author} {\bibfnamefont
  {A.}~\bibnamefont {Megrant}}, \bibinfo {author} {\bibfnamefont
  {Y.}~\bibnamefont {Chen}}, \bibinfo {author} {\bibfnamefont {J.}~\bibnamefont
  {Kelly}}, \bibinfo {author} {\bibfnamefont {V.}~\bibnamefont {Smelyanskiy}},
  \bibinfo {author} {\bibfnamefont {D.~A.}\ \bibnamefont {Abanin}},\ and\
  \bibinfo {author} {\bibfnamefont {P.}~\bibnamefont {Roushan}},\ }\href
  {https://doi.org/10.48550/ARXIV.2204.11372} {\bibinfo {title}
  {"noise-resilient majorana edge modes on a chain of superconducting qubits"}}
  (\bibinfo {year} {2022}),\ \Eprint {https://arxiv.org/abs/2204.11372}
  {arXiv:2204.11372 [quant-ph]} \BibitemShut {NoStop}%
\bibitem [{\citenamefont {Aghaee}\ \emph {et~al.}(2022)\citenamefont {Aghaee},
  \citenamefont {Akkala}, \citenamefont {Alam}, \citenamefont {Ali},
  \citenamefont {Ramirez}, \citenamefont {Andrzejczuk}, \citenamefont
  {Antipov}, \citenamefont {Aseev}, \citenamefont {Astafev}, \citenamefont
  {Bauer}, \citenamefont {Becker}, \citenamefont {Boddapati}, \citenamefont
  {Boekhout}, \citenamefont {Bommer}, \citenamefont {Hansen}, \citenamefont
  {Bosma}, \citenamefont {Bourdet}, \citenamefont {Boutin}, \citenamefont
  {Caroff}, \citenamefont {Casparis}, \citenamefont {Cassidy}, \citenamefont
  {Christensen}, \citenamefont {Clay}, \citenamefont {Cole}, \citenamefont
  {Corsetti}, \citenamefont {Cui}, \citenamefont {Dalampiras}, \citenamefont
  {Dokania}, \citenamefont {de~Lange}, \citenamefont {de~Moor}, \citenamefont
  {Saldaña}, \citenamefont {Fallahi}, \citenamefont {Fathabad}, \citenamefont
  {Gamble}, \citenamefont {Gardner}, \citenamefont {Govender}, \citenamefont
  {Griggio}, \citenamefont {Grigoryan}, \citenamefont {Gronin}, \citenamefont
  {Gukelberger}, \citenamefont {Heedt}, \citenamefont {Zamorano}, \citenamefont
  {Ho}, \citenamefont {Holgaard}, \citenamefont {Nielsen}, \citenamefont
  {Ingerslev}, \citenamefont {Krogstrup}, \citenamefont {Johansson},
  \citenamefont {Jones}, \citenamefont {Kallaher}, \citenamefont {Karimi},
  \citenamefont {Karzig}, \citenamefont {King}, \citenamefont {Kloster},
  \citenamefont {Knapp}, \citenamefont {Kocon}, \citenamefont {Koski},
  \citenamefont {Kostamo}, \citenamefont {Kumar}, \citenamefont {Laeven},
  \citenamefont {Larsen}, \citenamefont {Li}, \citenamefont {Lindemann},
  \citenamefont {Love}, \citenamefont {Lutchyn}, \citenamefont {Manfra},
  \citenamefont {Memisevic}, \citenamefont {Nayak}, \citenamefont {Nijholt},
  \citenamefont {Madsen}, \citenamefont {Markussen}, \citenamefont {Martinez},
  \citenamefont {McNeil}, \citenamefont {Mullally}, \citenamefont {Nielsen},
  \citenamefont {Nurmohamed}, \citenamefont {O'Farrell}, \citenamefont {Otani},
  \citenamefont {Pauka}, \citenamefont {Petersson}, \citenamefont {Petit},
  \citenamefont {Pikulin}, \citenamefont {Preiss}, \citenamefont {Perez},
  \citenamefont {Rasmussen}, \citenamefont {Rajpalke}, \citenamefont
  {Razmadze}, \citenamefont {Reentila}, \citenamefont {Reilly}, \citenamefont
  {Rouse}, \citenamefont {Sadovskyy}, \citenamefont {Sainiemi}, \citenamefont
  {Schreppler}, \citenamefont {Sidorkin}, \citenamefont {Singh}, \citenamefont
  {Singh}, \citenamefont {Sinha}, \citenamefont {Sohr}, \citenamefont
  {Stankevič}, \citenamefont {Stek}, \citenamefont {Suominen}, \citenamefont
  {Suter}, \citenamefont {Svidenko}, \citenamefont {Teicher}, \citenamefont
  {Temuerhan}, \citenamefont {Thiyagarajah}, \citenamefont {Tholapi},
  \citenamefont {Thomas}, \citenamefont {Toomey}, \citenamefont {Upadhyay},
  \citenamefont {Urban}, \citenamefont {Vaitiekėnas}, \citenamefont
  {Van~Hoogdalem}, \citenamefont {Viazmitinov}, \citenamefont {Waddy},
  \citenamefont {Van~Woerkom}, \citenamefont {Vogel}, \citenamefont {Watson},
  \citenamefont {Weston}, \citenamefont {Winkler}, \citenamefont {Yang},
  \citenamefont {Yau}, \citenamefont {Yi}, \citenamefont {Yucelen},
  \citenamefont {Webster}, \citenamefont {Zeisel},\ and\ \citenamefont
  {Zhao}}]{HopeMajorana}%
  \BibitemOpen
  \bibfield  {author} {\bibinfo {author} {\bibfnamefont {M.}~\bibnamefont
  {Aghaee}}, \bibinfo {author} {\bibfnamefont {A.}~\bibnamefont {Akkala}},
  \bibinfo {author} {\bibfnamefont {Z.}~\bibnamefont {Alam}}, \bibinfo {author}
  {\bibfnamefont {R.}~\bibnamefont {Ali}}, \bibinfo {author} {\bibfnamefont
  {A.~A.}\ \bibnamefont {Ramirez}}, \bibinfo {author} {\bibfnamefont
  {M.}~\bibnamefont {Andrzejczuk}}, \bibinfo {author} {\bibfnamefont {A.~E.}\
  \bibnamefont {Antipov}}, \bibinfo {author} {\bibfnamefont {P.}~\bibnamefont
  {Aseev}}, \bibinfo {author} {\bibfnamefont {M.}~\bibnamefont {Astafev}},
  \bibinfo {author} {\bibfnamefont {B.}~\bibnamefont {Bauer}}, \bibinfo
  {author} {\bibfnamefont {J.}~\bibnamefont {Becker}}, \bibinfo {author}
  {\bibfnamefont {S.}~\bibnamefont {Boddapati}}, \bibinfo {author}
  {\bibfnamefont {F.}~\bibnamefont {Boekhout}}, \bibinfo {author}
  {\bibfnamefont {J.}~\bibnamefont {Bommer}}, \bibinfo {author} {\bibfnamefont
  {E.~B.}\ \bibnamefont {Hansen}}, \bibinfo {author} {\bibfnamefont
  {T.}~\bibnamefont {Bosma}}, \bibinfo {author} {\bibfnamefont
  {L.}~\bibnamefont {Bourdet}}, \bibinfo {author} {\bibfnamefont
  {S.}~\bibnamefont {Boutin}}, \bibinfo {author} {\bibfnamefont
  {P.}~\bibnamefont {Caroff}}, \bibinfo {author} {\bibfnamefont
  {L.}~\bibnamefont {Casparis}}, \bibinfo {author} {\bibfnamefont
  {M.}~\bibnamefont {Cassidy}}, \bibinfo {author} {\bibfnamefont {A.~W.}\
  \bibnamefont {Christensen}}, \bibinfo {author} {\bibfnamefont
  {N.}~\bibnamefont {Clay}}, \bibinfo {author} {\bibfnamefont {W.~S.}\
  \bibnamefont {Cole}}, \bibinfo {author} {\bibfnamefont {F.}~\bibnamefont
  {Corsetti}}, \bibinfo {author} {\bibfnamefont {A.}~\bibnamefont {Cui}},
  \bibinfo {author} {\bibfnamefont {P.}~\bibnamefont {Dalampiras}}, \bibinfo
  {author} {\bibfnamefont {A.}~\bibnamefont {Dokania}}, \bibinfo {author}
  {\bibfnamefont {G.}~\bibnamefont {de~Lange}}, \bibinfo {author}
  {\bibfnamefont {M.}~\bibnamefont {de~Moor}}, \bibinfo {author} {\bibfnamefont
  {J.~C.~E.}\ \bibnamefont {Saldaña}}, \bibinfo {author} {\bibfnamefont
  {S.}~\bibnamefont {Fallahi}}, \bibinfo {author} {\bibfnamefont {Z.~H.}\
  \bibnamefont {Fathabad}}, \bibinfo {author} {\bibfnamefont {J.}~\bibnamefont
  {Gamble}}, \bibinfo {author} {\bibfnamefont {G.}~\bibnamefont {Gardner}},
  \bibinfo {author} {\bibfnamefont {D.}~\bibnamefont {Govender}}, \bibinfo
  {author} {\bibfnamefont {F.}~\bibnamefont {Griggio}}, \bibinfo {author}
  {\bibfnamefont {R.}~\bibnamefont {Grigoryan}}, \bibinfo {author}
  {\bibfnamefont {S.}~\bibnamefont {Gronin}}, \bibinfo {author} {\bibfnamefont
  {J.}~\bibnamefont {Gukelberger}}, \bibinfo {author} {\bibfnamefont
  {S.}~\bibnamefont {Heedt}}, \bibinfo {author} {\bibfnamefont {J.~H.}\
  \bibnamefont {Zamorano}}, \bibinfo {author} {\bibfnamefont {S.}~\bibnamefont
  {Ho}}, \bibinfo {author} {\bibfnamefont {U.~L.}\ \bibnamefont {Holgaard}},
  \bibinfo {author} {\bibfnamefont {W.~H.~P.}\ \bibnamefont {Nielsen}},
  \bibinfo {author} {\bibfnamefont {H.}~\bibnamefont {Ingerslev}}, \bibinfo
  {author} {\bibfnamefont {P.~J.}\ \bibnamefont {Krogstrup}}, \bibinfo {author}
  {\bibfnamefont {L.}~\bibnamefont {Johansson}}, \bibinfo {author}
  {\bibfnamefont {J.}~\bibnamefont {Jones}}, \bibinfo {author} {\bibfnamefont
  {R.}~\bibnamefont {Kallaher}}, \bibinfo {author} {\bibfnamefont
  {F.}~\bibnamefont {Karimi}}, \bibinfo {author} {\bibfnamefont
  {T.}~\bibnamefont {Karzig}}, \bibinfo {author} {\bibfnamefont
  {C.}~\bibnamefont {King}}, \bibinfo {author} {\bibfnamefont {M.~E.}\
  \bibnamefont {Kloster}}, \bibinfo {author} {\bibfnamefont {C.}~\bibnamefont
  {Knapp}}, \bibinfo {author} {\bibfnamefont {D.}~\bibnamefont {Kocon}},
  \bibinfo {author} {\bibfnamefont {J.}~\bibnamefont {Koski}}, \bibinfo
  {author} {\bibfnamefont {P.}~\bibnamefont {Kostamo}}, \bibinfo {author}
  {\bibfnamefont {M.}~\bibnamefont {Kumar}}, \bibinfo {author} {\bibfnamefont
  {T.}~\bibnamefont {Laeven}}, \bibinfo {author} {\bibfnamefont
  {T.}~\bibnamefont {Larsen}}, \bibinfo {author} {\bibfnamefont
  {K.}~\bibnamefont {Li}}, \bibinfo {author} {\bibfnamefont {T.}~\bibnamefont
  {Lindemann}}, \bibinfo {author} {\bibfnamefont {J.}~\bibnamefont {Love}},
  \bibinfo {author} {\bibfnamefont {R.}~\bibnamefont {Lutchyn}}, \bibinfo
  {author} {\bibfnamefont {M.}~\bibnamefont {Manfra}}, \bibinfo {author}
  {\bibfnamefont {E.}~\bibnamefont {Memisevic}}, \bibinfo {author}
  {\bibfnamefont {C.}~\bibnamefont {Nayak}}, \bibinfo {author} {\bibfnamefont
  {B.}~\bibnamefont {Nijholt}}, \bibinfo {author} {\bibfnamefont {M.~H.}\
  \bibnamefont {Madsen}}, \bibinfo {author} {\bibfnamefont {S.}~\bibnamefont
  {Markussen}}, \bibinfo {author} {\bibfnamefont {E.}~\bibnamefont {Martinez}},
  \bibinfo {author} {\bibfnamefont {R.}~\bibnamefont {McNeil}}, \bibinfo
  {author} {\bibfnamefont {A.}~\bibnamefont {Mullally}}, \bibinfo {author}
  {\bibfnamefont {J.}~\bibnamefont {Nielsen}}, \bibinfo {author} {\bibfnamefont
  {A.}~\bibnamefont {Nurmohamed}}, \bibinfo {author} {\bibfnamefont
  {E.}~\bibnamefont {O'Farrell}}, \bibinfo {author} {\bibfnamefont
  {K.}~\bibnamefont {Otani}}, \bibinfo {author} {\bibfnamefont
  {S.}~\bibnamefont {Pauka}}, \bibinfo {author} {\bibfnamefont
  {K.}~\bibnamefont {Petersson}}, \bibinfo {author} {\bibfnamefont
  {L.}~\bibnamefont {Petit}}, \bibinfo {author} {\bibfnamefont
  {D.}~\bibnamefont {Pikulin}}, \bibinfo {author} {\bibfnamefont
  {F.}~\bibnamefont {Preiss}}, \bibinfo {author} {\bibfnamefont {M.~Q.}\
  \bibnamefont {Perez}}, \bibinfo {author} {\bibfnamefont {K.}~\bibnamefont
  {Rasmussen}}, \bibinfo {author} {\bibfnamefont {M.}~\bibnamefont {Rajpalke}},
  \bibinfo {author} {\bibfnamefont {D.}~\bibnamefont {Razmadze}}, \bibinfo
  {author} {\bibfnamefont {O.}~\bibnamefont {Reentila}}, \bibinfo {author}
  {\bibfnamefont {D.}~\bibnamefont {Reilly}}, \bibinfo {author} {\bibfnamefont
  {R.}~\bibnamefont {Rouse}}, \bibinfo {author} {\bibfnamefont
  {I.}~\bibnamefont {Sadovskyy}}, \bibinfo {author} {\bibfnamefont
  {L.}~\bibnamefont {Sainiemi}}, \bibinfo {author} {\bibfnamefont
  {S.}~\bibnamefont {Schreppler}}, \bibinfo {author} {\bibfnamefont
  {V.}~\bibnamefont {Sidorkin}}, \bibinfo {author} {\bibfnamefont
  {A.}~\bibnamefont {Singh}}, \bibinfo {author} {\bibfnamefont
  {S.}~\bibnamefont {Singh}}, \bibinfo {author} {\bibfnamefont
  {S.}~\bibnamefont {Sinha}}, \bibinfo {author} {\bibfnamefont
  {P.}~\bibnamefont {Sohr}}, \bibinfo {author} {\bibfnamefont {T.}~\bibnamefont
  {Stankevič}}, \bibinfo {author} {\bibfnamefont {L.}~\bibnamefont {Stek}},
  \bibinfo {author} {\bibfnamefont {H.}~\bibnamefont {Suominen}}, \bibinfo
  {author} {\bibfnamefont {J.}~\bibnamefont {Suter}}, \bibinfo {author}
  {\bibfnamefont {V.}~\bibnamefont {Svidenko}}, \bibinfo {author}
  {\bibfnamefont {S.}~\bibnamefont {Teicher}}, \bibinfo {author} {\bibfnamefont
  {M.}~\bibnamefont {Temuerhan}}, \bibinfo {author} {\bibfnamefont
  {N.}~\bibnamefont {Thiyagarajah}}, \bibinfo {author} {\bibfnamefont
  {R.}~\bibnamefont {Tholapi}}, \bibinfo {author} {\bibfnamefont
  {M.}~\bibnamefont {Thomas}}, \bibinfo {author} {\bibfnamefont
  {E.}~\bibnamefont {Toomey}}, \bibinfo {author} {\bibfnamefont
  {S.}~\bibnamefont {Upadhyay}}, \bibinfo {author} {\bibfnamefont
  {I.}~\bibnamefont {Urban}}, \bibinfo {author} {\bibfnamefont
  {S.}~\bibnamefont {Vaitiekėnas}}, \bibinfo {author} {\bibfnamefont
  {K.}~\bibnamefont {Van~Hoogdalem}}, \bibinfo {author} {\bibfnamefont {D.~V.}\
  \bibnamefont {Viazmitinov}}, \bibinfo {author} {\bibfnamefont
  {S.}~\bibnamefont {Waddy}}, \bibinfo {author} {\bibfnamefont
  {D.}~\bibnamefont {Van~Woerkom}}, \bibinfo {author} {\bibfnamefont
  {D.}~\bibnamefont {Vogel}}, \bibinfo {author} {\bibfnamefont
  {J.}~\bibnamefont {Watson}}, \bibinfo {author} {\bibfnamefont
  {J.}~\bibnamefont {Weston}}, \bibinfo {author} {\bibfnamefont {G.~W.}\
  \bibnamefont {Winkler}}, \bibinfo {author} {\bibfnamefont {C.~K.}\
  \bibnamefont {Yang}}, \bibinfo {author} {\bibfnamefont {S.}~\bibnamefont
  {Yau}}, \bibinfo {author} {\bibfnamefont {D.}~\bibnamefont {Yi}}, \bibinfo
  {author} {\bibfnamefont {E.}~\bibnamefont {Yucelen}}, \bibinfo {author}
  {\bibfnamefont {A.}~\bibnamefont {Webster}}, \bibinfo {author} {\bibfnamefont
  {R.}~\bibnamefont {Zeisel}},\ and\ \bibinfo {author} {\bibfnamefont
  {R.}~\bibnamefont {Zhao}},\ }\href
  {https://doi.org/10.48550/ARXIV.2207.02472} {\bibinfo {title} {Inas-al hybrid
  devices passing the topological gap protocol}} (\bibinfo {year} {2022}),\
  \Eprint {https://arxiv.org/abs/2207.02472} {arXiv:2207.02472
  [cond-mat.mes-hall]} \BibitemShut {NoStop}%
\bibitem [{\citenamefont {{Maiellaro, Alfonso}}\ \emph
  {et~al.}(2018)\citenamefont {{Maiellaro, Alfonso}}, \citenamefont {{Romeo,
  Francesco}},\ and\ \citenamefont {{Citro, Roberta}}}]{ladder1}%
  \BibitemOpen
  \bibfield  {author} {\bibinfo {author} {\bibnamefont {{Maiellaro, Alfonso}}},
  \bibinfo {author} {\bibnamefont {{Romeo, Francesco}}},\ and\ \bibinfo
  {author} {\bibnamefont {{Citro, Roberta}}},\ }\bibfield  {title} {\bibinfo
  {title} {Topological phase diagram of a kitaev ladder},\ }\href
  {https://doi.org/10.1140/epjst/e2018-800090-y} {\bibfield  {journal}
  {\bibinfo  {journal} {Eur. Phys. J. Special Topics}\ }\textbf {\bibinfo
  {volume} {227}},\ \bibinfo {pages} {1397} (\bibinfo {year}
  {2018})}\BibitemShut {NoStop}%
\bibitem [{\citenamefont {Wakatsuki}\ \emph {et~al.}(2014)\citenamefont
  {Wakatsuki}, \citenamefont {Ezawa},\ and\ \citenamefont {Nagaosa}}]{ladder2}%
  \BibitemOpen
  \bibfield  {author} {\bibinfo {author} {\bibfnamefont {R.}~\bibnamefont
  {Wakatsuki}}, \bibinfo {author} {\bibfnamefont {M.}~\bibnamefont {Ezawa}},\
  and\ \bibinfo {author} {\bibfnamefont {N.}~\bibnamefont {Nagaosa}},\
  }\bibfield  {title} {\bibinfo {title} {Majorana fermions and multiple
  topological phase transition in kitaev ladder topological superconductors},\
  }\href {https://doi.org/10.1103/PhysRevB.89.174514} {\bibfield  {journal}
  {\bibinfo  {journal} {Phys. Rev. B}\ }\textbf {\bibinfo {volume} {89}},\
  \bibinfo {pages} {174514} (\bibinfo {year} {2014})}\BibitemShut {NoStop}%
\bibitem [{\citenamefont {Yang}\ \emph {et~al.}(2020)\citenamefont {Yang},
  \citenamefont {Perrin}, \citenamefont {Petrescu}, \citenamefont {Garate},\
  and\ \citenamefont {Le~Hur}}]{Yang_2020}%
  \BibitemOpen
  \bibfield  {author} {\bibinfo {author} {\bibfnamefont {F.}~\bibnamefont
  {Yang}}, \bibinfo {author} {\bibfnamefont {V.}~\bibnamefont {Perrin}},
  \bibinfo {author} {\bibfnamefont {A.}~\bibnamefont {Petrescu}}, \bibinfo
  {author} {\bibfnamefont {I.}~\bibnamefont {Garate}},\ and\ \bibinfo {author}
  {\bibfnamefont {K.}~\bibnamefont {Le~Hur}},\ }\bibfield  {title} {\bibinfo
  {title} {From topological superconductivity to quantum hall states in coupled
  wires\\},\ }\bibfield  {journal} {\bibinfo  {journal} {Physical Review B}\
  }\textbf {\bibinfo {volume} {101}},\ \href {https://doi.org/085116} {085116}
  (\bibinfo {year} {2020})\BibitemShut {NoStop}%
\bibitem [{\citenamefont {Katsura}\ \emph {et~al.}(2015)\citenamefont
  {Katsura}, \citenamefont {Schuricht},\ and\ \citenamefont
  {Takahashi}}]{Katsura_2015}%
  \BibitemOpen
  \bibfield  {author} {\bibinfo {author} {\bibfnamefont {H.}~\bibnamefont
  {Katsura}}, \bibinfo {author} {\bibfnamefont {D.}~\bibnamefont {Schuricht}},\
  and\ \bibinfo {author} {\bibfnamefont {M.}~\bibnamefont {Takahashi}},\
  }\bibfield  {title} {\bibinfo {title} {Exact ground states and topological
  order in interacting kitaev/majorana chains},\ }\bibfield  {journal}
  {\bibinfo  {journal} {Physical Review B}\ }\textbf {\bibinfo {volume} {92}},\
  \href {https://doi.org/10.1103/physrevb.92.115137}
  {10.1103/physrevb.92.115137} (\bibinfo {year} {2015})\BibitemShut {NoStop}%
\bibitem [{\citenamefont {Stoudenmire}\ \emph {et~al.}(2011)\citenamefont
  {Stoudenmire}, \citenamefont {Alicea}, \citenamefont {Starykh},\ and\
  \citenamefont {Fisher}}]{Alicea_interacting}%
  \BibitemOpen
  \bibfield  {author} {\bibinfo {author} {\bibfnamefont {E.~M.}\ \bibnamefont
  {Stoudenmire}}, \bibinfo {author} {\bibfnamefont {J.}~\bibnamefont {Alicea}},
  \bibinfo {author} {\bibfnamefont {O.~A.}\ \bibnamefont {Starykh}},\ and\
  \bibinfo {author} {\bibfnamefont {M.~P.}\ \bibnamefont {Fisher}},\ }\bibfield
   {title} {\bibinfo {title} {Interaction effects in topological
  superconducting wires supporting majorana fermions},\ }\href
  {https://doi.org/10.1103/PhysRevB.84.014503} {\bibfield  {journal} {\bibinfo
  {journal} {Phys. Rev. B}\ }\textbf {\bibinfo {volume} {84}},\ \bibinfo
  {pages} {014503} (\bibinfo {year} {2011})}\BibitemShut {NoStop}%
\bibitem [{\citenamefont {Ledermann}\ and\ \citenamefont
  {Le~Hur}(2000)}]{Ledermann_2000}%
  \BibitemOpen
  \bibfield  {author} {\bibinfo {author} {\bibfnamefont {U.}~\bibnamefont
  {Ledermann}}\ and\ \bibinfo {author} {\bibfnamefont {K.}~\bibnamefont
  {Le~Hur}},\ }\bibfield  {title} {\bibinfo {title} {Phases of the two-band
  model of spinless fermions in one dimension\\},\ }\href
  {https://doi.org/10.1103/physrevb.61.2497} {\bibfield  {journal} {\bibinfo
  {journal} {Physical Review B}\ }\textbf {\bibinfo {volume} {61}},\ \bibinfo
  {pages} {2497} (\bibinfo {year} {2000})}\BibitemShut {NoStop}%
\bibitem [{\citenamefont {Steinberg}\ \emph {et~al.}(2008)\citenamefont
  {Steinberg}, \citenamefont {Barak}, \citenamefont {Yacoby}, \citenamefont
  {Pfeiffer}, \citenamefont {West}, \citenamefont {Halperin},\ and\
  \citenamefont {Le~Hur}}]{Chargefract_leHur}%
  \BibitemOpen
  \bibfield  {author} {\bibinfo {author} {\bibfnamefont {H.}~\bibnamefont
  {Steinberg}}, \bibinfo {author} {\bibfnamefont {G.}~\bibnamefont {Barak}},
  \bibinfo {author} {\bibfnamefont {A.}~\bibnamefont {Yacoby}}, \bibinfo
  {author} {\bibfnamefont {L.~N.}\ \bibnamefont {Pfeiffer}}, \bibinfo {author}
  {\bibfnamefont {K.~W.}\ \bibnamefont {West}}, \bibinfo {author}
  {\bibfnamefont {B.~I.}\ \bibnamefont {Halperin}},\ and\ \bibinfo {author}
  {\bibfnamefont {K.}~\bibnamefont {Le~Hur}},\ }\bibfield  {title} {\bibinfo
  {title} {Charge fractionalization in quantum wires},\ }\bibfield  {journal}
  {\bibinfo  {journal} {Nature Phys 4, 116–119}\ }\href
  {https://doi.org/10.1038/nphys810%2Fs42005-021-00641-0} {} (\bibinfo {year}
  {2008})\BibitemShut {NoStop}%
\bibitem [{\citenamefont {Hutchinson}\ and\ \citenamefont
  {Le~Hur}(2021)}]{Hutchinson_2021}%
  \BibitemOpen
  \bibfield  {author} {\bibinfo {author} {\bibfnamefont {J.}~\bibnamefont
  {Hutchinson}}\ and\ \bibinfo {author} {\bibfnamefont {K.}~\bibnamefont
  {Le~Hur}},\ }\bibfield  {title} {\bibinfo {title} {Quantum entangled
  fractional topology and curvatures},\ }\bibfield  {journal} {\bibinfo
  {journal} {Commun Phys}\ }\textbf {\bibinfo {volume} {4}},\ \href
  {https://doi.org/144} {144} (\bibinfo {year} {2021})\BibitemShut {NoStop}%
\bibitem [{\citenamefont {Le~Hur}(2022{\natexlab{a}})}]{lehur_new}%
  \BibitemOpen
  \bibfield  {author} {\bibinfo {author} {\bibfnamefont {K.}~\bibnamefont
  {Le~Hur}},\ }\href {https://doi.org/10.48550/ARXIV.2209.15381} {\bibinfo
  {title} {Topological matter and fractional entangled geometry}} (\bibinfo
  {year} {2022}{\natexlab{a}}),\ \Eprint {https://arxiv.org/abs/2209.15381}
  {arXiv:2209.15381 [cond-mat.mes-hall]} \BibitemShut {NoStop}%
\bibitem [{\citenamefont {Sato}\ and\ \citenamefont {Ando}(2017)}]{Sato_2017}%
  \BibitemOpen
  \bibfield  {author} {\bibinfo {author} {\bibfnamefont {M.}~\bibnamefont
  {Sato}}\ and\ \bibinfo {author} {\bibfnamefont {Y.}~\bibnamefont {Ando}},\
  }\bibfield  {title} {\bibinfo {title} {Topological superconductors: a
  review},\ }\href {https://doi.org/10.1088/1361-6633/aa6ac7} {\bibfield
  {journal} {\bibinfo  {journal} {Reports on Progress in Physics}\ }\textbf
  {\bibinfo {volume} {80}},\ \bibinfo {pages} {076501} (\bibinfo {year}
  {2017})}\BibitemShut {NoStop}%
\bibitem [{\citenamefont {White}(1992)}]{PhysRevLett.69.2863}%
  \BibitemOpen
  \bibfield  {author} {\bibinfo {author} {\bibfnamefont {S.~R.}\ \bibnamefont
  {White}},\ }\bibfield  {title} {\bibinfo {title} {Density matrix formulation
  for quantum renormalization groups},\ }\href
  {https://doi.org/10.1103/PhysRevLett.69.2863} {\bibfield  {journal} {\bibinfo
   {journal} {Phys. Rev. Lett.}\ }\textbf {\bibinfo {volume} {69}},\ \bibinfo
  {pages} {2863} (\bibinfo {year} {1992})}\BibitemShut {NoStop}%
\bibitem [{\citenamefont {White}(1993)}]{PhysRevB.48.10345}%
  \BibitemOpen
  \bibfield  {author} {\bibinfo {author} {\bibfnamefont {S.~R.}\ \bibnamefont
  {White}},\ }\bibfield  {title} {\bibinfo {title} {Density-matrix algorithms
  for quantum renormalization groups},\ }\href
  {https://doi.org/10.1103/PhysRevB.48.10345} {\bibfield  {journal} {\bibinfo
  {journal} {Phys. Rev. B}\ }\textbf {\bibinfo {volume} {48}},\ \bibinfo
  {pages} {10345} (\bibinfo {year} {1993})}\BibitemShut {NoStop}%
\bibitem [{\citenamefont {Schollwöck}(2011)}]{Schollw_ck_2011}%
  \BibitemOpen
  \bibfield  {author} {\bibinfo {author} {\bibfnamefont {U.}~\bibnamefont
  {Schollwöck}},\ }\bibfield  {title} {\bibinfo {title} {The density-matrix
  renormalization group in the age of matrix product states},\ }\href
  {https://doi.org/10.1016/j.aop.2010.09.012} {\bibfield  {journal} {\bibinfo
  {journal} {Annals of Physics}\ }\textbf {\bibinfo {volume} {326}},\ \bibinfo
  {pages} {96} (\bibinfo {year} {2011})}\BibitemShut {NoStop}%
\bibitem [{\citenamefont {Affleck}(1986)}]{PhysRevLett.56.746}%
  \BibitemOpen
  \bibfield  {author} {\bibinfo {author} {\bibfnamefont {I.}~\bibnamefont
  {Affleck}},\ }\bibfield  {title} {\bibinfo {title} {Universal term in the
  free energy at a critical point and the conformal anomaly},\ }\href
  {https://doi.org/10.1103/PhysRevLett.56.746} {\bibfield  {journal} {\bibinfo
  {journal} {Phys. Rev. Lett.}\ }\textbf {\bibinfo {volume} {56}},\ \bibinfo
  {pages} {746} (\bibinfo {year} {1986})}\BibitemShut {NoStop}%
\bibitem [{\citenamefont {Tsvelik}(2003)}]{tsvelik_2003}%
  \BibitemOpen
  \bibfield  {author} {\bibinfo {author} {\bibfnamefont {A.~M.}\ \bibnamefont
  {Tsvelik}},\ }\href {https://doi.org/10.1017/CBO9780511615832} {\emph
  {\bibinfo {title} {Quantum Field Theory in Condensed Matter Physics}}},\
  \bibinfo {edition} {2nd}\ ed.\ (\bibinfo  {publisher} {Cambridge University
  Press},\ \bibinfo {year} {2003})\BibitemShut {NoStop}%
\bibitem [{\citenamefont {Gogolin}\ \emph {et~al.}(2004)\citenamefont
  {Gogolin}, \citenamefont {Nersesyan},\ and\ \citenamefont
  {Tsvelik}}]{Bosonization_and_strongly}%
  \BibitemOpen
  \bibfield  {author} {\bibinfo {author} {\bibfnamefont {A.~O.}\ \bibnamefont
  {Gogolin}}, \bibinfo {author} {\bibfnamefont {A.~A.}\ \bibnamefont
  {Nersesyan}},\ and\ \bibinfo {author} {\bibfnamefont {A.~M.}\ \bibnamefont
  {Tsvelik}},\ }\href@noop {} {\emph {\bibinfo {title} {Bosonization and
  Strongly Correlated Systems}}}\ (\bibinfo  {publisher} {Cambridge University
  Press},\ \bibinfo {year} {2004})\BibitemShut {NoStop}%
\bibitem [{\citenamefont {Itzykson}\ and\ \citenamefont
  {Drouffe}(1989)}]{itzykson_drouffe_1989}%
  \BibitemOpen
  \bibfield  {author} {\bibinfo {author} {\bibfnamefont {C.}~\bibnamefont
  {Itzykson}}\ and\ \bibinfo {author} {\bibfnamefont {J.-M.}\ \bibnamefont
  {Drouffe}},\ }\href {https://doi.org/10.1017/CBO9780511622779} {\emph
  {\bibinfo {title} {Statistical Field Theory}}},\ \bibinfo {series} {Cambridge
  Monographs on Mathematical Physics}, Vol.~\bibinfo {volume} {1}\ (\bibinfo
  {publisher} {Cambridge University Press},\ \bibinfo {year}
  {1989})\BibitemShut {NoStop}%
\bibitem [{\citenamefont {Le~Hur}(2022{\natexlab{b}})}]{Le_Hur_2022}%
  \BibitemOpen
  \bibfield  {author} {\bibinfo {author} {\bibfnamefont {K.}~\bibnamefont
  {Le~Hur}},\ }\bibfield  {title} {\bibinfo {title} {Global and local
  topological quantized responses from geometry, light, and time},\ }\href
  {https://doi.org/10.1103/physrevb.105.125106} {\bibfield  {journal} {\bibinfo
   {journal} {Physical Review B}\ }\textbf {\bibinfo {volume} {105}},\ \bibinfo
  {pages} {125106} (\bibinfo {year} {2022}{\natexlab{b}})}\BibitemShut
  {NoStop}%
\bibitem [{\citenamefont {Berry}(1984)}]{Berry}%
  \BibitemOpen
  \bibfield  {author} {\bibinfo {author} {\bibfnamefont {M.~V.}\ \bibnamefont
  {Berry}},\ }\bibfield  {title} {\bibinfo {title} {Quantal phase factors
  accompanying adiabatic changes},\ }\href
  {https://doi.org/10.1098/rspa.1984.0023} {\bibfield  {journal} {\bibinfo
  {journal} {Proceedings of the Royal Society A}\ }\textbf {\bibinfo {volume}
  {392}},\ \bibinfo {pages} {45} (\bibinfo {year} {1984})}\BibitemShut
  {NoStop}%
\bibitem [{\citenamefont {Kaufmann}\ \emph {et~al.}(2016)\citenamefont
  {Kaufmann}, \citenamefont {Li},\ and\ \citenamefont
  {Wehefritz-Kaufmann}}]{Kaufmann_2016}%
  \BibitemOpen
  \bibfield  {author} {\bibinfo {author} {\bibfnamefont {R.~M.}\ \bibnamefont
  {Kaufmann}}, \bibinfo {author} {\bibfnamefont {D.}~\bibnamefont {Li}},\ and\
  \bibinfo {author} {\bibfnamefont {B.}~\bibnamefont {Wehefritz-Kaufmann}},\
  }\bibfield  {title} {\bibinfo {title} {Notes on topological insulators},\
  }\href {https://doi.org/10.1142/s0129055x1630003x} {\bibfield  {journal}
  {\bibinfo  {journal} {Reviews in Mathematical Physics}\ }\textbf {\bibinfo
  {volume} {28}},\ \bibinfo {pages} {1630003} (\bibinfo {year}
  {2016})}\BibitemShut {NoStop}%
\bibitem [{\citenamefont {Kane}\ and\ \citenamefont {Mele}(2005)}]{Kane_2005}%
  \BibitemOpen
  \bibfield  {author} {\bibinfo {author} {\bibfnamefont {C.~L.}\ \bibnamefont
  {Kane}}\ and\ \bibinfo {author} {\bibfnamefont {E.~J.}\ \bibnamefont
  {Mele}},\ }\bibfield  {title} {\bibinfo {title} {${Z}_{2}$ topological order
  and the quantum spin hall effect},\ }\href
  {https://doi.org/10.1103/PhysRevLett.95.146802} {\bibfield  {journal}
  {\bibinfo  {journal} {Phys. Rev. Lett.}\ }\textbf {\bibinfo {volume} {95}},\
  \bibinfo {pages} {146802} (\bibinfo {year} {2005})}\BibitemShut {NoStop}%
\bibitem [{\citenamefont {Fu}\ \emph {et~al.}(2007)\citenamefont {Fu},
  \citenamefont {Kane},\ and\ \citenamefont {Mele}}]{Fu_2007}%
  \BibitemOpen
  \bibfield  {author} {\bibinfo {author} {\bibfnamefont {L.}~\bibnamefont
  {Fu}}, \bibinfo {author} {\bibfnamefont {C.~L.}\ \bibnamefont {Kane}},\ and\
  \bibinfo {author} {\bibfnamefont {E.~J.}\ \bibnamefont {Mele}},\ }\bibfield
  {title} {\bibinfo {title} {Topological insulators in three dimensions},\
  }\href {https://doi.org/10.1103/physrevlett.98.106803} {\bibfield  {journal}
  {\bibinfo  {journal} {Physical Review Letters}\ }\textbf {\bibinfo {volume}
  {98}},\ \bibinfo {pages} {106803} (\bibinfo {year} {2007})}\BibitemShut
  {NoStop}%
\bibitem [{\citenamefont {Chiu}\ \emph {et~al.}(2016)\citenamefont {Chiu},
  \citenamefont {Teo}, \citenamefont {Schnyder},\ and\ \citenamefont
  {Ryu}}]{Chiu_2016}%
  \BibitemOpen
  \bibfield  {author} {\bibinfo {author} {\bibfnamefont {C.-K.}\ \bibnamefont
  {Chiu}}, \bibinfo {author} {\bibfnamefont {J.~C.}\ \bibnamefont {Teo}},
  \bibinfo {author} {\bibfnamefont {A.~P.}\ \bibnamefont {Schnyder}},\ and\
  \bibinfo {author} {\bibfnamefont {S.}~\bibnamefont {Ryu}},\ }\bibfield
  {title} {\bibinfo {title} {Classification of topological quantum matter with
  symmetries},\ }\href {https://doi.org/10.1103/revmodphys.88.035005}
  {\bibfield  {journal} {\bibinfo  {journal} {Reviews of Modern Physics}\
  }\textbf {\bibinfo {volume} {88}},\ \bibinfo {pages} {035005} (\bibinfo
  {year} {2016})}\BibitemShut {NoStop}%
\bibitem [{\citenamefont {Altland}\ and\ \citenamefont
  {Zirnbauer}(1997)}]{Altland_1997}%
  \BibitemOpen
  \bibfield  {author} {\bibinfo {author} {\bibfnamefont {A.}~\bibnamefont
  {Altland}}\ and\ \bibinfo {author} {\bibfnamefont {M.~R.}\ \bibnamefont
  {Zirnbauer}},\ }\bibfield  {title} {\bibinfo {title} {Nonstandard symmetry
  classes in mesoscopic normal-superconducting hybrid structures},\ }\href
  {https://doi.org/10.1103/physrevb.55.1142} {\bibfield  {journal} {\bibinfo
  {journal} {Physical Review B}\ }\textbf {\bibinfo {volume} {55}},\ \bibinfo
  {pages} {1142} (\bibinfo {year} {1997})}\BibitemShut {NoStop}%
\bibitem [{\citenamefont {Kitaev}\ \emph {et~al.}(2009)\citenamefont {Kitaev},
  \citenamefont {Lebedev},\ and\ \citenamefont {Feigel'man}}]{Kitaev_2009}%
  \BibitemOpen
  \bibfield  {author} {\bibinfo {author} {\bibfnamefont {A.}~\bibnamefont
  {Kitaev}}, \bibinfo {author} {\bibfnamefont {V.}~\bibnamefont {Lebedev}},\
  and\ \bibinfo {author} {\bibfnamefont {M.}~\bibnamefont {Feigel'man}},\
  }\bibfield  {title} {\bibinfo {title} {Periodic table for topological
  insulators and superconductors},\ }in\ \href
  {https://doi.org/10.1063/1.3149495} {\emph {\bibinfo {booktitle} {{AIP}
  Conference Proceedings}}}\ (\bibinfo  {publisher} {{AIP}},\ \bibinfo {year}
  {2009})\BibitemShut {NoStop}%
\bibitem [{\citenamefont {Herviou}\ \emph {et~al.}(2017)\citenamefont
  {Herviou}, \citenamefont {Mora},\ and\ \citenamefont
  {Le~Hur}}]{Herviou_2017}%
  \BibitemOpen
  \bibfield  {author} {\bibinfo {author} {\bibfnamefont {L.}~\bibnamefont
  {Herviou}}, \bibinfo {author} {\bibfnamefont {C.}~\bibnamefont {Mora}},\ and\
  \bibinfo {author} {\bibfnamefont {K.}~\bibnamefont {Le~Hur}},\ }\bibfield
  {title} {\bibinfo {title} {Bipartite charge fluctuations in one-dimensional
  $\mathbb{Z}_{2}$ superconductors and insulators},\ }\bibfield  {journal}
  {\bibinfo  {journal} {Physical Review B}\ }\textbf {\bibinfo {volume} {96}},\
  \href {https://doi.org/121113(R)} {121113(R)} (\bibinfo {year}
  {2017})\BibitemShut {NoStop}%
\bibitem [{\citenamefont {Anderson}(1958)}]{PhysRev.112.1900}%
  \BibitemOpen
  \bibfield  {author} {\bibinfo {author} {\bibfnamefont {P.~W.}\ \bibnamefont
  {Anderson}},\ }\bibfield  {title} {\bibinfo {title} {Random-phase
  approximation in the theory of superconductivity},\ }\href
  {https://doi.org/10.1103/PhysRev.112.1900} {\bibfield  {journal} {\bibinfo
  {journal} {Phys. Rev.}\ }\textbf {\bibinfo {volume} {112}},\ \bibinfo {pages}
  {1900} (\bibinfo {year} {1958})}\BibitemShut {NoStop}%
\bibitem [{Note1()}]{Note1}%
  \BibitemOpen
  \bibinfo {note} {Ie. for each individual sphere}\BibitemShut {NoStop}%
\bibitem [{\citenamefont {Chen}(2023)}]{nonlocal1}%
  \BibitemOpen
  \bibfield  {author} {\bibinfo {author} {\bibfnamefont {W.}~\bibnamefont
  {Chen}},\ }\bibfield  {title} {\bibinfo {title} {Universal topological
  marker},\ }\href {https://doi.org/10.1103/PhysRevB.107.045111} {\bibfield
  {journal} {\bibinfo  {journal} {Phys. Rev. B}\ }\textbf {\bibinfo {volume}
  {107}},\ \bibinfo {pages} {045111} (\bibinfo {year} {2023})}\BibitemShut
  {NoStop}%
\bibitem [{\citenamefont {Molignini}\ \emph {et~al.}(2022)\citenamefont
  {Molignini}, \citenamefont {Lapierre}, \citenamefont {Chitra},\ and\
  \citenamefont {Chen}}]{nonlocal2}%
  \BibitemOpen
  \bibfield  {author} {\bibinfo {author} {\bibfnamefont {P.}~\bibnamefont
  {Molignini}}, \bibinfo {author} {\bibfnamefont {B.}~\bibnamefont {Lapierre}},
  \bibinfo {author} {\bibfnamefont {R.}~\bibnamefont {Chitra}},\ and\ \bibinfo
  {author} {\bibfnamefont {W.}~\bibnamefont {Chen}},\ }\href
  {https://doi.org/10.48550/ARXIV.2207.00016} {\bibinfo {title} {Measurement
  and topological quantum criticality of local and nonlocal chern markers}}
  (\bibinfo {year} {2022})\BibitemShut {NoStop}%
\bibitem [{\citenamefont {Fishman}\ \emph {et~al.}(2020)\citenamefont
  {Fishman}, \citenamefont {White},\ and\ \citenamefont
  {Stoudenmire}}]{https://doi.org/10.48550/arxiv.2007.14822}%
  \BibitemOpen
  \bibfield  {author} {\bibinfo {author} {\bibfnamefont {M.}~\bibnamefont
  {Fishman}}, \bibinfo {author} {\bibfnamefont {S.~R.}\ \bibnamefont {White}},\
  and\ \bibinfo {author} {\bibfnamefont {E.~M.}\ \bibnamefont {Stoudenmire}},\
  }\href {https://doi.org/10.48550/ARXIV.2007.14822} {\bibinfo {title} {The
  itensor software library for tensor network calculations}} (\bibinfo {year}
  {2020})\BibitemShut {NoStop}%
\bibitem [{Note2()}]{Note2}%
  \BibitemOpen
  \bibinfo {note} {For an extensive list of papers using the ITensor library
  see http://itensor.org/docs.cgi?page=papers}\BibitemShut {NoStop}%
\bibitem [{Note3()}]{Note3}%
  \BibitemOpen
  \bibinfo {note} {And the two Quantum Critical Points (QCPs), as will be
  discussed in detail in \ref {sec:Fractional}.}\BibitemShut {Stop}%
\bibitem [{\citenamefont {Fidkowski}\ and\ \citenamefont
  {Kitaev}(2010)}]{Fidkowski_2010}%
  \BibitemOpen
  \bibfield  {author} {\bibinfo {author} {\bibfnamefont {L.}~\bibnamefont
  {Fidkowski}}\ and\ \bibinfo {author} {\bibfnamefont {A.}~\bibnamefont
  {Kitaev}},\ }\bibfield  {title} {\bibinfo {title} {The effects of
  interactions on the topological classification of free fermion systems},\
  }\href {https://doi.org/10.1103/physrevb.81.134509} {\bibfield  {journal}
  {\bibinfo  {journal} {Physical Review B}\ }\textbf {\bibinfo {volume} {81}},\
  \bibinfo {pages} {134509} (\bibinfo {year} {2010})}\BibitemShut {NoStop}%
\bibitem [{\citenamefont {Le~Hur}(1999)}]{Karyn1999}%
  \BibitemOpen
  \bibfield  {author} {\bibinfo {author} {\bibfnamefont {K.}~\bibnamefont
  {Le~Hur}},\ }\bibfield  {title} {\bibinfo {title} {Critical ising modes in
  low-dimensional kondo insulators},\ }\href
  {https://doi.org/https://journals.aps.org/prb/abstract/10.1103/PhysRevB.60.9116}
  {\bibfield  {journal} {\bibinfo  {journal} {Phys. Rev. B}\ }\textbf {\bibinfo
  {volume} {60}},\ \bibinfo {pages} {9116} (\bibinfo {year}
  {1999})}\BibitemShut {NoStop}%
\bibitem [{\citenamefont {Calabrese}\ and\ \citenamefont
  {Cardy}(2004)}]{Calabrese_2004}%
  \BibitemOpen
  \bibfield  {author} {\bibinfo {author} {\bibfnamefont {P.}~\bibnamefont
  {Calabrese}}\ and\ \bibinfo {author} {\bibfnamefont {J.}~\bibnamefont
  {Cardy}},\ }\bibfield  {title} {\bibinfo {title} {Entanglement entropy and
  quantum field theory},\ }\href
  {https://doi.org/10.1088/1742-5468/2004/06/p06002} {\bibfield  {journal}
  {\bibinfo  {journal} {Journal of Statistical Mechanics: Theory and
  Experiment}\ }\textbf {\bibinfo {volume} {2004}},\ \bibinfo {pages} {P06002}
  (\bibinfo {year} {2004})}\BibitemShut {NoStop}%
\bibitem [{\citenamefont {Song}\ \emph {et~al.}(2010)\citenamefont {Song},
  \citenamefont {Rachel},\ and\ \citenamefont {Le~Hur}}]{Song_2010}%
  \BibitemOpen
  \bibfield  {author} {\bibinfo {author} {\bibfnamefont {H.~F.}\ \bibnamefont
  {Song}}, \bibinfo {author} {\bibfnamefont {S.}~\bibnamefont {Rachel}},\ and\
  \bibinfo {author} {\bibfnamefont {K.}~\bibnamefont {Le~Hur}},\ }\bibfield
  {title} {\bibinfo {title} {General relation between entanglement and
  fluctuations in one dimension},\ }\href
  {https://doi.org/10.1103/physrevb.82.012405} {\bibfield  {journal} {\bibinfo
  {journal} {Physical Review B}\ }\textbf {\bibinfo {volume} {82}},\ \bibinfo
  {pages} {012405} (\bibinfo {year} {2010})}\BibitemShut {NoStop}%
\bibitem [{\citenamefont {Sénéchal}(1999)}]{Senechal}%
  \BibitemOpen
  \bibfield  {author} {\bibinfo {author} {\bibfnamefont {D.}~\bibnamefont
  {Sénéchal}},\ }\href {https://doi.org/10.48550/ARXIV.COND-MAT/9908262}
  {\bibinfo {title} {An introduction to bosonization,}} (\bibinfo {year}
  {1999}),\ \Eprint {https://arxiv.org/abs/9908262} {arXiv:9908262
  [cond-mat.str-el]} \BibitemShut {NoStop}%
\bibitem [{\citenamefont {Giamarchi}(2004)}]{giamarchi2004quantum}%
  \BibitemOpen
  \bibfield  {author} {\bibinfo {author} {\bibfnamefont {T.}~\bibnamefont
  {Giamarchi}},\ }\href {https://books.google.fr/books?id=1MwTDAAAQBAJ} {\emph
  {\bibinfo {title} {Quantum Physics in One Dimension}}},\ International Series
  of Monographs on Physics\ (\bibinfo  {publisher} {Clarendon Press},\ \bibinfo
  {year} {2004})\BibitemShut {NoStop}%
\bibitem [{\citenamefont {Orignac}\ and\ \citenamefont
  {Giamarchi}(1998)}]{WeaklydisorderedSpinLadders}%
  \BibitemOpen
  \bibfield  {author} {\bibinfo {author} {\bibfnamefont {E.}~\bibnamefont
  {Orignac}}\ and\ \bibinfo {author} {\bibfnamefont {T.}~\bibnamefont
  {Giamarchi}},\ }\bibfield  {title} {\bibinfo {title} {Weakly disordered spin
  ladders},\ }\href {https://doi.org/10.1103/PhysRevB.57.5812} {\bibfield
  {journal} {\bibinfo  {journal} {Phys. Rev. B}\ }\textbf {\bibinfo {volume}
  {57}},\ \bibinfo {pages} {5812} (\bibinfo {year} {1998})}\BibitemShut
  {NoStop}%
\bibitem [{\citenamefont {Haldane}(1981)}]{Haldane}%
  \BibitemOpen
  \bibfield  {author} {\bibinfo {author} {\bibfnamefont {F.~D.~M.}\
  \bibnamefont {Haldane}},\ }\bibfield  {title} {\bibinfo {title} {'luttinger
  liquid theory' of one-dimensional quantum fluids. i. properties of the
  luttinger model and their extension to the general 1d interacting spinless
  fermi gas},\ }\href
  {https://doi.org/https://iopscience.iop.org/article/10.1088/0022-3719/14/19/010/meta}
  {\bibfield  {journal} {\bibinfo  {journal} {J. Phys. C: Solid State Phys}\
  }\textbf {\bibinfo {volume} {14}},\ \bibinfo {pages} {9116} (\bibinfo {year}
  {1981})}\BibitemShut {NoStop}%
\bibitem [{Note4()}]{Note4}%
  \BibitemOpen
  \bibinfo {note} {The choice in \cite {Herviou_2016} results in a single wire
  of a \protect \emph {doubled} number of sites $N$ at constant length $L =
  Na$. Thus, one finds $2a^{\prime } = a$, which results in a doubled
  interaction strength $g$.}\BibitemShut {Stop}%
\bibitem [{\citenamefont {Reeg}\ \emph {et~al.}(2018)\citenamefont {Reeg},
  \citenamefont {Dmytruk}, \citenamefont {Chevallier}, \citenamefont {Loss},\
  and\ \citenamefont {Klinovaja}}]{Olesia_andreev}%
  \BibitemOpen
  \bibfield  {author} {\bibinfo {author} {\bibfnamefont {C.}~\bibnamefont
  {Reeg}}, \bibinfo {author} {\bibfnamefont {O.}~\bibnamefont {Dmytruk}},
  \bibinfo {author} {\bibfnamefont {D.}~\bibnamefont {Chevallier}}, \bibinfo
  {author} {\bibfnamefont {D.}~\bibnamefont {Loss}},\ and\ \bibinfo {author}
  {\bibfnamefont {J.}~\bibnamefont {Klinovaja}},\ }\bibfield  {title} {\bibinfo
  {title} {Zero-energy andreev bound states from quantum dots in proximitized
  rashba nanowires},\ }\href {https://doi.org/10.1103/PhysRevB.98.245407}
  {\bibfield  {journal} {\bibinfo  {journal} {Phys. Rev. B}\ }\textbf {\bibinfo
  {volume} {98}},\ \bibinfo {pages} {245407} (\bibinfo {year}
  {2018})}\BibitemShut {NoStop}%
\bibitem [{\citenamefont {Prüfer~M.}(2020)}]{Zache_snapshots}%
  \BibitemOpen
  \bibfield  {author} {\bibinfo {author} {\bibfnamefont {K.~P.}\ \bibnamefont
  {Prüfer~M.}, \bibfnamefont {Zache T.V. et~al.}},\ }\bibfield  {title}
  {\bibinfo {title} {Experimental extraction of the quantum effective action
  for a non-equilibrium many-body system.},\ }\href
  {https://doi.org/10.1103/physrevb.103.214502} {\bibfield  {journal} {\bibinfo
   {journal} {Nat. Phys.}\ }\textbf {\bibinfo {volume} {16}},\ \bibinfo {pages}
  {1012} (\bibinfo {year} {2020})}\BibitemShut {NoStop}%
\bibitem [{\citenamefont {Zache}\ \emph {et~al.}(2020)\citenamefont {Zache},
  \citenamefont {Schweigler}, \citenamefont {Erne}, \citenamefont
  {Schmiedmayer},\ and\ \citenamefont {Berges}}]{ETcfs_1PI}%
  \BibitemOpen
  \bibfield  {author} {\bibinfo {author} {\bibfnamefont {T.~V.}\ \bibnamefont
  {Zache}}, \bibinfo {author} {\bibfnamefont {T.}~\bibnamefont {Schweigler}},
  \bibinfo {author} {\bibfnamefont {S.}~\bibnamefont {Erne}}, \bibinfo {author}
  {\bibfnamefont {J.}~\bibnamefont {Schmiedmayer}},\ and\ \bibinfo {author}
  {\bibfnamefont {J.}~\bibnamefont {Berges}},\ }\bibfield  {title} {\bibinfo
  {title} {Extracting the field theory description of a quantum many-body
  system from experimental data},\ }\href
  {https://doi.org/10.1103/PhysRevX.10.011020} {\bibfield  {journal} {\bibinfo
  {journal} {Phys. Rev. X}\ }\textbf {\bibinfo {volume} {10}},\ \bibinfo
  {pages} {011020} (\bibinfo {year} {2020})}\BibitemShut {NoStop}%
\bibitem [{\citenamefont {Fiete}\ \emph {et~al.}(2005)\citenamefont {Fiete},
  \citenamefont {Qian}, \citenamefont {Tserkovnyak},\ and\ \citenamefont
  {Halperin}}]{Momentum}%
  \BibitemOpen
  \bibfield  {author} {\bibinfo {author} {\bibfnamefont {G.~A.}\ \bibnamefont
  {Fiete}}, \bibinfo {author} {\bibfnamefont {J.}~\bibnamefont {Qian}},
  \bibinfo {author} {\bibfnamefont {Y.}~\bibnamefont {Tserkovnyak}},\ and\
  \bibinfo {author} {\bibfnamefont {B.~I.}\ \bibnamefont {Halperin}},\
  }\bibfield  {title} {\bibinfo {title} {Theory of momentum resolved tunneling
  into a short quantum wire},\ }\href
  {https://doi.org/https://doi.org/10.1103/PhysRevB.72.045315} {\bibfield
  {journal} {\bibinfo  {journal} {Phys. Rev. B}\ }\textbf {\bibinfo {volume}
  {72}},\ \bibinfo {pages} {045315} (\bibinfo {year} {2005})}\BibitemShut
  {NoStop}%
\bibitem [{\citenamefont {Le~Hur}(2006)}]{MomentumKLH}%
  \BibitemOpen
  \bibfield  {author} {\bibinfo {author} {\bibfnamefont {K.}~\bibnamefont
  {Le~Hur}},\ }\bibfield  {title} {\bibinfo {title} {The electron lifetime in
  luttinger liquids},\ }\href
  {https://doi.org/https://doi.org/10.1103/PhysRevB.74.165104} {\bibfield
  {journal} {\bibinfo  {journal} {Phys. Rev. B}\ }\textbf {\bibinfo {volume}
  {74}},\ \bibinfo {pages} {165104} (\bibinfo {year} {2006})}\BibitemShut
  {NoStop}%
\bibitem [{\citenamefont {{Le}~Hur}\ and\ \citenamefont
  {Rice}(2009)}]{KarynMaurice}%
  \BibitemOpen
  \bibfield  {author} {\bibinfo {author} {\bibfnamefont {K.}~\bibnamefont
  {{Le}~Hur}}\ and\ \bibinfo {author} {\bibfnamefont {T.~M.}\ \bibnamefont
  {Rice}},\ }\bibfield  {title} {\bibinfo {title} {Superconductivity close to
  the mott state: From condensed-matter systems to superfluidity in optical
  lattices},\ }\href
  {https://doi.org/https://doi.org/10.1016/j.aop.2009.02.004} {\bibfield
  {journal} {\bibinfo  {journal} {Annals of Physics}\ }\textbf {\bibinfo
  {volume} {324}},\ \bibinfo {pages} {1452} (\bibinfo {year}
  {2009})}\BibitemShut {NoStop}%
\bibitem [{\citenamefont {{Pfeuty}}(1970)}]{Pfeuty}%
  \BibitemOpen
  \bibfield  {author} {\bibinfo {author} {\bibfnamefont {P.}~\bibnamefont
  {{Pfeuty}}},\ }\bibfield  {title} {\bibinfo {title} {{The one-dimensional
  Ising model with a transverse field}},\ }\href
  {https://doi.org/10.1016/0003-4916(70)90270-8} {\bibfield  {journal}
  {\bibinfo  {journal} {Annals of Physics}\ }\textbf {\bibinfo {volume} {57}},\
  \bibinfo {pages} {79} (\bibinfo {year} {1970})}\BibitemShut {NoStop}%
\bibitem [{Note5()}]{Note5}%
  \BibitemOpen
  \bibinfo {note} {Ie. where no transition to \protect \emph {Mott} physics has
  occurred.}\BibitemShut {Stop}%
\bibitem [{\citenamefont {Pan}\ \emph {et~al.}(2021)\citenamefont {Pan},
  \citenamefont {Liu}, \citenamefont {Wimmer},\ and\ \citenamefont
  {Sarma}}]{Pan_2021}%
  \BibitemOpen
  \bibfield  {author} {\bibinfo {author} {\bibfnamefont {H.}~\bibnamefont
  {Pan}}, \bibinfo {author} {\bibfnamefont {C.-X.}\ \bibnamefont {Liu}},
  \bibinfo {author} {\bibfnamefont {M.}~\bibnamefont {Wimmer}},\ and\ \bibinfo
  {author} {\bibfnamefont {S.~D.}\ \bibnamefont {Sarma}},\ }\bibfield  {title}
  {\bibinfo {title} {Quantized and unquantized zero-bias tunneling conductance
  peaks in majorana nanowires: Conductance below and above $2e^{2}/h$},\
  }\bibfield  {journal} {\bibinfo  {journal} {Physical Review B}\ }\textbf
  {\bibinfo {volume} {103}},\ \href
  {https://doi.org/10.1103/physrevb.103.214502} {10.1103/physrevb.103.214502}
  (\bibinfo {year} {2021})\BibitemShut {NoStop}%
\bibitem [{\citenamefont {Chiara}\ \emph {et~al.}(2006)\citenamefont {Chiara},
  \citenamefont {Rizzi}, \citenamefont {Rossini},\ and\ \citenamefont
  {Montangero}}]{Chiara2006DensityMR}%
  \BibitemOpen
  \bibfield  {author} {\bibinfo {author} {\bibfnamefont {G.~D.}\ \bibnamefont
  {Chiara}}, \bibinfo {author} {\bibfnamefont {M.}~\bibnamefont {Rizzi}},
  \bibinfo {author} {\bibfnamefont {D.}~\bibnamefont {Rossini}},\ and\ \bibinfo
  {author} {\bibfnamefont {S.}~\bibnamefont {Montangero}},\ }\bibfield  {title}
  {\bibinfo {title} {Density matrix renormalization group for dummies},\
  }\href@noop {} {\bibfield  {journal} {\bibinfo  {journal} {Journal of
  Computational and Theoretical Nanoscience}\ }\textbf {\bibinfo {volume}
  {5}},\ \bibinfo {pages} {1277} (\bibinfo {year} {2006})}\BibitemShut
  {NoStop}%
\bibitem [{\citenamefont {Schollwöck}(2005)}]{Schollw_ck_2005}%
  \BibitemOpen
  \bibfield  {author} {\bibinfo {author} {\bibfnamefont {U.}~\bibnamefont
  {Schollwöck}},\ }\bibfield  {title} {\bibinfo {title} {The density-matrix
  renormalization group},\ }\href {https://doi.org/10.1103/revmodphys.77.259}
  {\bibfield  {journal} {\bibinfo  {journal} {Reviews of Modern Physics}\
  }\textbf {\bibinfo {volume} {77}},\ \bibinfo {pages} {259} (\bibinfo {year}
  {2005})}\BibitemShut {NoStop}%
\bibitem [{\citenamefont {von Delft}\ and\ \citenamefont
  {Schoeller}(1998)}]{https://doi.org/10.1002/andp.19985100401}%
  \BibitemOpen
  \bibfield  {author} {\bibinfo {author} {\bibfnamefont {J.}~\bibnamefont {von
  Delft}}\ and\ \bibinfo {author} {\bibfnamefont {H.}~\bibnamefont
  {Schoeller}},\ }\bibfield  {title} {\bibinfo {title} {Bosonization for
  beginners — refermionization for experts},\ }\href
  {https://doi.org/https://doi.org/10.1002/andp.19985100401} {\bibfield
  {journal} {\bibinfo  {journal} {Annalen der Physik}\ }\textbf {\bibinfo
  {volume} {510}},\ \bibinfo {pages} {225} (\bibinfo {year}
  {1998})}\BibitemShut {NoStop}%
\end{thebibliography}%

\end{document}